\documentclass[12pt]{article}
\usepackage[margin=1in]{geometry}
\usepackage[utf8]{inputenc}
\usepackage{authblk}
\usepackage{amsfonts, amssymb, amsthm, amsmath, mathrsfs, mathtools}
\usepackage{natbib, enumitem, setspace, multicol}
\usepackage{graphicx, float, caption, subcaption}
\usepackage{bm}
\usepackage{verbatim}
\usepackage{pdfpages}
\usepackage{fancyvrb}
\usepackage[colorlinks = true,
            linkcolor = blue,
            urlcolor  = blue,
            citecolor = blue,
            anchorcolor = blue]{hyperref}
\usepackage[linesnumbered,ruled]{algorithm2e}
\usepackage{algorithmic}
\usepackage{listings}
\usepackage{natbib}
\newtheorem{lemma}{Lemma}[section]

\newtheorem{theorem}{Theorem}[section]

\newtheorem{assm}{Assumption}[section]

\newtheorem*{result*}{Result}
\newtheorem*{remark*}{Remark}
\newtheorem*{claim*}{Claim}

\theoremstyle{definition}

\usepackage{bm}

\newcommand{\CSWLASSOBH}[1][]{\text{CSW}_{\text{LASSO}^{#1}}\text{--BH}}
\newcommand{\CSWdSTBH}[1][]{\text{CSW}_{\text{dST}^{#1}}\text{--BH}}
\newcommand{\CSWLASSOHolm}[1][]{\text{CSW}_{\text{LASSO}^{#1}}\text{--Holm}}
\newcommand{\CSWdSTHolm}[1][]{\text{CSW}_{\text{dST}^{#1}}\text{--Holm}}

\allowdisplaybreaks
\numberwithin{equation}{section}

\title{Co-sufficient weighting: leveraging sparsity for more powerful variable selection in the linear model}
\author{Souhardya Sengupta}
\author{Lucas Janson}
\affil{Department of Statistics, Harvard University}
\date{ }

\begin{document}
\maketitle

\begin{abstract}
    %One established way to gain power in multiple testing is to incorporate exogenous (quantitative) side information (e.g., an associated covariate for each hypothesis that is independent of the p-values) into weights on the . 
    We show that for the problem of controlled variable selection in the Gaussian linear model, informative and p-value-independent weights (for weighted multiple testing) can be derived entirely from sufficient statistics and a belief in sparsity using only the data itself and no external quantitative side information. %constructed entirely from the data itself without quantitative side information and instead using only , 
    \color{black}
    Using these weights, which we call \emph{co-sufficient weights} (CSW), in weighted versions of Holm's procedure and the Benjamini--Hochberg (BH) procedure results in substantially higher power than their unweighted analogues when the true coefficient vector is sparse, while retaining Holm's exact validity and BH's highly robust empirical validity regardless of sparsity. We demonstrate the superior performance of our proposed weighted methods compared to their unweighted analogues and other competing methods in a wide range of simulations and an application to HIV drug resistance.
    \color{black}
\end{abstract}

\section{Introduction}
 \label{sec:intro}

\subsection{Motivation}

Under the setting of the homoskedastic Gaussian linear model
\begin{align}
    \bm y \sim \mathcal N\left(\bm X\bm \beta, \sigma^2\bm I_n\right),
\end{align}
where $\bm X\in \mathbb R^{n\times d}$ is a full column-rank matrix, $\bm I_n$ denotes the $n\times n$ identity matrix and $\bm \beta \in \mathbb R^d,\sigma^2>0$ are unknown, this paper considers the problem of controlled variable selection, i.e., recovering the set of signal indices $\mathcal S = \{j:\beta_j\neq 0\}$ while controlling an error rate such as the familywise error rate (FWER) or false discovery rate (FDR). Variable selection in the linear model is one of the most canonical inference problems in science, with applications across many domains such as biology \citep{YAMASHITA20081414,gadetal2014} and economics \citep{belloni2012,KOZAK2020271}.

Often, 
analysts in these applications possess domain knowledge about properties of
%these applications present the analyst with side knowledge that induces 
%a belief about 
the coefficient vector $\bm \beta$, with a widely occurring example being knowledge of \emph{sparsity}, meaning that the analyst believes most entries of $\bm \beta$ are zero. %(or a form of approximate sparsity whereby a small fraction of the covariates explain most of the variance in $Y$ explained by $X$).
%, or equivalently, that most variables are noise. 
Such sparsity assumptions arise across many scientific domains, such as genetics \citep{Lucas_Carvalho_Wang_Bild_Nevins_West_2006}, econometrics \citep{annurev:/content/journals/10.1146/annurev-economics-061109-080451}, and chemistry \citep{https://doi.org/10.1002/cem.1418}. %\citep{Lucas_Carvalho_Wang_Bild_Nevins_West_2006,annurev:/content/journals/10.1146/annurev-economics-061109-080451,https://doi.org/10.1002/cem.1418}. 
Indeed, Section~\ref{sec:real_data} of this paper studies such a setting, where we apply variable selection methods to identify genetic mutations associated with resistance to an HIV drug.

In this paper, we show that such a structural belief, particularly sparsity, can be leveraged to obtain a more powerful variable selection procedure. A key aspect of this question is the inherent immeasurability of the structural belief---sparsity here reflects only the imprecise belief that 
a small fraction of the covariates account for the majority of the signal present in $\bm X$,
%\textit{only a few} variables are signals, 
and does not come with an explicit quantification, such as a known bound on the proportion of signal covariates or a fully specified prior distribution on $\bm\beta$. Thus, we develop methods that leverage sparsity in $\bm \beta$ solely to improve power. In particular, we propose a novel way to use the linear model to construct weights for use in multiple testing. With such weights, a certain weighted version of Holm's procedure controls the FWER regardless of sparsity, while a weighted version of BH, like unweighted BH,
%. Our FDR-controlling procedure, on the other hand, does not theoretically control the FDR, much like the standard BH procedure in this setting, 
exhibits robust empirical FDR control regardless of the sparsity level. Thus, an analyst can use our methods when they believe in sparsity (without having to explicitly quantify that belief) and will benefit when that belief is correct without having to worry about inflated errors if the belief is incorrect.

\subsection{Related Works}
Perhaps the simplest approach to variable selection in the linear model is to construct p-values for the individual hypotheses $H_j:\beta_j = 0$, for example using the $t$-test, and then plug these p-values into a multiple testing procedure. The literature on multiple testing is vast and we do not attempt to review it here (see \cite{Emmert-Streibetal2019,He2026} for recent reviews), though in Section~\ref{sec:weighted_mult_testing}, we will review some methods that are particularly relevant to our work.
% \textbf{the literature is vast; perhaps to canonical choice; 5 canonical papers; review papers; closure principle paper. One recent reviw is also fine---} 
Among the standard choices for multiple testing widely used for variable selection in the linear model are Holm's procedure \citep{holm} and the Benjamini--Hochberg procedure \citep{fdr}. However, these multiple testing algorithms have no mechanism for leveraging sparsity in $\bm \beta$ to improve power. %Multiple testing, in general, is a well discussed topic in the literature \textbf{cite}. In Section~\ref{sec:weighted_mult_testing}, we will also review some methods that are particularly relevant to our work.

Existing methods in the literature that leverage sparsity often do so at the cost of additional assumptions. For FWER control, one class of methods applies stability selection \citep{meinshausen2010,shahandsamworth2013} to the LASSO estimator \citep{LASSO}, a well-known sparse estimator of $\bm \beta$ that performs variable selection by shrinking some coefficients exactly to zero. In this framework, variables are selected according to how frequently they are included across repeated LASSO fits on subsamples of the data. Stability selection can then bound the FWER of the resulting procedure, but only under substantial assumptions on $\bm \beta$. Another class of methods \citep{wassermanandroeder2009,meinshausenetal2009,Schultheissetal2021} relies on sample splitting---a set of variables is selected on one part of the data, typically with the LASSO, and only the selected variables are then tested, with a multiplicity correction, on the remaining part. These methods, however, require additional assumptions ensuring that all signal variables are selected in the first step, which rarely happens in practice. \cite{gselletal2016} also discuss a LASSO-based method for FWER control, but at the cost of a similar assumption that the signals enter the LASSO path before any noise variable.

% To our knowledge, despite the continued practical 
% importance of FWER,
% this remains the only existing variable selection approach that targets FWER control while leveraging sparsity in $\bm \beta$.

% For FDR control, several works exploit properties of the LASSO to obtain FDR-controlling procedures. A notable example is the method of \citet{gselletal2016}, which controls the FDR under the assumption that signal variables enter the model before null variables along the LASSO path. This approach exploits sparsity through the effective ordering of variables induced by the LASSO. However, the assumption that all signal variables are selected before any null variables is often violated under general design matrices. A related method is SLOPE \citep{bogdanetal2015}, which replaces the LASSO's $\ell_1$-penalty with a sequence of penalties based on the ranks of the coefficients. While SLOPE achieves FDR control under orthogonal design matrices, this guarantee does not extend to more general designs. \citet{javanmardandjavadi2019} use debiased-LASSO-based statistics to propose an FDR-controlling procedure. However, their guarantees are asymptotic and rely on assumptions on the distributional tails of the design matrix rows, as well as on the proportion of signals.

For FDR control, several works also exploit properties of the LASSO to obtain FDR-controlling procedures; examples include \citep{bogdanetal2015,gselletal2016}. However, these methods again rely on the same ordering assumption that signals enter the LASSO path before noise variables. Another example is \citet{javanmardandjavadi2019}, who propose a multiple testing procedure for the linear model based on the debiased LASSO \citep{zhangandzhang2014,JMLR:v15:javanmard14a,debiasedLASSO}. Their primary objective, however, is to extend the BH procedure applied to $t$-test p-values to the $d>n$ setting; in particular, when $d\leq n$, their procedure reduces to the original BH procedure. To our knowledge, the only class of methods that achieves FDR control without such additional assumptions while explicitly leveraging sparsity is the fixed-X knockoffs framework \citep{knockoff}, along with subsequent refinements that improve its performance \citep{AS-LJ:2020, luo2022, renandbarber(2023), lee2024boosting}. Knockoff-based procedures introduce synthetic negative controls for each variable and construct importance measures by comparing each variable to its corresponding control. Sparsity can then be exploited through LASSO-based importance measures. However, this framework also has several limitations. It requires augmenting the design matrix with synthetic variables, effectively doubling its dimension; this can reduce power and, more importantly, imposes the constraint $n\ge 2d$ for fixed-X knockoffs to be applicable. The choice of knockoff variables also introduces ambiguity such that two analysts applying the same method to the same data could get different answers.
%, which may affect the reproducibility of discoveries. 
%Moreover, these methods do not naturally extend to FWER control \citep{LJ-WS:2016}.
More broadly, knockoffs belongs to a larger class of synthetic variable methods targeting FDR control
%based on the construction of synthetic null variables 
\citep{Wu01032007,junfdrdatasplit,gaussmirror,oatk,molinari2025falsediscoveryratecontrol} that rely on discrete comparisons between real and synthetic variables to make selections. Aside from fixed-X knockoffs, the methods in this class require additional assumptions beyond the Gaussian linear model to control the FDR, which we see in our simulations can result in significant violations. %Although we do not investigate the restrictiveness of these assumptions, we empirically assess the performance of some of these methods by comparing them with our proposed procedures.  %, preventing them from leveraging the wealth of which renders them

\subsection{Summary of our contributions}
\label{sec:contributions}

We develop novel methods for exploiting sparsity to improve the power of controlled variable selection in the linear model when $n> d$ (unlike knockoff-based methods, we do not require $n\geq 2d$). 
%All our methods apply to any full-column-rank design matrix with $n\geq d$; in particular, they are not constrained by the requirement $n\geq 2d$, as in knockoff-type methods. 
Our methods are weighted multiple testing procedures for the hypotheses $\{H_j:\beta_j=0\}_{j=1}^d$, where each hypothesis $H_j$ is assigned a weight $w_j$. Our primary contribution is a data-dependent choice of these weights, which we call \textit{co-sufficient weights} (CSW): we show that a belief in sparsity alone, without any additional side information, can be used to construct weights that are independent of their corresponding p-values and that tend to be larger for variables likely to be non-null. The construction leverages the sufficient statistic under each $H_j$ and combines it with a LASSO-type estimator to exploit sparsity.

Our two proposed multiple testing procedures incorporate these weights into weighted versions of Holm's procedure and the BH procedure, respectively, %giving rise to CSW-Holm and CSW-BH, respectively, 
and are substantially more powerful than their unweighted analogues when the true coefficient vector is sparse. The weighted version of Holm's procedure guarantees exact FWER control, regardless of sparsity. The weighted BH procedure, like the ordinary BH procedure in the general linear model setting, does not have strictly guarantee FDR control but empirically exhibits highly robust FDR control in practice. 
%Exact FDR control with our weights is possible by incorporating them into the dBH procedure \citep{dbh}, though at a prohibitively higher computational cost. This tradeoff is why we still recommend the weighted BH procedure over this alternative. 
Extensive simulations and an application to HIV drug resistance demonstrate our methods' power gains in sparse settings over methods that do not exploit sparsity, as well as their outperformance of other sparsity-exploiting methods.

% that our methods control their respective error rates, and are more powerful than existing variable selection procedures for the linear model across a variety of settings.

\section{Background: Weighted multiple testing}
\label{sec:weighted_mult_testing}

Our proposed variable selection procedures are constructed by applying weighted multiple testing with a novel, carefully chosen set of weights. We therefore begin with a review of weighted multiple testing procedures. Suppose that we have $d$ hypotheses $\{H_j\}_{j=1}^d$ to test, with corresponding p-values $\{p_j\}_{j=1}^d$, where $p_j$ is super-uniform under $H_j$; define $\mathcal H_0:= \{j:H_j \text{ is true}\}$ as the set of true nulls. In addition, suppose that each hypothesis $H_j$ has an associated weight $w_j$.% and, to begin with, assume the $w_j$ are non-random and $\sum_{j=1}^dw_j = 1$. 

\subsection{Methods for FWER control}
\label{sec:fwer_methods}
To begin with, assume the $w_j$ are non-random and $\sum_{j=1}^dw_j = 1$. A canonical example is the weighted analogue of the Bonferroni procedure, which rejects all hypotheses $j$ such that $p_j\leq \alpha w_j$, for a pre-specified nominal FWER level $\alpha\in (0,1)$. Larger weights for hypotheses that are not true translate to higher cutoffs, with weights larger than $1/d$ making such hypotheses easier to be discovered than under the unweighted Bonferroni procedure. The Bonferroni procedure can be uniformly improved by Holm's procedure \citep{holm}, which preserves FWER control and which likewise can be improved by a weighted analog \cite[Scheme 2]{holm}. In fact, a weighted version of Holm's procedure can be constructed as the closure \citep{marcusetal1976} of a weighted Bonferroni test. We describe this construction next as it motivates our FWER controlling procedure in Section~\ref{sec:fwer}.%If the weights are larger for non-null hypotheses, this procedure makes it easier for them to be discovered by lowering the cutoff. 

First, for any $I\subseteq [1:d]$, consider the %\textit{local} 
intersection hypothesis $H_{I}:=\cap_{j\in I} H_j$ and the weighted $I$-Bonferroni test that rejects $H_{I}$ if 
\begin{align}
    \label{eqn:local_rule}
    p_j \leq \alpha\cdot \frac{w_j}{\sum_{i\in I}w_i}\;\textrm{ for any }j \in I.
\end{align}
Then the closure of these weighted $I$-Bonferroni tests is a multiple testing procedure for the original $\{H_j\}_{j=1}^d$ that rejects $H_j$ if and only if $H_I$ is rejected whenever $j\in I$. This closure test is a weighted version of Holm's procedure and, because taking a closure uniformly improves the power \citep{romanoetal2009}, is at least as powerful than the original weighted Bonferroni procedure. Notably, weighted Holm's procedure admits a sequential representation that can be executed by examining at most $d$ subsets of indices $I$. %In fact, instead of examining the outcomes of all the local tests $H_I$, this closure procedure admits a much simpler, sequential representation as a weighted analogue of the celebrated Holm's procedure \citep{holm}; see \cite[Scheme 2]{holm} for this sequential representation that takes at most $d$ steps to execute.
% \footnote{Recall that the closure of a class of tests for the local intersection hypotheses $\left\{H_{I}=\cap_{j\in I}H_j: I\subseteq [1:d]\right\}$ rejects $H_i$ if and only if $H_{I}$ is rejected whenever $i\in I$.}\textbf{cite} procedure to local Bonferroni tests, which reject $\cap_{j\in I} H_j$ for any $I \subseteq [1:d]$ if $p_i < q\cdot \frac{w_i}{\sum_{j\in I}w_j}$, for some $i \in I$. The resulting procedure is a weighted analogue of the celebrated Holm's procedure \citep{holm}; like Holm's procedure, it admits a sequential representation (see \cite[Scheme 2]{holm}) that can be computed by examining at most $d$ local sets of indices $I$.

Recent work has extended validity of the weighted Bonferroni procedure to the case when the weights $w_j$ are random, under the following assumption (see \cite[Theorem 4.3]{ignatiadisetal2023}).
\begin{assm}\
    \label{assm:bonf_valid}
    %The following conditions holds:
    \begin{enumerate}[label = (\alph*)]
        \item $w_j$ is independent of $p_j$ for all $j\in \mathcal H_0$.
        \item $\mathbb E\left[\sum_{j\in \mathcal H_0}w_j\right] \leq 1$.
    \end{enumerate}
\end{assm}
 Unfortunately, it is not clear that the random-weighted Bonferroni procedure can be improved via the closure principle above under Assumption~\ref{assm:bonf_valid}, since Assumption~\ref{assm:bonf_valid}(a)'s independence condition will in general fail for the weighted $I$-Bonferroni tests due to the weight renormalization in Equation~\eqref{eqn:local_rule}. In Section~\ref{sec:csw}, we will show how to overcome this roadblock for the linear model, whose sufficient statistics can be leveraged to construct valid weights for these weighted $I$-Bonferroni tests. %show that this limitation can be bypassed. 
 This leads to a weighted version of Holm's procedure with random weights\footnote{As discussed in \cite{ignatiadisetal2023}, one can also construct a different weighted version of Holm's procedure when the weights are e-values. However, this assumption on the weights is restrictive and, in particular, does not hold in our case.}, which, like its non-random counterpart, admits a sequential representation, is computationally inexpensive, and is at least as powerful as the random-weighted Bonferroni procedure.

\subsection{Methods for FDR control}
\label{sec:fdr_methods}
%Weighted multiple testing procedures can also target FDR control. 
Among the earliest works in weighted multiple testing for FDR control is \cite{genovese2006}, which introduced a weighted version of the BH procedure by applying BH to the scaled p-values ${p_j / (d w_j)}$. The validity of this approach assumes non-random weights that sum to 1, but \cite{ignatiadisetal2023} show that validity also holds for random weights satisfying Assumption~\ref{assm:bonf_valid} and the following additional condition.
\begin{assm}\ 
    \label{assm:bh_validity}
    %The following conditions hold:
    \begin{enumerate}[label = (\alph*)]
        \item The set $\{(p_j,w_j)\}_{j \in \mathcal H_0}$ is independent of the set $\{(p_j,w_j)\}_{j \notin \mathcal H_0}$.% are independent of each other. 
        \item The $d$ pairs $(p_j,w_j)$ are mutually independent with each other across $j\in \mathcal H_0$.
    \end{enumerate}
\end{assm}
 % However, neither the classical BH procedure nor any of the weighted counterparts discussed above can theoretically control the FDR under the general dependence structure of linear regression p-values\footnote{Note that the classical BH can control FDR with dependent p-values, but the only known sufficient condition is when they exhibit a special positive dependence structure called PRDS \citep{BY2001}. This condition is again not generally satisfied by linear regression p-values.}. Hence, regardless of the construction of the weights, these procedures do not generally guarantee FDR control under the Gaussian linear model. For exact FDR control under arbitrary dependence, one can instead resort to the recently introduced dBH and related procedures \citep{dbh}. These methods calibrate a rejection threshold for each p-value in a way that ensures valid FDR control. This calibration relies on knowledge of a sufficient statistic under each null hypothesis $H_j$, which is well suited to the linear model, since such a sufficient statistic is readily available (see Section~\ref{sec:csw}). The authors also note that their framework can, in principle, incorporate weights, although this direction is not developed further. 
 However, neither the unweighted BH procedure nor its weighted counterpart controls the FDR under the general dependence structure of linear regression p-values \citep{dobriban2026benjaminihochbergprocedurefailcontrol,lei2026gaussiandependenceinflatebenjaminihochberg}\footnote{Note that the unweighted BH procedure controls the FDR with dependent p-values under a certain class of positive dependence structure called PRDS \citep{BY2001}, however this condition is not generally satisfied by linear regression p-values.}, which violate Assumption~\ref{assm:bh_validity} no matter the choice of weights. 
 %Regardless of the construction of the weights, the linear model p-values themselves generally do not allow for Assumption~\ref{assm:bh_validity} to be satisfied. 
 Yet the BH procedure is widely known to exhibit robust empirical FDR control under dependence structures encountered in practice, to the extent that many authors have conjectured its strict validity for correlated two-sided $z$-tests \citep{reiner-benaim2007,benjamini2010,sarkarandzhang2025}; in fact, only recently have examples of FDR violation been produced \cite{dobriban2026benjaminihochbergprocedurefailcontrol,lei2026gaussiandependenceinflatebenjaminihochberg}. Although Assumption~\ref{assm:bh_validity} does not seem to be necessary for practical FDR control, we emphasize that Assumption~\ref{assm:bonf_valid} very much is, including its item (a) (note that unweighted BH satisfies Assumption~\ref{assm:bonf_valid} with its weights $w_j=1/d$).
 %We believe that a crucial reason behind its robust empirical control is the fact that it satisfies Assumption~\ref{assm:bonf_valid} (with weights $w_j = 1/d$), particularly its item (a). 
 Without Assumption~\ref{assm:bonf_valid}(a), weights can be chosen to yield severely inflated errors; see Appendix~\ref{sec:counterexample} for an example that satisfies Assumption~\ref{assm:bonf_valid}(b) and Assumption~\ref{assm:bh_validity}, yet exhibits severe FDR violations. 
%  For example, under the global null with uniform null p-values, if one chooses $w_j = \frac{p_j}{\alpha}\mathbb I(p_j\leq t)$, then $p_j/(dw_j) = \alpha/d$ whenever $p_j\leq t$ and is wrongly rejected. 
% % For example, under the global null with $d$ uniform null p-values, if one chooses $w_j = \frac{1}{d\alpha}\mathbb I(p_j\leq \alpha)$ (which does satisfy Assumption~\ref{assm:bonf_valid}(a) and, if the $p_j$ are independent, Assumption~\ref{assm:bh_validity} as well), then it is easy to see that $H_j$ is rejected for any $p_j\le\alpha$, which happens for at least one $j$ with probability quickly approaching 1 as $d$ increases, yielding an FDR also approaching 1.
%  $p_j/(dw_j) = \alpha/d$ whenever $p_j\leq t$ and is wrongly rejected. Under global null the FDR is given by $\mathbb P(\exists j:p_j\leq t) = 1-(1-t)^d$. Choosing $t = \sqrt{2\alpha/d}$ shows $\mathbb P(\exists j:p_j\leq t)\to 1,$ as $d\to \infty$, while $\mathbb E_{H_j}[w_j] = 1$. Thus, for large enough $d$ this choice would cause severe FDR violation while still satisfying Assumption~\ref{assm:bonf_valid} item (b) and interestingly, Assumption~\ref{assm:bh_validity} as well.

Thus, since Assumption~\ref{assm:bh_validity} is unsatisfiable in general for linear regression and apparently unneccesary for practical FDR control, while Assumption~\ref{assm:bonf_valid} is critical for practical FDR control (and sufficient for strict FWER control as described in Section~\ref{sec:fwer_methods}), our focus in this paper will be on constructing weights that satisfy Assumption~\ref{assm:bonf_valid}. Indeed, as expected, we find in Section~\ref{sec:expt} that the weighted BH procedure of \cite{genovese2006}, with linear model p-values and using the weights we propose in Section~\ref{sec:methodology},
%(which are closely related to two-sided $z$-test p-values)
controls the FDR empirically quite robustly.
%shares this empirical robustness, while avoiding the computationally intensive step of conditional calibration.
 
 If strict FDR control in the linear model is desired, one can instead leverage conditional calibration \citep{dbh}. %These methods calibrate a rejection threshold for each p-value in a way that ensures valid FDR control. This calibration relies on knowledge of a sufficient statistic under each null hypothesis $H_j$, which is well suited to the linear model, since such a sufficient statistic is readily available (see Section~\ref{sec:csw}). 
 %The authors also note that their framework can, in principle, incorporate weights, although this direction is not developed further. 
%Although these methods provide exact guarantees, they are significantly more computationally intensive. In particular, the calibration step requires solving an equation whose evaluation at any point in its domain involves computing a conditional expectation. While \citet{dbh} provide computational shortcuts for faster root-finding, these shortcuts do not apply in our setting (with the weights), where the conditional expectation must instead be estimated using expensive Monte Carlo sampling. In Appendix~\ref{sec:app_dbh}, we provide further details explaining why this approach is practically infeasible in our setting.
Unfortunately, conditional calibration in general can come with a large computational cost due to the need to evaluate conditional expectations many times, and we find it to be prohibitive for our procedure; see Appendix~\ref{sec:app_dbh} for details.
 
% Despite lacking strict FDR control, the BH procedure is widely known to exhibit robust empirical FDR control under dependence structures encountered in practice, to the extent that many authors have conjectured its strict validity for correlated two-sided $z$-tests \citep{reiner-benaim2007,benjamini2010,sarkarandzhang2025}; in fact, only recently have examples of FDR violation been produced \cite{dobriban2026benjaminihochbergprocedurefailcontrol,lei2026gaussiandependenceinflatebenjaminihochberg}.
%for such counter-examples as well as a discussion on the robust control.
%In practice, however, the BH procedure applied to linear model p-values is widely known to exhibit robust empirical FDR control, despite lacking theoretical guarantees. 
% In particular, we find that the weighted BH procedure of \cite{genovese2006}, with linear model p-values and using the weights we propose in Section~\ref{sec:methodology},
% %(which are closely related to two-sided $z$-test p-values)
% similarly controls the FDR empirically quite robustly.
% %shares this empirical robustness, while avoiding the computationally intensive step of conditional calibration.

\subsection{Choice of weights}
For weighted multiple testing to improve power in practice, one needs a principled way to choose weights that both satisfy the above assumptions and tend to be larger for non-null variables. Much of the existing literature achieves this by translating external or quantitative side information into weights. Such side information may take the form of a Bayesian prior \citep{Dobriban_2015}, a group or hierarchical structure among the hypotheses \citep{hu2017,libarber2018}, an ordering of hypotheses by importance \citep{libarber2017}, or external covariates \citep{ignatiadis2021}. \cite{pfilter2017} provide a unified framework for incorporating many such types of side information into multiple testing. Related methods outside the weighted multiple-testing paradigm also leverage side information; for example, \cite{adapt,chao2021adaptgmm} do not directly construct weights, but instead use an iterative procedure to calibrate hypothesis-specific rejection thresholds based on the side information. These approaches, however, are not directly compatible with our setting, where no external or quantitative side information is available. Our only additional information is a belief that $\bm \beta$ is sparse, which is not in the form of a concrete, well-defined structure or the explicit quantitative or covariate-based side information required by these procedures. In the next section, we show that weights for powerful weighted multiple testing %weights satisfying the requirements stated above 
can nevertheless be constructed for the linear model without any external side information, by leveraging sparsity alone. 

\section{Powerful weighted multiple testing in the linear model}
\label{sec:methodology}
\subsection{Co-sufficient weighting}
\label{sec:csw}
Given the weighted multiple testing procedures reviewed in Section~\ref{sec:weighted_mult_testing}, our goal now reduces to constructing suitable weights for the linear model, which can then be plugged into these procedures. 
% To motivate our construction, we begin with FWER control, for which the weighted Holm's procedure is our preferred alternative. Recall also that it suffices to construct weights for the local weighted Bonferroni tests, whose closure yields the required Holm's procedure.

To construct these weights, we start with a non-negative variable-importance measure $\mathcal A_j(\bm y)$ for $\bm X_j$, where we suppress its notational dependence on the design $\bm X$ which is fixed throughout our discussion. When not specifically talking about any column index $j$, we write $\mathcal A$ to refer generically to a variable importance measure. We allow $\mathcal A$ to be a randomized function, i.e., its value can depend on external randomness independent of the data $(\bm y, \bm X)$ in addition to its inputs (e.g., it could use cross-validation with randomly chosen folds, or randomized algorithms like stochastic gradient descent or random forests).
%$\mathcal A(\bm y, \bm X;j)$ for $\bm X_j$, which we will often write as $\mathcal A_j(\bm y)$ to ease notation. 
Let $\bm S^{(j)} = (\bm X_{-j}^T\bm y, \|(\bm I_n - \bm P_{-j})\bm y\|^2)$ denote the minimal sufficient statistic for the linear model under $H_j$, where $\bm P_{-j}$ denotes the projection matrix onto the column space of $\bm X_{-j}$. See, for example, \cite[Appendix E.2]{luo2022} for a justification of sufficiency. Consider the following weights,
%We define our weights as
\begin{align*}
%\label{eqn:chosen_weights}
    w_j = \mathbb E_{H_j}\left[ \frac{\mathcal A_j(\bm y)}{\sum_{i=1}^d\mathcal A_i(\bm y)}\, \middle|\, \bm S^{(j)}\right], \qquad \forall j\in [1:d],
\end{align*}
% Because $w_{\mathcal A, j}$ is a function of $\bm S^{(j)}$, it is independent of any p-value $p_j$ for testing $H_j$ that is itself independent of $\bm S^{(j)}$ under $H_j$; for example, the $t$-test p-value satisfies this property.
%Unless otherwise stated, from here on, the notation $w_j$ will always refer to the above choice of weights. 
where we assume the convention that $0/0 = 0$ for the ratio inside the expectation. Unfortunately it will not generally be tractable to compute the above conditional expectation exactly, so we can instead consider its Monte Carlo estimate,
\begin{align}
\label{eqn:plug_in_w}
    \hat w_j = \frac{1}{B}\sum_{b=1}^B\frac{\mathcal A_j(\tilde{\bm y}^{(j)}_{b})}
     {\sum_{i = 1}^d\mathcal A_i(\tilde{\bm y}^{(j)}_{b})},
     \qquad \forall j\in [1:d],
    %=
    % \frac{1}{M}\sum_{r=1}^M
    % \frac{\mathcal A(\tilde{\bm y}_{r};j)}
    % {\sum_{i\in I}\mathcal A(\tilde{\bm y}_{r};i)},
    % \qquad \forall j\in I,
\end{align}
where the $\tilde{\bm y}^{(j)}_{b}$ are sampled i.i.d. from the distribution of $\bm y\mid \bm S^{(j)}$ under $H_j$, conditionally independently of $\bm y$; this so called \emph{co-sufficient} sampling \citep{CSS} is easily carried out for the linear model \citep{luo2022} and justifies calling the $\{\hat{w}_j\}_{j=1}^d$ \emph{co-sufficient weights} (CSW). Note that $\hat{w}_j$ has the same expectation as $w_j$ and is conditionally independent of $\bm y$ given $\bm S^{(j)}$. 

The key to CSW is that $\hat w_j$ only depends on the data through $\bm S^{(j)}$, making it easy to establish Assumption~\ref{assm:bonf_valid} because linear model p-values typically satisfy the following assumption:
%Next, we show that under a mild assumption on the p-values, co-sufficient weights satisfy Assumption~\ref{assm:bonf_valid}.
%Next, we assume a sufficient condition on the p-values $\{p_j\}_{j=1}^d$ which ensures Assumption~\ref{assm:bonf_valid} is satisfied for our weights, as Theorem~\ref{thm:validity_weights} establishes.
\begin{assm}
    \label{assm:p_val_suff}
    For any $j\in[1:d]$, the p-value $p_j$ for testing $H_j$ is independent of $\bm S^{(j)}$ under $H_j$.
\end{assm}
Assumption~\ref{assm:p_val_suff} is satisfied, e.g., by standard linear model $t$-test p-values, and makes the proof of the following theorem straightforward.
\begin{theorem}
    \label{thm:validity_weights}
    If the p-values $\{p_j\}_{j=1}^d$ satisfy Assumption~\ref{assm:p_val_suff}, then the weights $\{\hat w_j\}_{j=1}^d$ defined in Equation~\eqref{eqn:plug_in_w} satisfy Assumption~\ref{assm:bonf_valid}.
\end{theorem}
\begin{proof}
    Because $\hat w_{j}$ is conditionally independent of the data given $\bm S^{(j)}$, it is independent of any p-value $p_j$ that is itself independent of $\bm S^{(j)}$ under $H_j$. Furthermore, we have that 
    \begin{align*}
        \mathbb E\left[\sum_{j\in \mathcal H_0}\hat w_j\right] =& \mathbb E\left[\sum_{j\in \mathcal H_0}w_j\right] = \mathbb E\left[\sum_{j\in \mathcal H_0}\mathbb E_{H_j}\left[ \frac{\mathcal A_j(\bm y)}{\sum_{i=1}^d\mathcal A_i(\bm y)} \,\middle|\, \bm S^{(j)}\right]\right] = \sum_{j\in \mathcal H_0}\mathbb E_{H_j}\left[ \frac{\mathcal A_j(\bm y)}{\sum_{i=1}^d\mathcal A_i(\bm y)} \right]\\
        =& \mathbb E\left[\sum_{j\in \mathcal H_0}\frac{\mathcal A_j(\bm y)}{\sum_{i=1}^d\mathcal A_i(\bm y)}\right]\leq 1,
    \end{align*}
    thereby showing that both the conditions of Assumption~\ref{assm:bonf_valid} are satisfied.
\end{proof}
Theorem~\ref{thm:validity_weights} holds for any choice of $\mathcal{A}$, but we next suggest appealing choices for high power under sparsity and tractable computation.
%Note that Assumption~\ref{assm:p_val_suff} on the p-values is not stringent, and in particular is satisfied by standard linear model $t$-test p-values. %The above weights also naturally motivate the construction of weights for the local Bonferroni test; we discuss more on this in Section~\ref{sec:fwer}. 

%Finally, because the weights for $H_j$ are based on the conditional distribution of $\bm y\mid \bm S^{(j)}$ under $H_j$---sampling from which is commonly called \textit{co-sufficient} sampling in the literature \citep{CSS}---we call them \textit{co-sufficient weights}. We next discuss practical and powerful ways to implement these weights.
%consider the task of efficiently evaluating these weights followed by choice of the variable importance measures $\{\mathcal A_j\}_{j=1}^d$.%, followed by the task of efficiently evaluating the conditional expectation in \eqref{eqn:chosen_weights}.

\subsection{Sparsity-aware weights}
\label{sec:reccomendaion}

One design objective in choosing $\mathcal{A}_j$ is for $\hat{w}_j$ to have small conditional variance given $\bm S^{(j)}$, so that $B$ can be as small as possible, saving computation (note that if this conditional variance were zero, $\hat{w}_j=w_j$ for any $B\ge 1$). While it is not clear how to fully achieve this, it turns out we can get close by choosing $\mathcal{A}_j$ to only depend on $\bm y$ through $\bm S^{(j)}$. 
This makes $\hat{w}_j$'s numerator deterministic given $\bm S^{(j)}$, and although each denominator term with $i\neq j$ will be a function of $\bm S^{(i)}$ (and hence not deterministic given $\bm S^{(j)}$), $\mathcal{A}$ functions that measure variable importance in the $i$th variable generally do not vary much when resampling $\bm y$ conditional on $\bm S^{(j)}$, as such resampling holds $\bm X_i^T \bm y$ fixed. In fact, for the LASSO-based $\mathcal{A}$ we recommend in Section~\ref{sec:reccomendaion_LASSO}, we found in Appendix~\ref{sec:weights_variability} that the conditional variance of $\hat{w}_j$ is close to zero even for $B=1$. Appendix~\ref{sec:weights_variability} does show that the more scalable soft-thresholding-based $\mathcal{A}$ we propose in Section~\ref{sec:reccomendaion_dST} can be marginally but nontrivially improved by $B>1$, this $\mathcal{A}$'s main goal is computational efficiency and $B=1$ still performs well and is of course the fastest choice of $B$. Thus, for the remainder of the paper, we will always consider $B=1$, and in particular drop the subscript from $\tilde{\bm y}^{(j)}$ going forward.
To design a $\mathcal{A}$ that only depends on $\bm S^{(j)}$, we observe that the first $d-1$ components of $\bm S^{(j)}$, $\bm X_{-j}^T\bm y$, transformed by $(\bm X_{-j}^T\bm X_{-j})^{-1/2}$ to $\bm z^{(j)} := (\bm X_{-j}^T\bm X_{-j})^{-1/2} \bm X_{-j}^T\bm y$, follow a homoskedastic Gaussian linear model with coefficient vector $\bm \beta$ and a $(d-1)\times d$ design matrix given by $\bm W^{(j)} := (\bm X_{-j}^T\bm X_{-j})^{-1/2}\bm X_{-j}^T\bm X$:
\begin{align}
\label{eq:linmod}
    \bm z^{(j)} \sim \mathcal N\left(\bm W^{(j)}\bm \beta, \sigma^2\bm I_{d-1}\right).
\end{align}
%To leverage the sparsity of $\bm \beta$, we incorporate an $\ell_1$-penalization with maximum likelihood estimation based on the distribution of $\bm z$.
%it is natural to use an $\ell_1$-penalized maximum likelihood estimate of $\beta_j$ as a variable importance measure. 
As $\bm z^{(j)}$ is a function of $\bm S^{(j)}$, a natural way to construct $\mathcal{A}$ to only depend on $\bm S^{(j)}$ is to have it estimate $\beta_j$ from the above linear model~\eqref{eq:linmod}. Note this neglects the last component of $\bm S^{(j)}$, $\|(\bm I_n - \bm P_{-j})\bm y\|$, from informing $\mathcal{A}$; we justify this decision in Appendix~\ref{sec:app_conditioning_signal}. 

Equation~\eqref{eq:linmod} reveals a failure mode for any method trying to extract signal about $H_j$ from $\bm X_{-j}^T\bm y$: if $\bm X_j$ is orthogonal to the remaining columns of $\bm X$, then $\beta_j$ is unidentfiable from $\bm X_{-j}^T\bm y$, and in particular, $\bm W^{(j)}_j = \bm 0$. Thus, non-zero inner products between columns are critical for CSW to perform well, as these inner products are how useful information is transmitted to the weights. That said, columns having a zero inner product \emph{in expectation} are not a problem for CSW, and indeed Section~\ref{sec:expt} shows excellent performance even when the columns of $\bm X$ are all uncorrelated with one another.
%many simulations in which we show CSW's strong performance, , and $\beta_j$ is not identifiable from the linear model~\eqref{eq:linmod}. Consequently, no variable importance measure $\mathcal A_j$ based on $\bm z^{(j)}$ can be informative about $\beta_j$. Henceforth, we assume that no column of $\bm X$ corresponding to a hypothesis of interest is orthogonal to all the remaining columns. This assumption accommodates, for example, designs with an intercept and centered covariates, provided that we are interested in only multiple testing the effects of those covariates. In such cases, one should compute the p-values and weights only for the columns of interest and apply our procedures to this subset.

%Exact orthogonality can nevertheless arise in specialized settings, such as orthogonally designed experiments. Outside such settings, however, we regard a column being perfectly orthogonal to all others as a stylized occurrence, particularly in observational data. We therefore do not view this restriction as a substantial limitation for the applications motivating our work.

In the remainder of this subsection we propose concrete choices of $\mathcal{A}$ that estimate $\beta_j$ in the linear model \eqref{eq:linmod}.

\subsubsection{Our recommendation: a LASSO-based importance measure}
\label{sec:reccomendaion_LASSO}
To obtain our recommended importance measure, for $\lambda> 0$, define
\begin{align}
\label{eqn:lasso_objective}
\begin{aligned}
    \hat{\bm \beta}^{\lambda,j}
    &=
    \underset{\bm \beta}{\arg\min}\left(
    \frac{1}{2(d-1)}
    \|\bm z^{(j)} - \bm W^{(j)}\bm \beta\|^2
    +\lambda\sum_{i\neq j}|\beta_i|
    \right).
\end{aligned}
\end{align}
This is the LASSO estimator applied to the linear model \eqref{eq:linmod}, except $\beta_j$ is excluded from the penalty term.
%an $\ell_1$-penalized likelihood estimator of $\bm \beta$ based on the likelihood of $\bm z^{(j)}$ that exploits the sparsity in $\bm \beta$, while leaving the $j^{\mathrm{th}}$ coefficient (which our variable importance measure targets) unregularized. 
We make this modification because regularizing $\beta_j$ induces a positive probability of $\hat w_j$ being zero, in which case $H_j$ cannot be rejected by weighted multiple testing,
%which is detrimental for a signal $\beta_j$ as it completely removes the possibility of being discovered, 
no matter how small its p-value $p_j$ is, and indeed we found penalizing $\beta_j$ to hurt power empirically. 
%Likewise, we empirically found such a regularized version to deliver lower power than our recommended unregularized version.
% This is because regularizing it induces a non-zero probability of the $j^{\mathrm{th}}$ estimate being 0. Note that because $\mathcal A(;j)$ is a function of $\bm S^{(j)}$, it factors out from the conditional expectation in $w_j$; as a result, $w_j$ also has a non-zero probability of attaining 0 exactly. Having a zero weight for a signal variable is detrimental, as it completely removes the possibility of it being discovered. 
$\lambda$ can be chosen by cross-validation\footnote{We use a 10-fold cross-validation by default.\label{common_footnote}}, and we denote this choice by $\hat\lambda_j$, leading to the natural variable importance measure proposal
\[
    \mathcal A_j^{\rm LASSO}(\bm y) = \left\lvert\hat\beta^{\hat \lambda_j, j}_j\right\rvert.
\]
Note that while $\hat{\bm \beta}^{\lambda,j}$ for a fixed $\lambda$ is a function only of $\bm S^{(j)}$, selecting $\hat\lambda_j$ by cross-validation in general introduces randomness in the choice of the folds so that $\hat{\bm\beta}^{\hat \lambda_j, j}$, and hence $\mathcal{A}_j^{\text{LASSO}}$, does not strictly satisfy our design objective of being exclusively a function of $\bm S^{(j)}$, though we found this extra randomness to make little difference. 
%For the choice of $\lambda$, we recommend cross-validation based on the objective in \eqref{eqn:lasso_objective}, using response $\bm z$ and design matrix $\bm W$. Denote this cross-validated choice by $\hat \lambda_j$. 
%Note $\hat\lambda_j$ is a function of only $\bm S^{(j)}$ and the randomness in cross-validation's fold splitting.
%from the cross-validation procedure. 
    %Although this choice of $\hat \lambda_j$ slightly deviates from our aim of having $\mathcal A(;j)$ as a function of $\bm S^{(j)}$, we still found it to work well in practice. Particularly, since the randomness introduced by the cross-validation step is independent of the data, the proof of Theorem~\ref{thm:validity_weights} remains unchanged.
    %Although we initially defined $\mathcal A_j$ as a function of the sufficient statistic, we can also allow it to depend on such external randomness. Since this randomness is independent of the data and in particular, is not affected by conditioning on $\bm S^{(j)}$, the proof of Theorem~\ref{thm:validity_weights} remains unchanged. Thus, from this point on, we extend the definition of $\mathcal A_j$ to allow dependence on external randomness, and define our recommended variable importance measure as
%     Thus, our recommended variable importance measure is:
% \[
%     \mathcal A_j^{\rm LASSO}(\bm y) = \left\lvert\hat\beta^{\hat \lambda_j, j}_j\right\rvert  
% \]

Computing $\hat{w}_j$ with $\mathcal A_j^{\text{LASSO}}$ requires $d$ cross-validated LASSO fits, and computing $\{\hat{w}_j\}_{j=1}^d$ requires $d^2$ such fits, which can be computationally expensive when $d$ is large. This cost can be mitigated, however, by the fact that once the samples $\left\{\tilde{\bm y}^{(j)}\stackrel{H_j}{\sim} \bm y\mid \bm S^{(j)}: 1\leq j\leq d \right\}$ have been generated for each hypothesis ${H_j}$, the resulting $d^2$ optimization problems are independent and can be run in parallel. 
%Parallelization can therefore substantially reduce the runtime of the proposed procedure. 
As another way to reduce computational cost, we next propose another importance measure that still leverages sparsity in $\bm \beta$ and can be computed much more efficiently due to shared computation across the $d^2$ fits.
%In Section~\ref{sec:faster_compute_csw}, we discuss tricks for faster computation of $\mathcal A_{\rm LASSO}(\bm y;j)$ across $j\in [1:d]$.
 %(with tricks for faster evaluation of the latter weights discussed in Appendix~\ref{sec:faster_compute_csw})

\subsubsection{Faster alternative: Importance measure based on distilled soft-thresholding}
\label{sec:reccomendaion_dST}
% The LASSO-based measure of Section~\ref{sec:reccomendaion_LASSO} can be computationally expensive: evaluating a single weight $w_j$ requires computing the importance measure $d$ times, so the number of cross-validated LASSO runs needed for all $d$ weights scales as $d^2$. To address this, we propose an alternative variable importance measure that still exploits the sparsity in $\bm\beta$, but at a much lower computational cost.

Closely related to the LASSO estimator with penalty $\lambda\ge 0$ is the soft-thresholding estimator, defined as applying the soft-thresholding function $\beta\mapsto {\rm sign}(\beta)\left(|\beta|-\lambda\right)_{+}$ coordinate-wise to the ordinary least squares (OLS) estimate. In fact, under orthogonal design, soft-thresholding matches the LASSO exactly.
% Closely related to the LASSO estimator with penalty $\lambda$ is the soft-thresholding estimator, defined as follows. For any $(\bm y, \bm X)$, one starts from a base estimator $\hat{\bm \beta}$---for the linear model, the OLS estimator is the canonical choice---and applies the soft-thresholding function $\eta_{\rm ST}(\beta, \lambda) = {\rm sign}(\beta)\left(|\beta|-\lambda\right)_{+}$ coordinate-wise. Indeed, under an orthogonal design the LASSO estimator coincides exactly with the soft-thresholded OLS estimator. 
Because $\bm W^{(j)}$ has more columns than rows, 
the OLS estimate for the linear model \eqref{eq:linmod} is not well-defined\footnote{Even if we chose a well-defined estimate like ridgeless least squares to soft-threshold, we would still want to avoid soft-thresholding the $j$th coordinate for the same reason we do not penalize it in Equation~\eqref{eqn:lasso_objective}. But then the variable importance measure that only takes the $j$th coordinate of the estimate would be entirely unaffected by the soft-thresholding and thus in no way leverage its sparse bias.}, so
%to make the OLS estimate well-defined and also to remove the regularization on $\beta_j$, 
we instead
%Following the discussion in Section~\ref{sec:reccomendaion_LASSO}, when used to compute weights this measure is applied to the response $\bm z$ and the design matrix $\bm W$. This, however, raises two difficulties. First, $\bm W$ is $(d-1)\times d$ dimensional, so a unique OLS solution does not exist. Second, under the full soft-thresholding scheme the resulting importance measure is zero with non-zero probability, which---as noted in Section~\ref{sec:reccomendaion_LASSO}---is detrimental\footnote{Simply leaving the $j$th variable out of the thresholding step just retains the base OLS estimator, which is not helpful, as this estimator does not leverage sparsity.}.
% does not help either, since the resulting estimator would then coincide with the base estimator.}.
%To bypass these difficulties, we instead 
consider a distilled \citep{dcrt} version of soft-thresholding as follows. We first compute the soft-thresholded estimator $\hat{\bm \beta}^{\lambda,-j}_{\rm ST}$ from $\bm z^{(j)}$ and $\bm W^{(j)}_{-j}$ (which, unlike $\bm W^{(j)}$, is invertible). The importance measure is then the magnitude of the univariate OLS coefficient of regressing the residual $\bm r_j = \bm z^{(j)} - \bm W^{(j)}_{-j}\hat{\bm \beta}^{\lambda,-j}_{\rm ST}$ on $\bm W^{(j)}_j$, and is given by $\frac{\left\lvert {\bm r_j^T\bm W^{(j)}_j}\right\rvert }{\|\bm W^{(j)}_j\|^2}$. The intuition is as follows: if the $j^{\rm th}$ variable is \textit{important}, then $\bm W^{(j)}_{-j}$ alone cannot explain the signal in $\bm z^{(j)}$, so the residual $\bm r_j$ retains signal attributable to $\bm W^{(j)}_j$ and should be correlated with it.  %Such a statistic has also been discussed in \cite{dcrt}.
%The measure quantifies the strength of this correlation.

It remains to choose the threshold $\lambda$. Since the procedure can ultimately be viewed as a prediction problem for the residual $\bm r_j$ from $\bm W^{(j)}_j$, we select $\lambda$ by cross-validation\footref{common_footnote}, minimizing the prediction mean-squared-error of this problem; further details are given in Appendix~\ref{sec:cv_st}. Denoting the resulting choice by $\hat\lambda^{\rm ST}_j$, the distilled soft-thresholded importance measure is
\[
    \mathcal A_j^{\rm dST}(\bm y) = \frac{\left\lvert\left(\bm z^{(j)} - \bm W^{(j)}_{-j}\hat{\bm \beta}^{\hat\lambda^{\rm ST}_j,-j}_{\rm ST}\right)^T\bm W^{(j)}_j\right\rvert}{\|\bm W^{(j)}_j\|^2}.
\]
Note that evaluating $\mathcal A_j^{\rm dST}(\bm y)$ for all $j$ requires computing $\bm W^{(j)}$ together with the OLS fit of $\bm z^{(j)}$ on $\bm W^{(j)}_{-j}$ for every $j$. In Appendix~\ref{sec:faster_compute_csw} we show that this can be accelerated substantially via faster matrix updates, avoiding a separate matrix inverse and square root for each $j$. Similar speed-ups apply to the cross-validation step, for the sub-matrices restricted to the training and testing folds. Indeed, we found $\mathcal{A}^{\text{dST}}$ to be at least 20 times as fast as $\mathcal{A}^{\text{LASSO}}$ even for $d=100$; see Section~\ref{sec:runtime}.
%In fact, we found $\mathcal A^{\rm dST}$ to be substantially more computationally efficient than $\mathcal A^{\rm LASSO}$; in Section~\ref{sec:runtime} we find that even for $d=100$, constructing weights using $\mathcal A^{\rm dST}$ can be 20 times faster than when using $\mathcal A^{\rm LASSO}$.

% in Section~\ref{sec:runtime} that weights computed using $\mathcal A^{\rm dST}$ can run in less than $5\%$ of the time it takes $\mathcal A^{\rm LASSO}$ to run for $d$ as small as $d=100$.

\subsubsection{Other importance measures}
\label{sec:powered_methods}
The set of possible importance measures one could consider is enormous, but we have focused on simple and natural choices based on sparse coefficient estimators in the linear model. Even just for these choices, one may wonder why it is best to take the magnitude of the coefficient estimate, as opposed to, say, its square, or square root, or any other monotone transformation, any of which generally results in a different CSW. We explore the power of $\left(\mathcal A_j^{\text{LASSO}}\right)^m$ for various choices of exponentiation $m$ in Appendix~\ref{sec:exponentiated_simulations}, finding no single choice dominates the others. We found that higher exponents tend to have relatively higher power in sparser settings and relatively lower power in denser settings, and overall we felt that using $\mathcal A^{\text{LASSO}}$ (i.e., $m=1$) provides a good balance of power gain in sparse settings and limited power loss in dense settings, compared to unweighted multiple testing.

\subsection{Putting it all together: New multiple testing methods}
\subsubsection{FWER control}
\label{sec:fwer}
Our recommended procedure to leverage our proposed weights is a weighted version of Holm's procedure and is summarized in Algorithm~\ref{alg:wholm} below. We state its validity in Theorem~\ref{thm:wholm_validity} and prove it in Appendix~\ref{sec:proof_wholm_validity}.

\begin{theorem}
    \label{thm:wholm_validity}
    Suppose the p-values in the input of Algorithm~\ref{alg:wholm} satisfy Assumption~\ref{assm:p_val_suff}. Then for any $\alpha \in (0,1)$, Algorithm~\ref{alg:wholm} controls the FWER at level $\alpha$.
\end{theorem}

\begin{algorithm}[h]
    \caption{CSW-Holm}
    \label{alg:wholm}
    \textbf{Input: }p-values: $p_j$ for $H_j,\forall j\in [1:d]$, Importance measure: $\mathcal A$, Nominal FWER: $\alpha$.\\
    \textbf{Output: }A subset of $[1:d]$.\\
    % Obtain p-values: $p_j$ for $H_j$, for all $j\in [1:d]$.\\
    Initialize $\hat R\leftarrow\emptyset$\\
    Initialize \textit{$\texttt{counter}$}\\
    %\While{not stopped {\rm\textbf{and}} $|\hat R|<d$}{
    Sample $\tilde{\bm y}^{(j)}\stackrel{H_j}{\sim}  \bm y\mid\bm S^{(j)}$, for all $j\in [1:d]$.\\
    \For{\texttt{counter} in $1$ to $d$}{
        Define $\hat w^{(\hat R^c)}_j = \frac{\mathcal A_j(\tilde{\bm y}^{(j)})}{\sum_{i\notin \hat R}\mathcal A_i(\tilde{\bm y}^{(j)})}, \forall j \notin \hat R$.\\
        $j^*\leftarrow \underset{j\notin \hat R}{\arg\min} \frac{p_j}{\hat w^{(\hat R^c)}_j}$.\\
        \If{$p_{j^*}/\hat w^{(\hat R^c)}_{j^*}\leq \alpha$}{
            $\hat R\leftarrow \hat R\cup\{j^*\}$
        }
        \Else{
            Stop
        }
    }
    Return $\hat R$.
\end{algorithm}

The proof in Appendix~\ref{sec:proof_wholm_validity} relies on the fact that Algorithm~\ref{alg:wholm} is the sequential representation of the closure of the weighted $I$-Bonferroni test, discussed in Section~\ref{sec:fwer_methods}, with weights
\begin{align}
\label{eqn:chosen_local_weights}
\begin{aligned}
    % &w^{(I)}_j = \mathbb E_{H_j}\left[ \frac{\mathcal A_j(\bm y)}{\sum_{i\in I}\mathcal A_i(\bm y)} \middle| \bm S^{(j)}\right], \qquad \forall j\in I\\
    % & 
    \hat w^{(I)}_j
    =
    \frac{\mathcal A_j(\tilde{\bm y}^{(j)})}
    {\sum_{i\in I}\mathcal A_i(\tilde{\bm y}^{(j)})},~\textrm{where }~\tilde{\bm y}^{(j)}\stackrel{H_j}{\sim}\bm y\mid \bm S^{(j)},
    \qquad \forall j\in I,
\end{aligned}
\end{align}
that satisfy Assumption~\ref{assm:bonf_valid}. Recall from the discussion in Section~\ref{sec:fwer_methods} that it is generally unclear how to construct $I$-specific weights satisfying Assumption~\ref{assm:bonf_valid} from global weights satisfying Assumption~\ref{assm:bonf_valid}, but the CSW construction admits a natural way to do so. Indeed, Algorithm~\ref{alg:wholm} is not simply a CSW version of an existing weighted Holm's procedure, since our weights are not e-values and $\hat{w}_j^{(I)}$ above is not simply an $I$-renormalization of the global weight $\hat{w}_j$.
\subsubsection{FDR control}
 \label{sec:fdr}
% \begin{algorithm}[h]
%     \caption{Recommended FDR controlling procedure..}
%     \label{alg:wbh}
%     \textbf{Input: }Data: $(\bm y, \bm X)$, Nominal FDR: $\alpha$.\\
%     \textbf{Output: }A subset of discovered indices.\\
%     Obtain p-values: $p_j$ for $H_j$, for all $j\in [1:d]$.\\
%     Compute $\hat {w}_{\mathcal A,j}$ for all $1\leq j \leq d$.\\
%     Define $\tilde p_j:=p_j/(\hat {w}_{\mathcal A,j}d)$ for all $1\leq j \leq d$.\\
%     Run BH procedure at level $\alpha$ on the set of modified p-values $\{\tilde p_j\}_{1\leq j \leq d}$.\\
%     Return the rejection set $\hat R$.
% \end{algorithm}
As described in Section~\ref{sec:weighted_mult_testing}, our recommended procedure for FDR control in the linear model, following \cite{genovese2006}, applies the Benjamini--Hochberg procedure \citep{fdr} to the scaled p-values $\left\{\frac{p_j}{d\hat w_j}\right\}_{j=1}^d$, where the weights $\hat w_j$ are as given in Equation~\eqref{eqn:plug_in_w}. We call this resulting procedure, the Co-sufficient Weighted Benjamini--Hochberg (CSW-BH) procedure.
% We summarize this procedure in Algorithm~\ref{alg:wbh}.  

As noted previously, like the BH procedure applied to linear model p-values, this procedure lacks strict FDR control but exhibits robust empirical FDR control; we present empirical evidence for this in Section~\ref{sec:empirical_FDR} and, as discussed in Section~\ref{sec:fdr_methods}, attribute it to CSW weights satisfying Assumption~\ref{assm:bonf_valid}. 
%In fact, similar to the discussion for the usual BH procedure in Section~\ref{sec:fdr_methods}, our weighted version, despite not satisfying Assumption~\ref{assm:bh_validity}, crucially satisfies Assumption~\ref{assm:bonf_valid}. We similarly expect this fact to be contributing significantly to its empirical robustness.
% Importantly, both procedures lack fail the violate the standard sufficient conditions for FDR control in the same way---both satisfy Assumption~\ref{assm:bonf_valid} but not Assumption~\ref{assm:bh_validity}, where the usual BH procedure can be viewed as a weighted procedure with weights $1/d$. We believe that Assumption~\ref{assm:bonf_valid} plays a crucial role in their empirical FDR control and without it, weights can be chosen to yield severely inflated errors. For example, choosing weights proportional to $1/p_j$ would violate Assumption~\ref{assm:bonf_valid} and lead to substantial FDR inflation.
%Also, as noted earlier in Section~\ref{sec:fdr_methods}, o
One could also use conditional calibration (discussed earlier in Section~\ref{sec:fdr_methods}) to obtain an exact FDR-controlling procedure that leverages these weights. However, we find this approach to be computationally prohibitive; see Appendix~\ref{sec:app_dbh} for details.

% Finally, observe that both Algorithm~\ref{alg:wholm} and Algorithm~\ref{alg:wbh} require the computation of individual p-values for testing the hypotheses $H_j$. The canonical choice is the usual $t$-test p-values; however, we explore other choices below.

\subsubsection{The choice of p-values}
\label{sec:pval_choice}
Although this paper primarily focuses on obtaining informative weights for the linear model, the power of the resulting weighted multiple testing procedures also depends on the choice of p-values for testing $\{H_j\}_{j=1}^d$, the canonical choice for which is the usual $t$-test p-values. However, we recommend instead $\ell$-test p-values \citep{sengupta2024elltestleveragingsparsitygaussian}, which leverage sparsity in $\beta$ (in a conceptually similar but operationally very different way to co-sufficient weighting) to improve over the $t$-test while still satisfying Assumption~\ref{assm:p_val_suff}. In particular, our implementation in Sections~\ref{sec:expt}--\ref{sec:real_data} of both the CSW procedures described above uses $\ell$-test p-values.
% Recently, however, \cite{sengupta2024elltestleveragingsparsitygaussian} introduced the $\ell$-test, which, like our approach, leverages sparsity in $\bm \beta$ to improve single-coefficient inference over the traditional $t$-test. In this paper, we therefore recommend the $\ell$-test p-values. These p-values are also satisfy Assumption~\ref{assm:p_val_suff} and are therefore compatible with our multiple testing procedures.

\section{Experiments}
\label{sec:expt}
We consider an $n\times d$ design matrix $\bm X$ whose rows are drawn from $\mathcal N_d(\bm 0, \bm \Sigma(\rho))$ and columns are normalized, where $\bm\Sigma(\rho)$ has the Toeplitz structure $\bm\Sigma(\rho)_{ij} = \rho^{|i-j|}$. The coefficient vector $\bm \beta$ is zero except at $k$ randomly chosen signal indices, each assigned a non-zero value (the mechanisms for which are described below). The response is generated as $\bm y \sim \mathcal N(\bm X\bm \beta, \bm I_n)$. We vary four aspects of this setting:
\begin{itemize}
    \item \textbf{Inter-variable correlation:} we take $\rho \in \{0, 0.5, 0.9\}$, corresponding to covariate independence, medium correlation, and high correlation, respectively.
    \item \textbf{Aspect ratio:} we set $n/d$ to either $1.5$ or $3$. %Particular choices of $n$ and $d$ are discussed below.
    \item \textbf{Signal generation:} we draw the signal entries of $\bm \beta$ either from ${\rm Unif}\{+A,-A\}$ (discrete) or from $\mathcal N(0,A^2)$ (continuous); the amplitude $A$ is specified below.
    \item \textbf{Simulation scale: }we consider small-scale and large-scale simulations, corresponding to $d=100$ and $d=1000$ respectively.
\end{itemize}
Each of our simulation setting involves varying one parameter while holding the others fixed, according to one of the following three regimes: (a) fixing the amplitude $A$ and varying the sparsity $k$; (b) fixing $k/d = 5\%$ (high sparsity) and varying $A$; and (c) fixing $k/d = 50\%$ (low sparsity) and varying $A$. The only exception is an additional MCC setting considered for $d=1000$, which we describe shortly. Unless otherwise specified, we use nominal FWER and FDR levels of 0.05 and 0.1, respectively.

For $d=100$, we consider all combinations of the choices of $\rho$, $n/d$, signal generation, and simulation regimes ((a), (b), or (c)) described above. For $d=1000$, however, we restrict attention to a subset of settings, primarily to reduce computational cost. Specifically, we fix $n/d = 3$ (since the knockoff-based methods are not applicable when $n/d = 1.5$), take $\rho \in \{0,0.9\}$, and consider regimes (a) and (b) for each choice of $\rho$. In the $d=100$ simulations, these four settings capture the broad performance trends exhibited by our methods across the remaining settings. We therefore focus on these settings for $d=1000$ as well, with the expectation that they remain representative of the broader performance patterns. Additionally, we consider a setting in which the design matrix follows the MCC design of \citet{luo2022}. This setting exhibits the whiteout phenomenon \citep{li2021whiteout}, under which the knockoff+ procedure performs poorly, and serves as a key example demonstrating the advantage of the cknockoff procedure. This setting features $n=3000$, $d=1000$, and $k=10$, and  compares the methods across a range of nominal FDR levels; for further details, see \cite[Section 5.1]{luo2022}. Altogether, we consider 40 simulation settings for the FWER-controlling methods and 41 settings for the FDR-controlling methods (each setting varies one data-generating aspect, so the total number of distinct data-generating distributions is around 200). Throughout, we report empirical error rates (FWER and FDR) and average power averaged over 500 independent replicates for $d=100$ and 200 independent replicates for $d=1000$.

Next, we list the methods that we compare. For FWER control, we consider four procedures: CSW-Holm using (1) $\mathcal A^{\rm LASSO}$ ($\CSWLASSOHolm$) and (2) $\mathcal A^{\rm dST}$ ($\CSWdSTHolm$), both applied to the $\ell$-test p-values; and Holm's procedure applied to the (3) $\ell$-test ($\ell$-Holm) and (4) $t$-test ($t$-Holm) p-values. For FDR control, we compare 11 procedures: CSW-BH using (1) $\mathcal A^{\rm LASSO}$ ($\CSWLASSOBH$) and (2) $\mathcal A^{\rm dST}$ ($\CSWdSTBH$), both applied to the $\ell$-test p-values; (3) knockoffs and (4) knockoffs+ \citep{knockoff}; (5) cknockoff and (6) cknockoff$^\ast$ \citep{luo2022}; the BH procedure applied to the (7) $\ell$-test ($\ell$-BH) and (8) $t$-test ($t$-BH) p-values; (9) Gaussian mirrors (GM) \citep{gaussmirror}; (10) multiple data-splitting (MDS) \citep{junfdrdatasplit}; and (11) OATK \citep{oatk}. For the $d=1000$ simulations, we include only $\CSWdSTHolm$ and $\CSWdSTBH$ among the CSW procedures due to their greater computational efficiency.

The full set of results for all methods across all the settings described above is deferred to Appendix~\ref{sec:detailed_simulation_results}; we present a selected subset in the main text. In particular, we fix $n/d = 3$ (and also additionally consider the $n/d = 1.5$ case for FDR simulations for reasons described in Section~\ref{sec:fdr_powers} below) and focus on regime (a) to understand the effect of varying the sparsity level $k$. Note that regimes (b) and (c) zoom in on two specific levels of sparsity and, as we found, exhibit trends already seen at the corresponding sparsity levels in regime (a). We also focus only on discrete signal generation, as the continuous signal regime exhibits similar trends.

% We compare the following FDR-controlling procedures: co-sufficient weighting based on (1) LASSO weights ($\CSWLASSOBH$) and (2) distilled soft-thresholding weights ($\CSWdSTBH$) on $\ell$-test p-values; the (3) knockoffs and (4) knockoffs+ procedures of \cite{knockoff}; the (5) cknockoff and (6) cknockoff$^\ast$ procedures of \cite{luo2022}; the BH procedure applied to the (7) $\ell$-test ($\ell$-BH) and (8) $t$-test ($t$-BH) $p$-values; (9) the Gaussian mirrors procedure (GM) of \cite{gaussmirror}; (10) the multiple data-splitting procedure (MDS) of \cite{junfdrdatasplit}; and (11) the OATK procedure of \cite{oatk}. For FWER control, we compare (1) $\CSWLASSOHolm$, (2) $\CSWdSTHolm$, and Holm's procedure applied to the (3) $\ell$-test ($\ell$-Holm) and (4) $t$-test ($t$-Holm) $p$-values.

\subsection{FWER controlling simulations}
\label{sec:main_fwer}
\begin{figure}[!ht]
    \centering
    \includegraphics{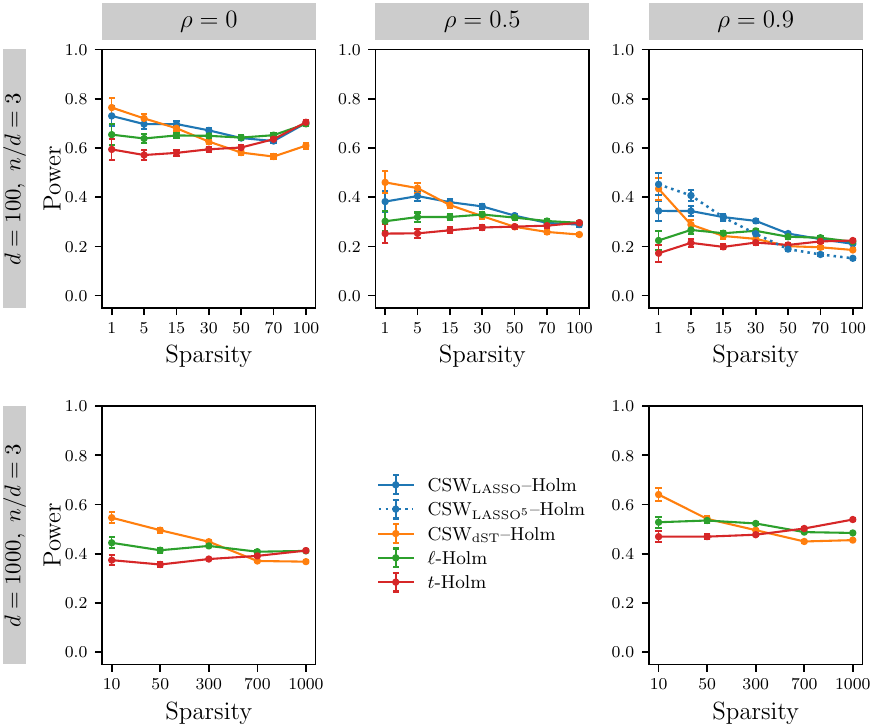}
    \caption{The power of various FWER controlling procedures under simulation regime (a), discrete signal generation and $n/d = 3$. The choice of the amplitude $A$ is described in Appendix~\ref{sec:detailed_simulation_results}. The error bars represent plus or minus two standard errors.}
    \label{fig:fwer_main_paper}
\end{figure}
We present the select set of power curves of the various FWER-controlling methods described above in Figure~\ref{fig:fwer_main_paper}. Because all the methods provably control the FWER, we do not additionally present their empirical FWER here; these results are available in Appendix~\ref{sec:detailed_simulation_results}.
For the $d=100$ simulations, we find that $\CSWLASSOHolm$ and $\CSWdSTHolm$ outperform the unweighted procedures in the sparse regime. In particular, as shown in the top-rightmost panel of Figure~\ref{fig:fwer_main_paper}, $\CSWLASSOHolm$ and $\CSWdSTHolm$ achieve up to $54\%$ and $94\%$ higher power, respectively, than the best-performing non-CSW procedure.  As $k$ increases, the gap between the CSW power curves narrows, with $\CSWLASSOHolm$ performing almost indistinguishably from the unweighted procedures in the denser regimes, while $\CSWdSTHolm$ slightly underperforms. This behavior of $\CSWdSTHolm$ also carries over to the $d=1000$ regime.

\subsection{FDR controlling simulations}
\label{sec:main_fdr}
Among the FDR-controlling methods, we retain only $\CSWLASSOBH$, $\CSWdSTBH$, knockoffs+, cknockoff$^\ast$ (which uniformly dominates cknockoff), $\ell$-BH, and $t$-BH for our main text plots. We exclude GM, OATK, knockoffs, and MDS from the main text plots because we found them to exhibit poor empirical FDR control in a number of settings. For example, GM's empirical FDR reaches 0.67 in the top-middle panel of Figure~\ref{fig:appendix_1_discrete_error_fdr}, MDS's empirical FDR reaches 0.20 in the bottom-right panel of Figure~\ref{fig:appendix_large_error_fdr}, and OATK's and knockoffs' empirical powers in Figure~\ref{fig:appendix_1_normal_error_fdr} reach 0.40 and 0.37, respectively; recall the nominal FDR is 0.1, so all these violations are at least 100\% above the nominal level. Notably, all of these methods exhibit their worst FDR violations in sparse settings, which is our primary motivating regime, and arguably, is widely encountered in scientific problems where variable selection is of interest. Additionally, in many of the sparse regimes where MDS does empirically control the FDR, its power is nearly zero even while other methods that control FDR achieve powers $\ge 0.5$ in the same settings; see, e.g., Figure~\ref{fig:appendix_1_discrete_power_fdr}.
%for the MDS procedure, we find that in the sparsest settings where it does exhibit empirical FDR control, it in fact, makes almost no discoveries.
% \textbf{say that the other methods have their worst violation in the sparsest setting. }
%The full set of simulation results, including all omitted methods, is available in Appendix~\ref{sec:detailed_simulation_results}. 

\subsubsection{Summary of empirical FDRs}
\label{sec:empirical_FDR}
\begin{figure}[!ht]
    \centering
    \includegraphics{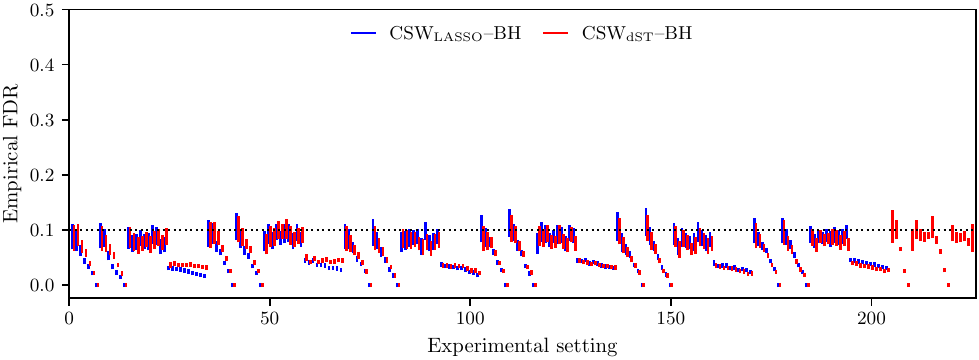}
    \caption{Empirical FDRs of $\CSWLASSOBH$ and $\CSWdSTBH$ across all the simulations considered in this paper. The x-axis represents an enumeration of the different experimental settings with nominal FDR 0.1 and the settings corresponding to the large-scale regime with $d=1000$ appear on the rightmost end. The error bars represent plus or minus two standard errors. The horizontal black dotted line represents the nominal FDR level of 0.1.}
    \label{fig:empirical_fdr}
\end{figure}

%As discussed in Section~\ref{sec:fdr}, our proposed CSW procedures do not enjoy strict FDR control. However, as argued previously, we expect them to exhibit robust empirical control. This is indeed what we observe. 
Figure~\ref{fig:empirical_fdr} summarizes the empirical FDR of $\CSWLASSOBH$ and $\CSWdSTBH$ across all the data-generating processes we simulated, and we find that the empirical FDRs of both methods are within Monte Carlo error of being bounded by the nominal FDR level. The figure excludes only the simulations from the MCC design corresponding to nominal FDR levels different from $0.1$; these results are available in Appendix Figure~\ref{fig:appendix_mcc_fdr}, where we likewise observe empirical FDR control.

\subsubsection{Powers}
\label{sec:fdr_powers}
\begin{figure}[!ht]
    \centering
    \includegraphics{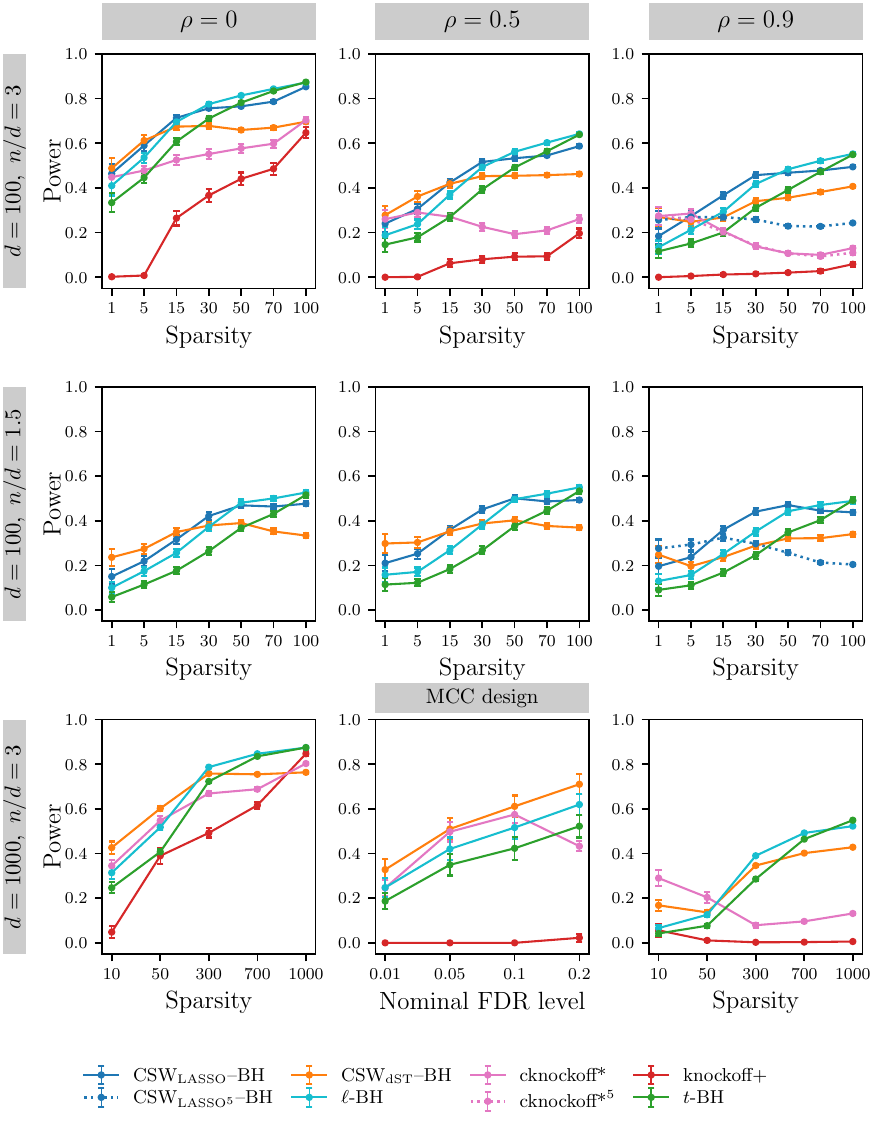}
    \caption{The power of various FDR controlling procedures under simulation regime (a) and discrete signal generation. The choice of the amplitude $A$ is described in Appendix~\ref{sec:detailed_simulation_results}. The error bars represent plus or minus two standard errors.}
    \label{fig:fdr_main_paper}
\end{figure}

Turning to the power of FDR-controlling procedures, Figure~\ref{fig:fdr_main_paper} considers the same settings as Figure~\ref{fig:fwer_main_paper} plus the simulation settings with $n/d = 1.5$ for $d=100$ (where the knockoff-based procedures are not applicable) and the MCC design in the $d=1000$ regime.
The first two rows of Figure~\ref{fig:fdr_main_paper} show that, for $d=100$, $\CSWLASSOBH$ outperforms nearly all non-CSW procedures across the regimes considered, achieving up to $51\%$ higher power than the best-performing non-CSW procedure, as highlighted in the rightmost panel of the middle row. There are, however, two exceptions. First, when the coefficient vector is dense ($\geq 50\%$), $\CSWLASSOBH$ is slightly outperformed by the unweighted BH procedures. Second, in the sparsest regime with $\rho = 0.9$ (upper-left panel), cknockoff$^\ast$ slightly outperforms $\CSWLASSOBH$. However, this does not appear to represent a fundamental weakness of the CSW approach, as the same behavior is not observed for $\CSWdSTBH$. We also include a curve for $\CSWLASSOBH[5]$ (formed by using $\mathcal A_{\rm LASSO}^5$ as the importance measure, as discussed in Section~\ref{sec:powered_methods}), which shows that even a slight variation of $\CSWLASSOBH$ can close this gap. For comparison, we also include a curve for cknockoff$^\ast$ using a LASSO$^5$ variable importance measure, showing that this modification does not meaningfully change the performance of cknockoff$^\ast$.

The takeaway for $\CSWdSTBH$ throughout Figure~\ref{fig:fdr_main_paper} is similar to that for $\CSWLASSOBH$ except $\CSWdSTBH$ seems to have relatively higher power than $\CSWLASSOBH$ in sparse regimes and relatively lower power in dense regimes. Indeed $\CSWdSTBH$ can achieve more than twice the power of the best performing non-CSW method in the sparsest setting of the leftmost panel in the middle row. In the bottom-right panel of Figure~\ref{fig:fdr_main_paper}, however, $\CSWdSTBH$ is moderately outperformed in sparse settings by cknockoff$^*$, though in the denser parts of that panel, $\CSWdSTBH$ outperforms cknockoff$^*$ by a wide margin.
%exhibits a similar overall trend throughout , although it underperforms the vanilla BH procedures by a larger margin in the high-density regimes. 
%It slightly outperforms $\CSWLASSOBH$ in sparse regimes except when the inter-variable correlation is very high, as in the $\rho = 0.9$ setting, while underperforming it in denser regimes. These trends hold in both the $n/d = 3$ and $n/d = 1.5$ cases and, for $\CSWdSTBH$, extend to the analogous settings considered in the $d=1000$ simulations. 
In the MCC design of \citet{luo2022}, $\CSWdSTBH$ outperforms all other methods including \citet{luo2022}'s method cknockoff$^*$.

\subsection{Runtime comparison}
\label{sec:runtime}
To understand the computational efficiency of the procedures based on $\mathcal A^{\rm dST}$ over $\mathcal A^{\rm LASSO}$, we recorded the runtimes of $\CSWdSTHolm$ and $\CSWLASSOHolm$ for the simulations in the top-left panel of Figure~\ref{fig:fwer_main_paper}, averaging them across replicates for each sparsity level. We found that, across all sparsity levels considered, constructing weights by normalizing $\mathcal A^{\rm dST}$ is at least 20 times faster than when using $\mathcal A^{\rm LASSO}$. Although it was not computationally feasible to use $\mathcal A^{\rm LASSO}$ in the large-scale simulations with $d=1000$, we expect the computational advantage of $\mathcal A^{\rm dST}$ to be even more pronounced in this setting.

\section{Application to the HIV drug-resistance dataset}
\label{sec:real_data}

The HIV drug resistance dataset \citep{hivdata} contains data corresponding to 16 regressions whose covariates are the presence of certain genetic mutations and whose outcomes are a measure of resistance to a particular drug. % These 16 drugs fall into three groups, namely protease inhibitors (PIs), nucleoside reverse transcriptase
%inhibitors (NRTIs), and nonnucleoside reverse transcriptase inhibitors (NNRTIs).
For each of the 16 regressions we follow the same pre-processing steps as \cite{knockoff} and compare our CSW methods with a selection of the methods considered in our simulations, informed by those simulations. In particular, in our FDR simulations, cknockoff$^\ast$ is the most powerful among the knockoffs-based procedures that exhibit consistent empirical FDR control, $\ell$-BH is at least as powerful as $t$-BH throughout, and OATK, MDS, and GM did not exhibit consisten empirical FDR control, so for FDR control we compare $\CSWLASSOBH$ and $\CSWdSTBH$ with only cknockoff$^\ast$ and $\ell$-BH, at a nominal FDR level of 0.1. 
%Among the knockoffs-based procedures, cknockoff$^\ast$ is the most powerful without exhibiting FDR violations according to our findings in Section~\ref{sec:expt}, while $\ell$-BH is at least as powerful as $t$-BH throughout. We did not include OATK, MDS and GM, which we found to exhibit severe FDR violations in various settings. 
Similarly, for FWER control we compare $\CSWLASSOHolm$ and $\CSWdSTHolm$ with only $\ell$-Holm, at a nominal FWER level of 0.05.

Previous analyses \citep{knockoff,dbh} have aggregated the discoveries to a gene-level and also compared the discovered set of genes against a set of mutations identified in the treatment-selected mutations (TSM) panel of \cite{hivdata}, and we refer the reader to \cite[Section 4]{knockoff} for further details. In Table~\ref{tab:hiv_disc} we tabulate the fraction of columns discovered by each of the methods across all 16 regressions, as well as the fraction of gene-level discoveries. In Figures~\ref{fig:hiv_fdr} and \ref{fig:hiv_fwer} of the appendix, we present the number of genes present in the TSM list discovered by each of the methods for each of the drugs. 
For the FDR controlling procedures we generally find that cknockoff$^\ast$ produces far fewer discoveries than the other three methods, which perform similarly. For FWER control, all three procedures also perform similarly.

\begin{table}[h]
    \centering
    \begin{tabular}{|c|c|c|}
    \hline
    \hline
    \textbf{Method} & \textbf{$\%$ of raw discoveries} & \textbf{$\%$ of gene-level discoveries}\\
    \hline 
    \hline
    $\CSWLASSOBH$ (FDR) & 11.62$\%$ & 21.87$\%$ \\
    $\CSWdSTBH$ (FDR) & 10.99$\%$  & 20.94$\%$ \\
    cknockoff$^\ast$ & 4.83$\%$  & 10.79$\%$ \\
    $\ell$-BH & 12.01$\%$ & 22.87$\%$ \\
    \hline
    $\CSWLASSOHolm$ (FWER) & 6.90$\%$ &13.15$\%$ \\
    $\CSWdSTHolm$ (FWER) & 6.87$\%$& 13.22$\%$\\
    $\ell$-Holm & 6.52$\%$& 12.51$\%$\\
    \hline
    \end{tabular}
    \caption{Percentage of total columns (raw discoveries) and genes discovered by all the methods across all the 16 regressions.}
    \label{tab:hiv_disc}
\end{table}

Thus, in this application where we expect sparsity to hold, our weighted procedures yield a similar number of rejections as their unweighted counterparts. We argue that a primary reason for this behavior is that these regressions contain several strong signals but relatively few discoverable weaker ones. The standard nominal FDR and FWER levels that we use are sufficiently high that both weighted and unweighted methods discover the strong potential signals, leaving any differences in performance to be driven primarily by the weaker signals. Since the quality of our weights also depends on signal strength, the weighted procedures offer little advantage over their unweighted counterparts in identifying these weaker potential signals at the nominal FWER and FDR levels of 0.05 and 0.1, respectively.

\begin{figure}[!h]
    \centering
    \includegraphics{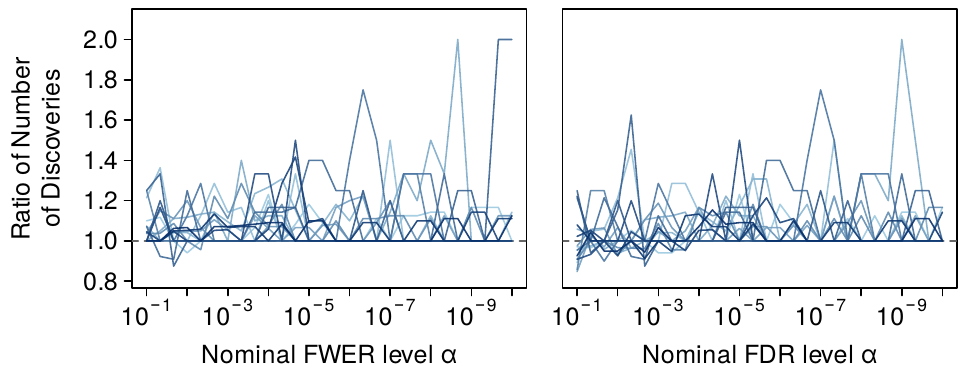}
    \caption{The ratio of number of discoveries of $\CSWLASSOHolm$ to that of $\ell$-Holm (left) and that of $\CSWLASSOBH$ to that of $\ell$-BH (right) across various levels of nominal FWER (left) and FDR (right) for all the 16 regressions from the HIV drug-resistance dataset.}
    \label{fig:ratio_hiv}
\end{figure}

To illustrate this phenomenon better and highlight the benefits of weighting, in Figure~\ref{fig:ratio_hiv}, we consider smaller nominal FWER and FDR levels and plot, for each of the 16 regressions, the ratio of the number of discoveries made by our corresponding weighted procedures to those made by their unweighted counterparts. We observe that all 16 curves in both panels are heavily biased above 1, often substantially so, particularly at smaller nominal error rates. At these more stringent levels, even the stronger potential signals become harder to discover, and weighting offers a clear advantage---our procedures often yield more discoveries, sometimes as many as twice those of their unweighted counterparts. We complement Figure~\ref{fig:ratio_hiv} with Appendix~\ref{sec:weighting_hiv}, where we present figures showing the exact numbers of discoveries made by these methods across varying nominal error rates. These figures show the benefits of weighting are real but can be more or less pronounced depending on the distribution of signal strengths relative to the nominal level of error control.

% To better illustrate the benefit of weighting, in Appendix~\ref{sec:weighting_hiv}, we summarize the number of discoveries by these procedures by tightening the nominal error levels so that even the stronger signals become more difficult to discover. We find that, at smaller nominal FDR and FWER levels, the weighted methods can still make additional rejections while the unweighted procedures almost never discover more variables than the unweighted procedures.

\section{Discussion}
\label{sec:discussion}
% CSW testing-based methods use the sufficient statistic under $H_j$ to leverage sparsity in $\bm \beta$ and construct weights $w_j$ that can then be fed into downstream multiple testing procedures. We found that these procedures outperform existing methods in their respective error-control regimes under a wide range of settings with sparse $\bm \beta$. We discussed two choices of obtaining these weights by normalizing a LASSO-based and a soft-thresholded distilled variable importance measure. 

% On a high level, this work highlights the novel insight that one can use the sufficient statistic to construct weights that can leverage structural information that can be hard to quantify otherwise, like sparsity. This opens up the provision of using other kinds of structural information for improved testing; for example, smoothness in $\bm \beta$. Many further extensions remain for future work. For example, the $\CSWLASSOBH$ and $\CSWdSTHolm$ can hugely benefit from a faster implementation of obtaining all the weights. A part of this is implementational and one can use customized-warm starts and paralellization to compute the LASSO runs required for obtaining the weights more efficiently. Another interesting direction could be extending CSW-testing beyond the Gaussian linear models; for example to models that are asymptotically Gaussian.
Co-sufficient weighting uses, for each $j$, the sufficient statistic under $H_j$ to leverage sparsity in $\bm \beta$ and construct a weight $w_j$. These weights can then be fed into downstream multiple testing procedures that outperform existing methods in their respective error-control regimes across a wide range of settings with sparse $\bm \beta$, while suffering at most only a slight sacrifice in power in dense regimes. We discussed two choices for obtaining these weights by normalizing either our recommended LASSO-based or a computationally faster distilled soft-thresholding variable importance measure.

 %For instance, $\CSWLASSOBH$ achieves up to $159\%$ higher power than $t$-BH (Figure~\ref{fig:appendix_0_discrete_power_fdr}, middle row, leftmost panel), while $\CSWLASSOHolm$ achieves up to $127\%$ higher power than $t$-Holm (Figure~\ref{fig:appendix_0_discrete_power_fwer}, middle row, leftmost panel).

%Interestingly, despite addressing different problems, these methods share a fundamental principle: they exploit the independence of $p_j$ and $\bm S^{(j)}$ under $H_j$ to leverage information about the signal in $\bm S^{(j)}$, thereby improving p-value based inference using this information without affecting existing validity. While our CSW procedures use this information to construct weights, \citep{sengupta2024elltestleveragingsparsitygaussian,DP-ea:2025} argue that their tests implicitly exploit the sufficient statistics to infer the direction of the signals and improve inference on coefficients.

We hope future work can address two limitations of the CSW approach: First, $\CSWLASSOBH$ and $\CSWLASSOHolm$ could benefit substantially from a faster implementation for obtaining all the weights. Part of this is implementational, and one could use customized warm starts and parallelization to perform the LASSO runs required to obtain the weights more efficiently. Second, BH (with or without weights) is known to be conservative in dense regimes, but we did not explore integrating CSW into a weighted analogue of Storey--BH \citep{storey} to address this.
At a higher level, our work suggests the possibility of exploiting other forms of structural information, such as smoothness in $\bm \beta$, to improve multiple testing.  %Another interesting direction is to extend CSW testing beyond Gaussian linear models, for example, to models that are asymptotically Gaussian.

Stepping back further, this work highlights how sufficient statistics can be used to construct weights that leverage structural information, such as sparsity, that may otherwise be difficult to quantify. Our work thus belongs to a broader line of research exploiting sparsity to improve inference in the linear model \citep{sengupta2024elltestleveragingsparsitygaussian,DP-ea:2025} without sacrificing validity. Despite addressing different problems, these methods share a fundamental principle: they exploit the independence of $p_j$ and $\bm S^{(j)}$ under $H_j$ to leverage information about the signal in $\bm S^{(j)}$, thereby improving p-value based inference using this information. In fact, our recommended p-values are the sparsity-leveraging p-values of \citep{sengupta2024elltestleveragingsparsitygaussian}, whose benefits compound with ours, together yielding even more substantial power gains over conventional multiple testing procedures than either alone.

\section{Acknowledgments}
\label{sec:acknowledgments}
LJ and SS were partially supported by DMS-2045981.  The authors also acknowledge the use of Claude Opus 5.5 and GPT Sol 5.6 for proof reading and generating the content of Appendix~\ref{sec:proof_second_component_dist} and generating the counterexample in Appendix~\ref{sec:counterexample}, and for assistance literature search, code development, and for polishing the writing of this paper. All AI-generated outputs have been manually checked and edited by the authors, who remain solely responsible for its content.

\bibliography{refs}
\appendix
\section{Counter-example for FDR violation of the weighted BH procedure}
\label{sec:counterexample}

Consider $d$ independent uniform null p-values $\{p_j\}_{j=1}^d$, and choose weights $w_j = \frac{p_j}{\alpha}\mathbb I(p_j\leq \sqrt{2\alpha/d})$. These weights satisfy Assumption~\ref{assm:bonf_valid}(b), since
\[
    \mathbb E[w_j] = \frac{1}{\alpha}\int_0^{\sqrt{2\alpha/d}}p\,dp = \frac{1}{\alpha}\times \frac{2\alpha}{2d} = \frac{1}{d},
\]
and hence, $\mathbb E\left[\sum_{j=1}^d w_j\right] = 1$. Furthermore, $\{(w_j,p_j)\}_{j=1}^d$ satisfy Assumption~\ref{assm:bh_validity}(a) vacuously, since there are no non-null hypotheses. Assumption~\ref{assm:bh_validity}(b) is also satisfied, since each $w_j$ depends only on $p_j$, and hence, the pairs $\{(w_j,p_j)\}_{j=1}^d$ are independent across $j$.

Now, for any $j$ such that $p_j\leq \sqrt{2\alpha/d}$, we have $p_j/(dw_j) = \alpha/d$, which is the smallest possible cutoff of the BH procedure. Hence, any such hypothesis is rejected by the weighted BH procedure. Since all hypotheses are null, the FDR is therefore
\begin{align*}
    \mathbb P\left(\exists j:p_j\leq \sqrt{\frac{2\alpha}{d}}\right) = 1-\left(1-\sqrt{\frac{2\alpha}{d}}\right)^d\approx 1-\exp\left(-\sqrt{2\alpha d}\right)\to 1,\quad d\to \infty.
\end{align*}
Thus, for sufficiently large $d$, the FDR of the weighted BH procedure can be made arbitrarily close to 1, despite the weights satisfying Assumption~\ref{assm:bonf_valid}(b) and Assumption~\ref{assm:bh_validity}.

\section{Weighted dBH using co-sufficient weights}
\label{sec:app_dbh}
The dependence-adjusted Benjamini--Hochberg (dBH) procedure of \citet{dbh} modifies the usual BH procedure to guarantee theoretical FDR control under arbitrary $p$-value dependence, and in particular applies to the linear model. The vanilla BH procedure \citep{fdr} at nominal FDR level $\alpha$ finds the largest index $\hat i$ such that $p_{(\hat i)}\leq\frac{\alpha\hat i}{d}$, and rejects any hypothesis $H_j$ with $p_j\leq \hat \tau$, where $\hat \tau = \frac{\alpha\hat i}{d}$. The dBH procedure instead assigns each $H_j$ its own cutoff $\hat \tau_j$, chosen so that its \textit{FDR-contribution} is controlled at $\alpha/d$:
\begin{align*}
    \mathbb E_{H_j}\left[\frac{\mathbb I(p_j\leq \hat \tau_j)}{\hat R_j}\,\middle|\,\bm S^{(j)}\right]\leq \frac{\alpha}{d},
\end{align*}
where $\hat R_j$ is an estimate of $|\mathcal R\cup \{j\}|$ and $\mathcal R$ is the final rejection set. One plausible choice calibrates against a base BH procedure by setting $\hat R_j = |\hat{\mathcal R}^{{\rm BH}(\gamma \alpha)}\cup \{j\}|$, where $\hat{\mathcal R}^{{\rm BH}(\gamma \alpha)}$ is the rejection set of BH applied to the $p$-values at level $\gamma \alpha$; the authors use $\gamma = 0.9$ under arbitrary dependence. A hypothesis $H_j$ is then rejected if it passes two steps: (1) $p_j\leq \hat\tau_j$, and (2) a randomized pruning step applied to the hypotheses rejected in step (1). The pruning step depends only on the $\hat R_j$ of the step-(1) rejections and on externally generated noise, and it can only shrink the step-(1) rejection set (and hence is detrimental to the power); it is invoked only when the $\hat R_j$ of a rejected index $j$ fails to lower bound the eventual $|\mathcal R\cup \{j\}|$.

In \cite[Section 7.3]{dbh}, the authors note that a weighted variant of dBH is also valid: instead of controlling the above quantity at $\alpha/d$, one controls it at $\alpha\tilde w_j$, where $\tilde w_j$ is a function of $\bm S^{(j)}$ satisfying $\mathbb E[\sum_{j\in \mathcal H_0}\tilde w_j]\leq 1$ and $\mathcal H_0$ is the set of null indices. This makes our weights $\{\hat w_j\}_{j=1}^d$ an excellent candidate for use with dBH in the linear model, where an informative choice of weights should boost power just as it did for the weighted Bonferroni procedure. For completeness, we state the validity of the resulting procedure in the following theorem, whose proof follows from \cite[Theorem 2 and Remark 2.1]{dbh}
\begin{theorem}
    \label{thm:fdr_validity}
    Suppose p-values $\{p_j\}_{j=1}^d$ satisfy Assumption~\ref{assm:p_val_suff}. Then, for all $\alpha\in (0,1)$, the weighted dBH procedure described above, run at nominal FDR level $\alpha$ with weights $\{\hat w_j\}_{j=1}^d$, controls the FDR at level $\alpha$.
\end{theorem}
As with dBH, the weighted procedure must calibrate its underlying base weighted BH procedure; we can use $\hat R_j = |\hat{\mathcal R}^{{\rm BH}(\hat{\bm w};\gamma \alpha)}\cup \{j\}|$, where $\hat{\mathcal R}^{{\rm BH}(\hat{\bm w};\gamma \alpha)}$ is the rejection set of the weighted BH procedure at nominal FDR $\gamma \alpha$ with weight $\hat w_j$ for $H_j$. To leverage sparsity, a further natural choice is to take $\{p_j\}_{j=1}^d$ to be the $\ell$-test $p$-values. Under these choices, the cutoff calibration takes the form
\begin{align}
    \label{eqn:calibration}
    \mathbb E_{H_j}\left[\frac{\mathbb I(p_j/(dw_j)\leq \hat \tau_j)}{\hat R_j}\,\middle|\,\bm S^{(j)}\right]\leq \alpha \hat w_j.
\end{align}
In practice, the left-hand side has no closed form and must be evaluated by Monte Carlo sampling. Calibrating $\hat\tau_j$ to reasonable accuracy requires estimating this expectation quite precisely, hence many Monte Carlo draws, and recomputing the weights for each draw is computationally very expensive. The original paper \cite{dbh} proposes homotopy algorithms for evaluating this expectation and calibrating the base BH procedure with $t$-test $p$-values, but these do not extend to the weighted case.\footnote{One can calibrate the base BH procedure to obtain $\hat R_j$, while retaining the weights only on the right-hand side of Equation~\ref{eqn:calibration}; we found, however, that this variant frequently triggers the randomized pruning step, significantly affecting its power.} We therefore leave a computationally feasible version of this weighted analog to future work.

\section{Empirical exploration of weights variability}
\label{sec:weights_variability}
To empirically assess the variability of our weights, we conduct the following experiment. For a given response $\bm y$, design matrix $\bm X$, and signal index $j$, we assess the conditional variability of $\hat w_j$ by generating $M$ realizations of the weight defined in \eqref{eqn:plug_in_w} with $B=1$, using $M$ conditionally independent draws of $\tilde{\bm y}^{(j)}$ from the conditional distribution of $\bm y \mid \bm S^{(j)}$ under $H_j$. Denote these realizations by $\{\hat w_j^{(1)},\cdots, \hat w_j^{(M)}\}$. We then estimate the coefficient of variation by taking the ratio of the sample standard deviation to the sample mean of these weights, denoting the resulting quantity by $\omega_{j,M}$, which can then be averaged across independently generated data sets.

We estimate this measure under four selected configurations from Section~\ref{sec:expt}: Setting (a) with $n=300$, $d=100$, discrete signal generation, $k\in \{1,5\}$, and $\rho \in \{0,0.9\}$. Using $M=50$, we summarize the estimated values of $\omega_{j,M}$ in Table~\ref{tab:weights_cv}. We find low estimated coefficient of variation for our recommended weights, based on $\mathcal A^{\rm LASSO}$, supporting our claim of low variability in Section~\ref{sec:reccomendaion}.

Although the coefficient of variation is higher for our computationally efficient alternative based on $\mathcal A^{\rm dST}$, we do not investigate this phenomenon further or attempt to modify the procedure. This is because we found that the power benefits of increasing $B$ for this importance measure is at most modest, and non-zero in only the dense regimes. Furthermore, the simulations in Section~\ref{sec:expt} and Appendix~\ref{sec:detailed_simulation_results} already demonstrate the power gain of the procedure corresponding to $B=1$, especially under sparsity, and increasing $B$ would only increase its computational expense.

\begin{table}[h]
    \centering
    \begin{tabular}{|c|c|c|c|}
    \hline
        $\rho$ & $k$ & Estimated $\omega_{j,M}$ using $\mathcal A^{\rm LASSO}$ & Estimated $\omega_{j,M}$ using $\mathcal A^{\rm dST}$\\
    \hline
        0 & 1& 0.06 & 0.22\\
        0 & 5& 0.05 & 0.26 \\
        0.9 &1 & 0.04 & 0.1 \\
        0.9 &5 & 0.03 & 0.55\\
    \hline
    \end{tabular}
    \caption{Estimates of $\omega_{j,M}$ based on weights formed using both the importance measures $\mathcal A^{\rm LASSO}$ and $\mathcal A^{\rm dST}$. We report the mean estimate from 200 independent replicates of each setting.}
    \label{tab:weights_cv}
\end{table}

\section{Why \texorpdfstring{$\mathcal A_j$~} ddoes not depend on the last component of \texorpdfstring{$\bm S^{(j)}$} ~}
\label{sec:app_conditioning_signal}

In Section~\ref{sec:reccomendaion}, we restricted our variable importance measures to depend only on the first $d-1$ components of the sufficient statistic $\bm S^{(j)} = (\bm X_{-j}^T\bm y, \|(\bm I_n - \bm P_{-j})\bm y\|^2)$. In this section, we provide further justification for excluding the last component.

Recall that our weights are defined as
\[
    w_j = \mathbb E_{H_j}\left[\frac{\mathcal A_j(\bm y)}{\sum_{i=1}^d \mathcal A_i(\bm y)}\middle| \bm S^{(j)}\right]
    = \mathbb E_{H_j}\left[\frac{\mathcal A_j(\tilde{\bm y}^{(j)})}{\sum_{i=1}^d \mathcal A_i(\tilde{\bm y}^{(j)})}\middle| \bm S^{(j)}\right],
\]
where $\tilde{\bm y}^{(j)}\sim \bm y\mid \bm S^{(j)}$, under $H_j$. \cite[Proposition E.1]{luo2022} gives a closed-form representation of this draw: One can realize $\tilde{\bm y}^{(j)}$ as
\begin{align}
\label{eqn:y_tilde_sample}
    \tilde{\bm y}^{(j)} = \bm P_{-j}\bm y + \hat \sigma_j \bm V^{(j)}\bm u,
\end{align}
where $\bm V^{(j)}$ is any orthogonal matrix for ${\rm Col.sp.}(\bm X_{-j})^{\perp}$ and $\bm u \sim {\rm Unif}(\mathbb S^{n-d})$ is drawn independently of the data, where ${\rm Col.sp.}$ returns the column space of a matrix.
Now, let us expand our notations and let $\bm S^{(j)}(\bm y)$ denote the sufficient statistic under $H_j$ computed from the response $\bm y$. By construction, for every $j$,
\[
    \bm S^{(j)}(\tilde{\bm y}^{(j)}) = \bm S^{(j)}(\bm y).
\]
For $i\neq j$, however,
\[
    \bm S^{(i)}(\tilde{\bm y}^{(j)})
    = \left(\bm X_{-i}^T\tilde{\bm y}^{(j)},
    \|(\bm I_n - \bm P_{-i})\tilde{\bm y}^{(j)}\|^2\right).
\]
The following lemma, proved in Appendix~\ref{sec:proof_second_component_dist}, provides a useful representation of the last component.

\begin{lemma}
\label{lem:second_component_dist}
Define
\[
    \bm e_i = \frac{(\bm I_n - \bm P_{-i})\bm X_i}{\|(\bm I_n - \bm P_{-i})\bm X_i\|},
    \qquad
    \rho_{ij} = \bm e_i^T\bm e_j.
\]
Suppose $\bm V^{(j)}$ is chosen such that its first column is $\bm e_j$ and its remaining columns form an orthonormal basis for the orthogonal complement of ${\rm Col.sp.}(\bm X)$. Then,
\[
    \|(\bm I_n - \bm P_{-i})\tilde{\bm y}^{(j)}\|^2
    = (\bm e_i^T\bm P_{-j}\bm y + \rho_{ij}\hat \sigma_j u_1)^2
    + (1-u_1^2)\hat \sigma_j^2.
\]
\end{lemma}

We next consider a heuristic approximation of this expression. Since $u_1$ is the first component of a vector $\bm u$ distributed uniformly on $\mathbb S^{n-d}$, it concentrates around zero as $n-d$ grows,
%In particular, $u_1^2 = O_p((n-d+1)^{-1})$, 
so that for an $n$ large enough compared to $d$, $1-u_1^2\approx 1$. Then Lemma~\ref{lem:second_component_dist} suggests the approximation
\[
      \|(\bm I_n - \bm P_{-i})\tilde{\bm y}^{(j)}\|^2 \approx (\bm e_i^T\bm P_{-j}\bm y + \rho_{ij}\hat \sigma_j u_1)^2 + \hat \sigma_j^2.
\]
This approximation suggests that the last component of the sufficient statistic for any variable $i\neq j$, evaluated at $\tilde{\bm y}^{(j)}$, approximately consists of the last component of $\bm S^{(j)}$, namely $\hat\sigma_j^2$, together with an additional non-negative term. Importantly, this latter term is also influenced by the signal strength $\beta_j$.

% More precisely, the distribution of this term is characterized by
% \[
%     \bm e_i^T\bm P_{-j}\bm y \sim \mathcal N(\|(\bm I_n - \bm P_{-i})\bm X_i\|\beta_i-\bm e_i^T\bm e_j \|(\bm I_n - \bm P_{-j})\bm X_j\|\beta_j, \sigma^2(1-(\bm e_i^T\bm e_j)^2)).
% \]

Thus, the signal associated with $\beta_j$ can \textit{bleed into} the last component of $\bm S^{(i)}(\tilde{\bm y}^{(j)})$ for other variables $i\neq j$, through a term that may grow with $|\beta_j|$ and is at least as large as the last component of $\bm S^{(j)}$. Consequently, if an importance measure $\mathcal A_j$ depends on the last component of the sufficient statistic in such a way that larger values of this component yield greater importance, then a strong signal at $j$ may increase not only $\mathcal A_j(\tilde{\bm y}^{(j)})$, but also $\mathcal A_i(\tilde{\bm y}^{(j)})$ for $i\neq j$, potentially to an even greater extent because of the additional square term that can also depend on $|\beta_j|$. Since our weights are obtained by normalizing these importance measures, this effect can reduce the relative weight assigned to $j$, thereby weakening the ability of the weights to distinguish signal variables from null variables. This motivates our choice to exclude the last component of the sufficient statistic when constructing our importance measures.

\section{Tricks for faster computation}
\label{sec:faster_compute}
\subsection{Faster computation of the co-sufficient weights}
\label{sec:faster_compute_csw}
We first discuss high-level tricks that speed up the computation of the co-sufficient weights for both $\mathcal A^{\rm LASSO}$ and $\mathcal A^{\rm dST}$. Two things that can be sped up are the computation of $(\bm X_{-j}^T\bm X_{-j})^{-1/2}$ and $\tilde{\bm y}^{(j)}$ for all $j\in[1:d]$, as follows.
\begin{itemize}
    \item \textbf{Computing $(\bm X_{-j}^T\bm X_{-j})^{-1/2}$ for all $j$.} We use the Cholesky square root rather than the principal square root, since the former admits the computational speed-ups described next.

    First compute $\bm G = \bm X^T\bm X$, its inverse $\bm H = \bm G^{-1}$, and the Cholesky square root $\bm L$ of $\bm G^{-1}$. Then, for any $j$, the Schur-complement formula gives $(\bm X_{-j}^T\bm X_{-j})^{-1} = \bm H_{-j,-j} - \bm v_j\bm v_j^T$, where
    \[
        \bm v_j = \frac{\bm H_{-j,j}}{\sqrt{H_{jj}}}.
    \]
    Efficient computations \citep[Section 6.5.4]{golubandvanloan2013} then yield the Cholesky square root of $(\bm X_{-j}^T\bm X_{-j})^{-1} = \bm H_{-j,-j} - \bm v_j\bm v_j^T$ from $\bm L$. Together, these steps make the Cholesky square root of $(\bm X_{-j}^T\bm X_{-j})^{-1}$ much faster to compute across $j$. We denote this Cholesky square root by $\bm L_j$, so that $\bm L_j\bm L_j^T = (\bm X_{-j}^T\bm X_{-j})^{-1}$, and set $(\bm X_{-j}^T\bm X_{-j})^{-1/2} := \bm L_j^T$ throughout. This choice gives $\bm z^{(j)} = \bm L_j^T\bm X_{-j}^T\bm y$ and $\bm W^{(j)}_{-j} = \bm L_j^{-1}$.
    \item \textbf{Computing $\tilde{\bm y}^{(j)}$ for all $j$.} By \eqref{eqn:y_tilde_sample},
    \[
        \tilde{\bm y}^{(j)} = \bm P_{-j}\bm y + \|(\bm I_n - \bm P_{-j})\bm y\|\, \bm V^{(j)}\bm u,
    \]
    where $\bm P_{-j} = \bm X_{-j}(\bm X_{-j}^T\bm X_{-j})^{-1}\bm X_{-j}^T$, the matrix $\bm V^{(j)}$ is an orthonormal basis for the orthogonal complement of the column space of $\bm X_{-j}$, and $\bm u \sim {\rm Unif}(\mathbb S^{n-d})$. Here the bottlenecks are $(\bm X_{-j}^T\bm X_{-j})^{-1}$ and $\bm V^{(j)}$. The former can again be computed from $\bm G^{-1}$ using the Schur-complement formula, exactly as in the previous item. For $\bm V^{(j)}$, one computes a single orthonormal basis $\bm V^{\perp}$ for the orthogonal complement of the column space of $\bm X$; then, for any $j$, $\bm V^{(j)}$ is obtained by prepending the column $\frac{(\bm I_n - \bm P_{-j})\bm X_j}{\|(\bm I_n - \bm P_{-j})\bm X_j\|}$ to $\bm V^{\perp}$. The vector $(\bm I_n - \bm P_{-j})\bm X_j = \bm X_j - \bm P_{-j}\bm X_j$ can then be evaluated from the quantities already computed above.
\end{itemize}
These tricks alone yield significant speed-ups in computing all $d$ of the $\bm z^{(j)}$ and $\bm W^{(j)}$. Beyond them, evaluating $\mathcal A^{\rm LASSO}$ requires running cross-validated LASSO on
\[
    \left((\bm X_{-i}^T\bm X_{-i})^{-1/2}\bm X_{-i}^T\tilde{\bm y}^{(j)},\ (\bm X_{-i}^T\bm X_{-i})^{-1/2}\bm X_{-i}^T\bm X\right)
\]
for all $i$ and $j$. In principle one could use warm starts to avoid rerunning LASSO from scratch for each $i$ and $j$, but we could not find a reliable way to do so with the standard \texttt{glmnet()} implementation in \textbf{R}, and we leave this to future work.

The computation of $\mathcal A^{\rm dST}$ across $i$ and $j$, on the other hand, can be sped up substantially. For fixed $\lambda$, $i$, and $j$, computing $\mathcal A^{\rm dST}_i(\tilde{\bm y}^{(j)})$ requires the OLS fit of $(\bm X_{-i}^T\bm X_{-i})^{-1/2}\bm X_{-i}^T\tilde{\bm y}^{(j)}$ on $\bm W^{(i)}_{-i} = (\bm X_{-i}^T\bm X_{-i})^{-1/2}\bm X_{-i}^T\bm X_{-i}$, whose bottleneck is the inverse $({\bm W^{(i)}_{-i}}^T\bm W^{(i)}_{-i})^{-1} = (\bm X_{-i}^T\bm X_{-i})^{-1}$ which can again be computed from $\bm G^{-1}$ via the Schur-complement formula across $i$. The denominator of $\mathcal A^{\rm dST}_i(\tilde{\bm y}^{(j)})$, namely $\|\bm W^{(i)}_i\|^2 = \|\bm P_{-i}\bm X_i\|^2$, has the same bottleneck---the inversion of $\bm X_{-i}^T\bm X_{-i}$---and hence benefits similarly from the above computation. Thus, once $\bm G^{-1}$ is computed, it can be shared across all evaluations of $\mathcal A^{\rm dST}_i(\tilde{\bm y}^{(j)})$. Next, we describe the cross-validation step used to select $\lambda$ for $\mathcal A^{\rm dST}$.

\subsection{Cross-validation for computing the distilled soft-thresholding measure}
\label{sec:cv_st}

To compute $\mathcal A_i^{\rm dST}(\bm y)$ for a given $i$, we select $\hat\lambda_i^{\rm ST}$ by cross-validation. Specifically, we partition $[1:d-1]$ into $K=10$ folds, sequentially holding out one fold for testing and using the remaining folds for training. At each iteration, $(\bm z^{(i)},\bm W^{(i)})$ is partitioned into training and testing sets, denoted by $(\bm z_{\rm tr}^{(i)},\bm W_{\rm tr}^{(i)})$ and $(\bm z_{\rm tst}^{(i)},\bm W_{\rm tst}^{(i)})$, respectively. Because $\bm W_{{\rm tr},-i}^{(i)}$ has fewer rows than columns, we cannot compute the usual OLS estimator to soft-threshold. Thus, we replace it with the ridgeless least-squares fit:
\begin{align}
    \label{eqn:min_norm_dst_cv}
    \arg\min_{\bm \beta}~\|\bm \beta\|^2\quad\textrm{subject to}\quad\bm W^{(i)}_{{\rm tr},-i}\bm \beta = \bm z_{\rm tr}^{(i)}.
\end{align}

We then obtain the $\lambda$-soft-thresholded version of this estimate, denoted by $\hat{\bm \beta}_{\rm ST, tr}^{\lambda,-i}$, and the OLS coefficient from regressing the residual $\bm z_{\rm tr}^{(i)} - \bm W_{{\rm tr},-i}^{(i)}\hat{\bm \beta}_{\rm ST, tr}^{\lambda,-i}$ on $\bm W^{(i)}_{{\rm tr},i}$, denoted by $\hat\gamma^{\lambda}_{{\rm tr},i}$. We evaluate the prediction error on the held-out fold as
\begin{align}
\label{eqn:lambda_mse}
    \left\|\left( \bm z_{\rm tst}^{(i)} - \bm W_{{\rm tst},-i}^{(i)}\hat{\bm \beta}_{\rm ST, tr}^{\lambda,-i}\right) - \hat\gamma^{\lambda}_{{\rm tr},i}\bm W^{(i)}_{{\rm tst},i}\right\|^2.
\end{align}
Averaging this quantity across the $K$ folds yields the cross-validated mean-squared error (MSE) for a given $\lambda$. We select the $\lambda$ that minimizes this MSE.

Next, note that the standard solution to \eqref{eqn:min_norm_dst_cv} is ${\bm W^{(i)}_{{\rm tr},-i}}^+\bm z_{\rm tr}^{(i)}$, where $\bm A^+$ denotes the Moore--Penrose pseudoinverse of a matrix $\bm A$. However, computing the pseudoinverse can be expensive. We instead propose the following computationally cheaper alternative.

Recall the Cholesky square root $\bm L_i$ of $(\bm X_{-i}^T\bm X_{-i})^{-1}$ from Appendix~\ref{sec:faster_compute_csw}, which is already available before cross-validation, and define $\bm Q = \bm L_i^T\bm L_i$. Since $\bm W_{-i}^{(i)} = \bm L_i^{-1}$, the reparameterization $\bm \theta = \bm W_{-i}^{(i)}\bm \beta$ gives $\bm \beta = \bm L_i\bm \theta$, so \eqref{eqn:min_norm_dst_cv} is equivalent to
\begin{align}
\label{eqn:min_norm_dst_cv2}
     \arg\min_{\bm \theta}~\|\bm L_i\bm \theta\|^2\quad\textrm{subject to}\quad \bm \theta_{\rm tr} = \bm z^{(i)}_{\rm tr}.
\end{align}
To solve this problem, observe that
\begin{align*}
    \|\bm L_i\bm \theta\|^2
    &= \bm \theta^T\bm Q \bm \theta\\
    &= \bm \theta_{\rm tr}^T\bm Q_{\rm tr, tr}\bm \theta_{\rm tr} + \bm \theta_{\rm tst}^T\bm Q_{\rm tst, tst}\bm \theta_{\rm tst} + 2\bm \theta_{\rm tr}^T\bm Q_{\rm tr,tst}\bm \theta_{\rm tst}\\
    &= {\bm z_{\rm tr}^{(i)}}^T\bm Q_{\rm tr, tr}\bm z_{\rm tr}^{(i)} + \bm \theta_{\rm tst}^T\bm Q_{\rm tst, tst}\bm \theta_{\rm tst} + 2{\bm z_{\rm tr}^{(i)}}^T\bm Q_{\rm tr,tst}\bm \theta_{\rm tst}.
\end{align*}
The first term is constant under the constraint in \eqref{eqn:min_norm_dst_cv2}. Hence, the solution is $\bm \theta^* = (\bm z^{(i)}_{\rm tr}, \bm \theta^*_{\rm tst})$, where
\[
    \bm \theta^*_{\rm tst} = \arg\min_{\bm \theta_{\rm tst}} \left\{\bm \theta_{\rm tst}^T\bm Q_{\rm tst, tst}\bm \theta_{\rm tst} + 2{\bm z_{\rm tr}^{(i)}}^T\bm Q_{\rm tr,tst}\bm \theta_{\rm tst}\right\} = - \bm Q_{\rm tst, tst}^{-1}\bm Q_{\rm tst, tr}\bm z^{(i)}_{\rm tr}.
\]
Thus, the solution to \eqref{eqn:min_norm_dst_cv} is $\bm \beta^* = \bm L_i\bm \theta^*$. Importantly, computing $\bm \theta^*_{\rm tst}$ only requires solving the linear system $\bm Q_{\rm tst, tst}\bm \theta_{\rm tst} = -\bm Q_{\rm tst, tr}\bm z^{(i)}_{\rm tr}$, which can be done efficiently using a Cholesky-based solver without explicitly computing $\bm Q_{\rm tst, tst}^{-1}$.

Our implementation exploits further computational savings. For each fixed $i$, we use the same training-test fold splits across all $j$ when computing $\mathcal A_i^{\rm dST}(\tilde{\bm y}^{(j)})$. Consequently, $\bm Q_{\rm tst,tr}$ and $\bm Q_{\rm tst,tst}$, along with the Cholesky decomposition of $\bm Q_{\rm tst,tst}$ required to solve the linear system for $\bm \theta^*_{\rm tst}$, can be computed once for each $i$ and fold split and reused across all $j$.

Moreover, once this ridgeless least-squares fit is available, evaluating the MSE in \eqref{eqn:lambda_mse} for different values of $\lambda$ requires only soft-thresholding this solution. Thus, evaluating the cross-validation criterion over a grid of $\lambda$ values incurs no additional computationally expensive matrix operations.

Finally, for each $i$ and $j$, we choose a grid of 50 equally spaced $\lambda$ values on the logarithmic scale between $b^{(i,j)}_{\max}$ and $10^{-3}b^{(i,j)}_{\max}$, where
\[
    b^{(i,j)}_{\max} = \max_{k}~\left\lvert \hat\beta_k^{(i,j)} \right\rvert,\qquad \hat\beta_k^{(i,j)} = (\bm X_{-i}^T\bm X_{-i})^{-1}\bm X_{-i}^T\tilde{\bm y}^{(j)}.
\]
\section{Proofs}
\subsection{Proof of Theorem~\ref{thm:wholm_validity}}
\label{sec:proof_wholm_validity}
\begin{proof}
    Consider any subset $I\subseteq [1:d]$ and define the weights
\[
    \hat w_{j}^{(I)} = \frac{\mathcal A_j(\tilde{\bm y}^{(j)})}{\sum_{i\in I}\mathcal A_i(\tilde{\bm y}^{(j)})},
\]
where $\tilde{\bm y}^{(j)}$ is sampled from the conditional distribution of $\bm y\mid \bm S^{(j)}$ under $H_j$. Note that the p-values $\{p_j\}_{j\in I}$ and the weights $\{\hat w^{(I)}_j\}_{j\in I}$ satisfy Assumption~\ref{assm:bonf_valid} by Theorem~\ref{thm:validity_weights}. Hence, it follows from \cite[Theorem 4.3]{ignatiadisetal2023} that the weighted $I$-Bonferroni test using weights $\{\hat w^{(I)}_j\}_{j\in I}$ is valid. Thus, by the closure principle \citep{marcusetal1976} (discussed in Section~\ref{sec:fwer_methods}), the closure of these Bonferroni tests is a valid FWER-controlling procedure. We will show that Algorithm~\ref{alg:wholm} is precisely this closure.

First, suppose that Algorithm~\ref{alg:wholm} rejects an index $i\in [1:d]$. We will show that this index is also rejected by the closure. Let $\emptyset=\hat R_0\subset \hat R_1 \subset \cdots \subset \hat R_K$ denote the rejection sets of Algorithm~\ref{alg:wholm}, which makes $K\geq 1$ rejections, and let $i = j_m$ for some $1\leq m\leq K$. Now consider any set $I$ containing $i$ and define $m^* = \min\{r: j_r\in I\}$. This minimum is well-defined since the set is nonempty, as it contains at least $m$. Because $j_{m^*}$ was rejected by Algorithm~\ref{alg:wholm}, we have
\[
    p_{j_{m^*}} \leq \alpha \hat w_{j_{m^*}}^{(\hat R^c_{m^* - 1})}.
\]
By the definition of $m^*$, we also have $j_r \notin I$ for all $r<m^*$ (if any; otherwise, the statement holds vacuously). Hence, $\hat R_{m^* - 1}$ is disjoint from $I$, implying that $I\subseteq \hat R_{m^* - 1}^c$. By the definition of the weights, this yields
\[
    \hat w_{j_{m^*}}^{(\hat R^c_{m^* - 1})} = \frac{\mathcal A_{j_{m^*}}(\tilde{\bm y}^{(j_{m^*})})}{\sum_{i\in \hat R^c_{m^* - 1}}\mathcal A_i(\tilde{\bm y}^{(j_{m^*})})} \leq  \frac{\mathcal A_{j_{m^*}}(\tilde{\bm y}^{(j_{m^*})})}{\sum_{i\in I}\mathcal A_i(\tilde{\bm y}^{(j_{m^*})})} =  \hat w_{j_{m^*}}^{(I)}.
\]
Consequently, $p_{j_{m^*}} \leq \alpha \hat w_{j_{m^*}}^{(I)}$. Since $j_{m^*}\in I$, the weighted $I$-Bonferroni test rejects $H_I$. As this holds for every $I$ containing $i$, the closure also rejects $i$.

Conversely, suppose that Algorithm~\ref{alg:wholm} does not reject some index $i$. We will show that the closure does not reject $i$ either. For this, it suffices to identify a set $I$ containing $i$ such that the weighted $I$-Bonferroni test does not reject $H_I$. Since Algorithm~\ref{alg:wholm} does not reject $i$, we must have $i\in \hat R_{K}^c$ and $K<d$. Define $I = \hat R_K^c$, so that none of the indices in $I$ has been rejected by Algorithm~\ref{alg:wholm}. By the stopping rule of the algorithm, for every $j\in I$, we must have $p_j> \alpha \hat w_j^{(I)}$; otherwise, $j$ would have been rejected by Algorithm~\ref{alg:wholm}. Thus, the weighted $I$-Bonferroni test does not reject $H_I$. Since $i\in I$, the closure does not reject $i$ either. This completes the proof of equivalence.
\end{proof}

\subsection{Proof of Lemma~\ref{lem:second_component_dist}}
\label{sec:proof_second_component_dist}
\begin{proof}
Let $\bm V^\perp$ be an orthonormal basis for the orthogonal
complement of ${\rm Col.sp.}(\bm X)$. Since
\[
{\rm Col.sp.}(\bm X_{-i})^\perp
= \operatorname{span}(\bm e_i)\oplus
{\rm Col.sp.}(\bm X)^\perp,
\]
with $\bm e_i$ defined as in the statement of Lemma~\ref{lem:second_component_dist}, we have
\[
\bm I_n-\bm P_{-i}
=\bm e_i\bm e_i^\top+\bm V^\perp(\bm V^\perp)^\top.
\]
Moreover, by the choice of $\bm V^{(j)}$,
\[
\tilde{\bm y}^{(j)}
=\bm P_{-j}\bm y+
\hat\sigma_j\bm e_j u_1+
\hat\sigma_j\bm V^\perp\bm u_{-1},
\]
where $\bm u_{-1}=(u_2,\ldots,u_{n-d+1})^\top$.
Consequently, using the orthogonality of $\bm e_i$ and
${\rm Col.sp.}(\bm V^\perp)$,
\begin{align*}
\|(\bm I_n-\bm P_{-i})\tilde{\bm y}^{(j)}\|^2
&=\bigl(\bm e_i^\top\bm P_{-j}\bm y+
\rho_{ij}\hat\sigma_j u_1\bigr)^2
+\hat\sigma_j^2\|\bm u_{-1}\|^2\\
&=\bigl(\bm e_i^\top\bm P_{-j}\bm y+
\rho_{ij}\hat\sigma_j u_1\bigr)^2
+(1-u_1^2)\hat\sigma_j^2,
\end{align*}
where the last equality follows from $\|\bm u\|=1$.
\end{proof}

\section{Further details on HIV drug resistance dataset analysis}
\subsection{Genetic mutations identified by the various procedures}
\label{sec:gene_hiv}
\begin{figure}[!h]
    \centering
    \includegraphics{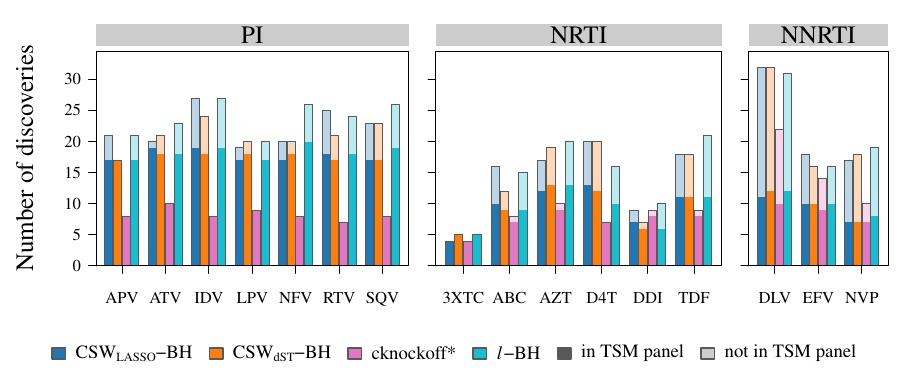}
    \caption{Number of genetic mutations identified by the various FDR controlling methods for various drug-specific regressions (marked on the x-axis) at nominal level 0.1 and their split based on presence/absence in the TSM list.}
    \label{fig:hiv_fdr}
\end{figure}

\begin{figure}[!h]
    \centering
    \includegraphics{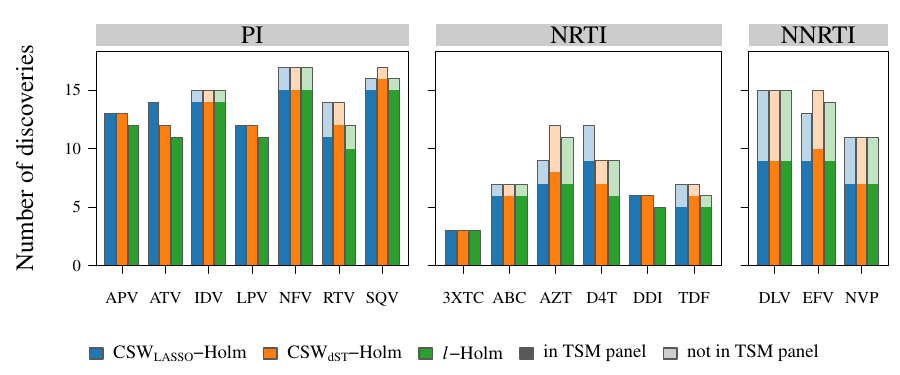}
    \caption{Number of genetic mutations identified by the various FWER controlling methods for various drug-specific regressions (marked on the x-axis) at nominal level 0.05 and their split based on presence/absence in the TSM list.}
    \label{fig:hiv_fwer}
\end{figure}

Figures~\ref{fig:hiv_fdr} and \ref{fig:hiv_fwer} summarize the genetic mutations identified by the FDR-controlling procedures (at a nominal level of 0.1) and the FWER-controlling procedures (at a nominal level of 0.05), respectively, considered in Section~\ref{sec:real_data} for each of the 16 drug-specific regressions. These drugs are classified into three groups: protease inhibitors (PIs), nucleoside reverse transcriptase inhibitors (NRTIs), and nonnucleoside reverse transcriptase inhibitors (NNRTIs). The results in both figures are organized according to these groups.

As alluded to in Section~\ref{sec:real_data}, many prior works \citep{knockoff,dbh,luo2022} have also created such a summary, which is obtained by aggregating the raw discoveries of these methods to a gene-level summary. As also has been done in prior works, we compare these genetic discoveries against a pre-published list of influential genes from the TSM list. As we observed in Section~\ref{sec:real_data}, for each of these regressions the number of discoveries by our weighted and the vanilla unweighted procedures are very similar.

\subsection{Further exploring the role of weighting}
\label{sec:weighting_hiv}
\begin{figure}[!h]
    \centering
    \includegraphics[width=1.0\textwidth]{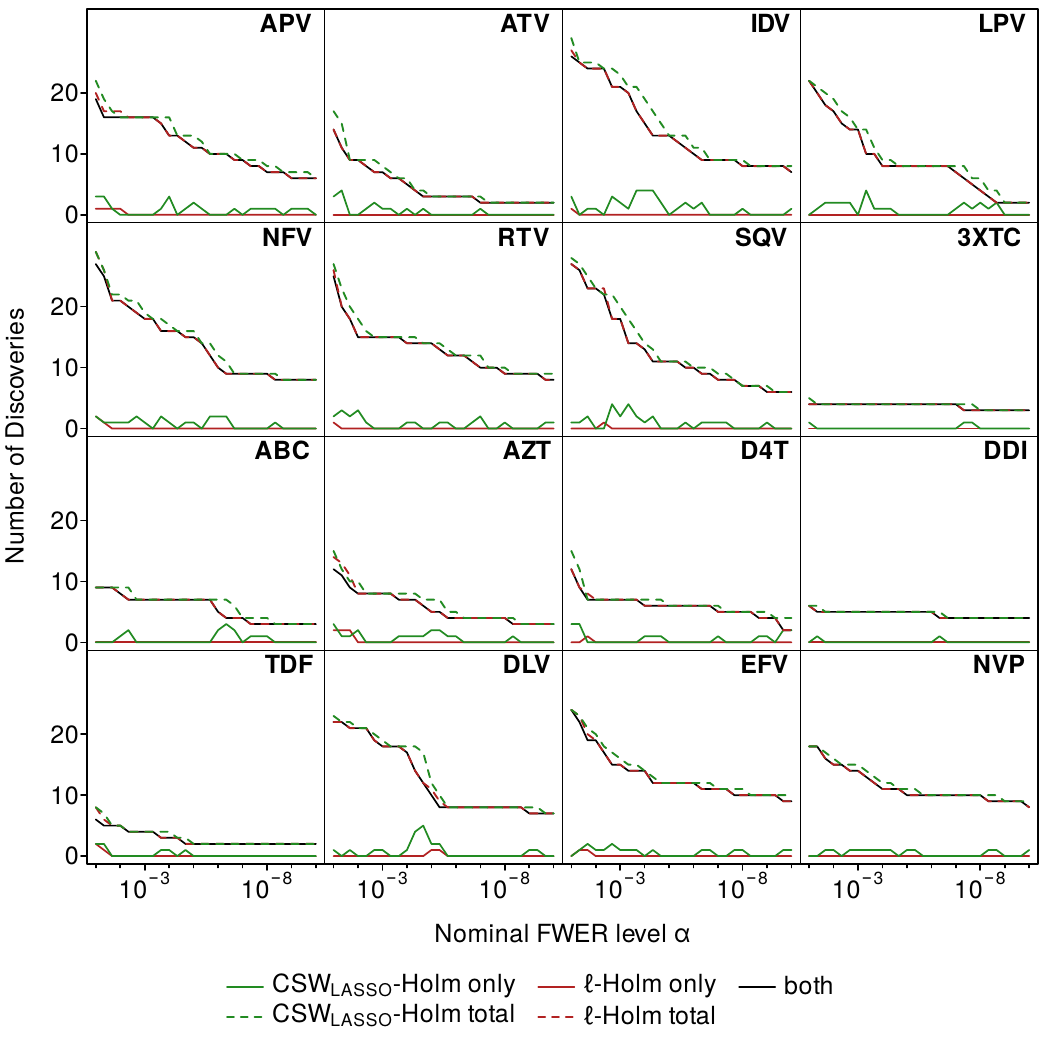}
    \caption{The number of discoveries by $\CSWLASSOHolm$ and $\ell$-Holm for varying levels of nominal FWER for all the 16 regressions.}
    \label{fig:hiv_holm_overlap}
\end{figure}

\begin{figure}[!h]
    \centering
    \includegraphics[width=1.0\textwidth]{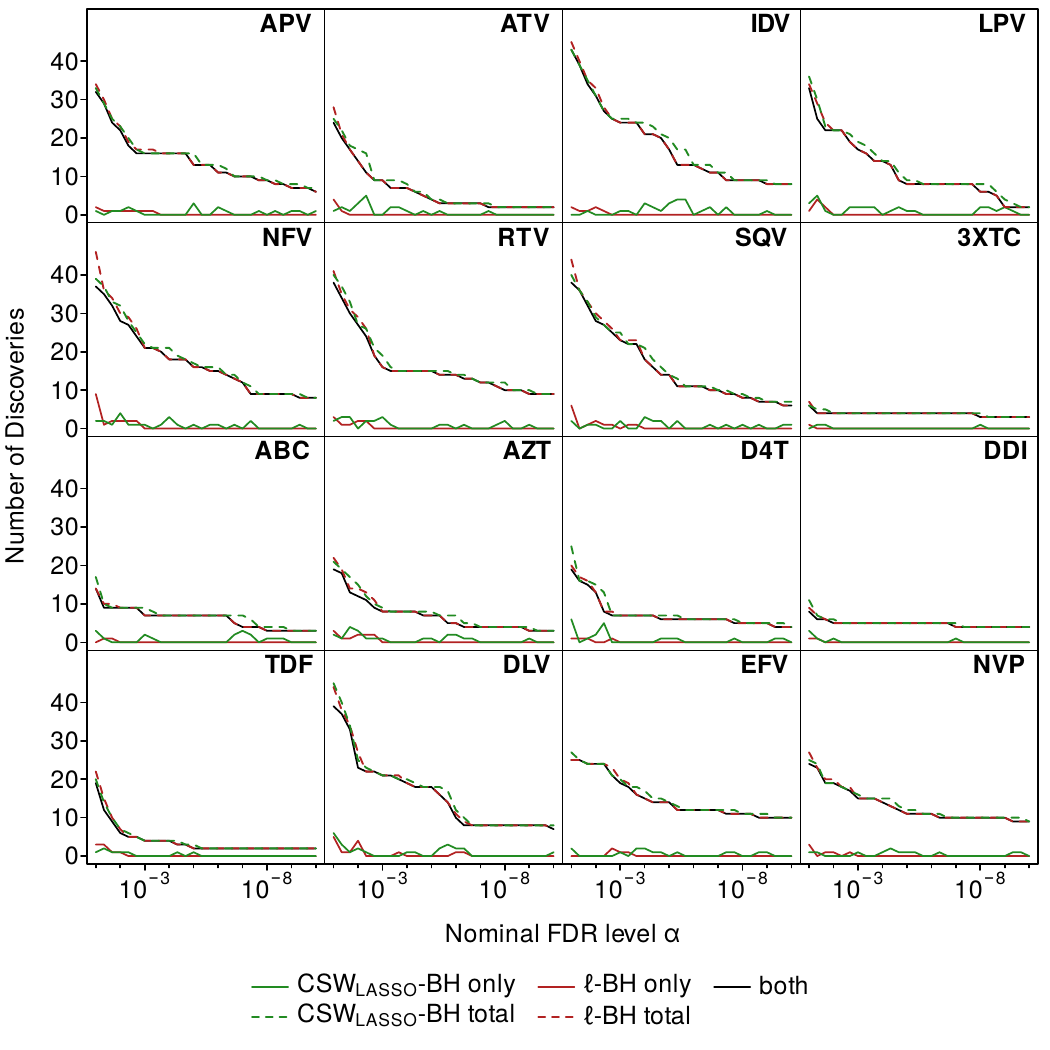}
    \caption{The number of discoveries by $\CSWLASSOBH$ and $\ell$-BH for varying levels of nominal FDR for all the 16 regressions.}
    \label{fig:hiv_bh_overlap}
\end{figure}

In Figures~\ref{fig:hiv_holm_overlap} and \ref{fig:hiv_bh_overlap}, we explore the number of discoveries for each of these procedures when the nominal FWER and FDR, respectively, are reduced. Particularly, for each of the 16 regressions, we track three quantities for these different nominal error levels: the common number of discoveries made by both the methods, the number of discoveries by the weighted procedure that is not discovered by the corresponding unweighted procedure, and the number of discoveries not discovered by the weighted procedure that are discovered by the corresponding unweighted method. As discussed in Section~\ref{sec:real_data}, these curves also demonstrate that our weighted procedures often make additional discoveries over their unweighted counterparts for tightening levels of the error rates, thereby demonstrating the benefit of weighting.

\section{Further details on simulations}
\subsection{Detailed simulation results}
\label{sec:detailed_simulation_results}
Here we present the complete set of empirical results for the simulation settings described in Section~\ref{sec:expt}. Methods that do not exhibit consistent empirical FDR control across settings---knockoffs, GM, MDS and OATK---are shown with dashed lines. In the $n/d = 1.5$ regime, the knockoff-based procedures (knockoffs, knockoffs+, cknockoff and cknockoff$^\ast$) are omitted, as they require $n/d \geq 2$. In the $d = 1000$ regime, $\CSWLASSOHolm$ and $\CSWLASSOBH$ are omitted for computational reasons. 

We next describe the choice of the amplitude $A$. These amplitudes were calibrated so that the power of the methods under comparison lies in an interesting range: not too close to 0 or 1. Accordingly, $A$ is increased in settings that are intrinsically harder.

For regimes (b) and (c), the values of $A$ are shown on the x-axes of the corresponding figures. For regime (a), $A$ is held fixed across sparsity levels and set to
\[
    A = 3.5\times \kappa_{\rm design}\times \kappa_{\rm signal}\times \kappa_{\rm FWER},
\]
where
\begin{align*}
    \kappa_{\rm design} &= \begin{cases}
        1, & \rho = 0,\\
        1.3, & \rho = 0.5,\ n/d = 1.5,\\
        1, & \rho = 0.5,\ n/d = 3,\\
        3, & \rho = 0.9,\ n/d = 1.5,\\
        2.2, & \rho = 0.9,\ n/d = 3,
    \end{cases}\qquad
    \kappa_{\rm signal} = \begin{cases}
        1, & \text{discrete signal generation},\\
        1.5, & \text{continuous signal generation},
    \end{cases}
\end{align*}
and $\kappa_{\rm FWER} = 1.3$ for the FWER simulations and $1$ for the FDR simulations. There are two exceptions to this rule. First, for the FWER simulations with $\rho = 0.9$ and $d = 1000$, we set $A = 15$ directly (recall that for $d = 1000$ we only consider discrete signal generation). Second, for the MCC design, every non-null coefficient is set to $+A$ with $A = 4.61$, for both the FDR and FWER simulations; \cite[Section 5.1]{luo2022} chose this value so that the BH procedure at nominal FDR level $0.2$ has power $0.5$.

\begin{figure}[h]
    \centering
    \includegraphics{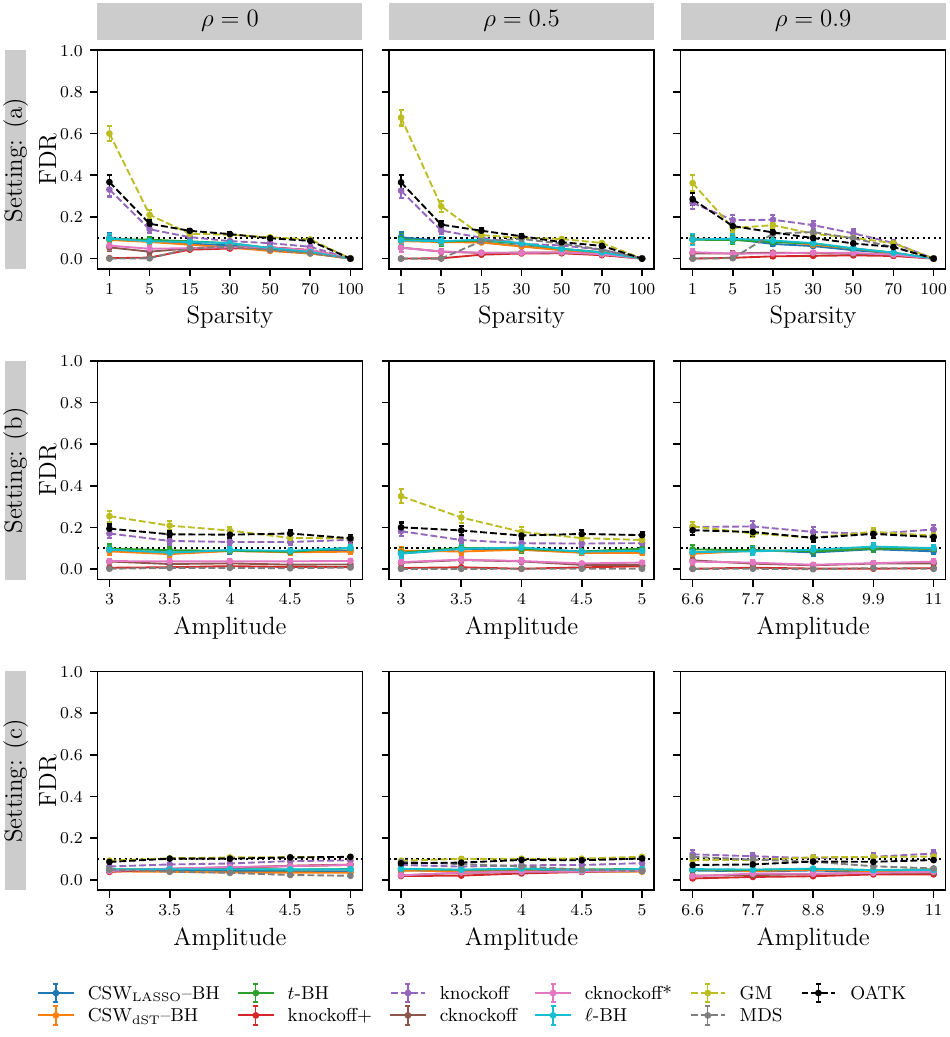}
    \caption{The empirical FDR of various FDR controlling procedures under discrete signal generation, $n/d = 3$, $d = 100$ and simulation regimes (a), (b) and (c). The error bars represent plus or minus two standard errors.}
    \label{fig:appendix_1_discrete_error_fdr}
\end{figure}

\begin{figure}[h]
    \centering
    \includegraphics{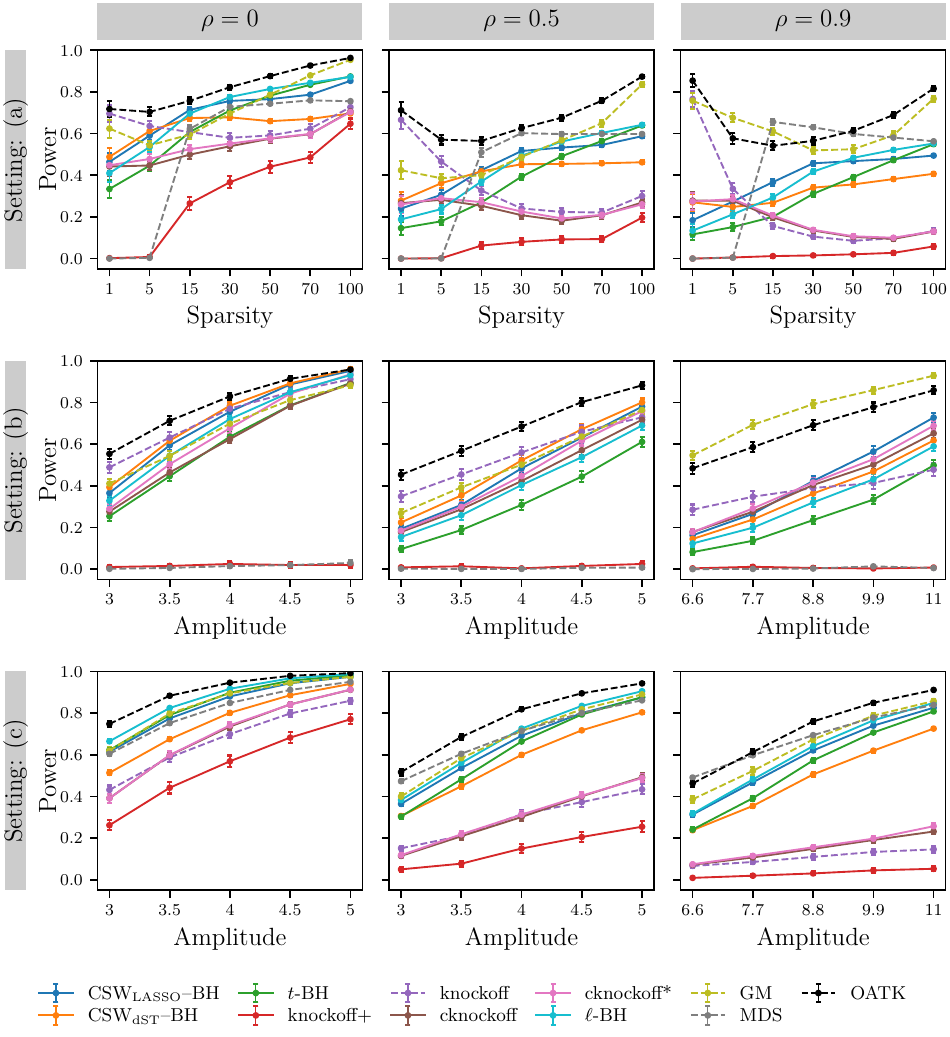}
    \caption{The empirical power of various FDR controlling procedures under discrete signal generation, $n/d = 3$, $d = 100$ and simulation regimes (a), (b) and (c). The error bars represent plus or minus two standard errors.}
    \label{fig:appendix_1_discrete_power_fdr}
\end{figure}

\begin{figure}[h]
    \centering
    \includegraphics{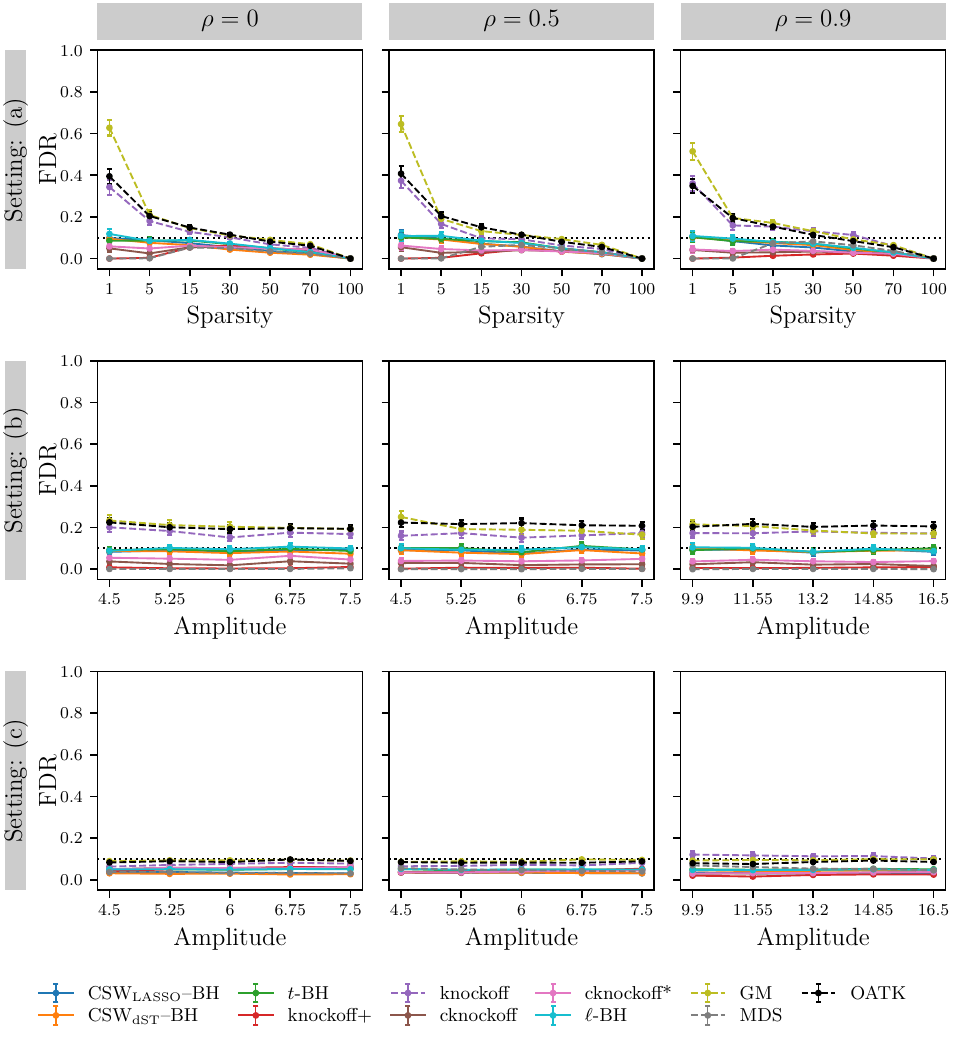}
    \caption{The empirical FDR of various FDR controlling procedures under continuous signal generation, $n/d = 3$, $d = 100$ and simulation regimes (a), (b) and (c). The error bars represent plus or minus two standard errors.}
    \label{fig:appendix_1_normal_error_fdr}
\end{figure}

\begin{figure}[h]
    \centering
    \includegraphics{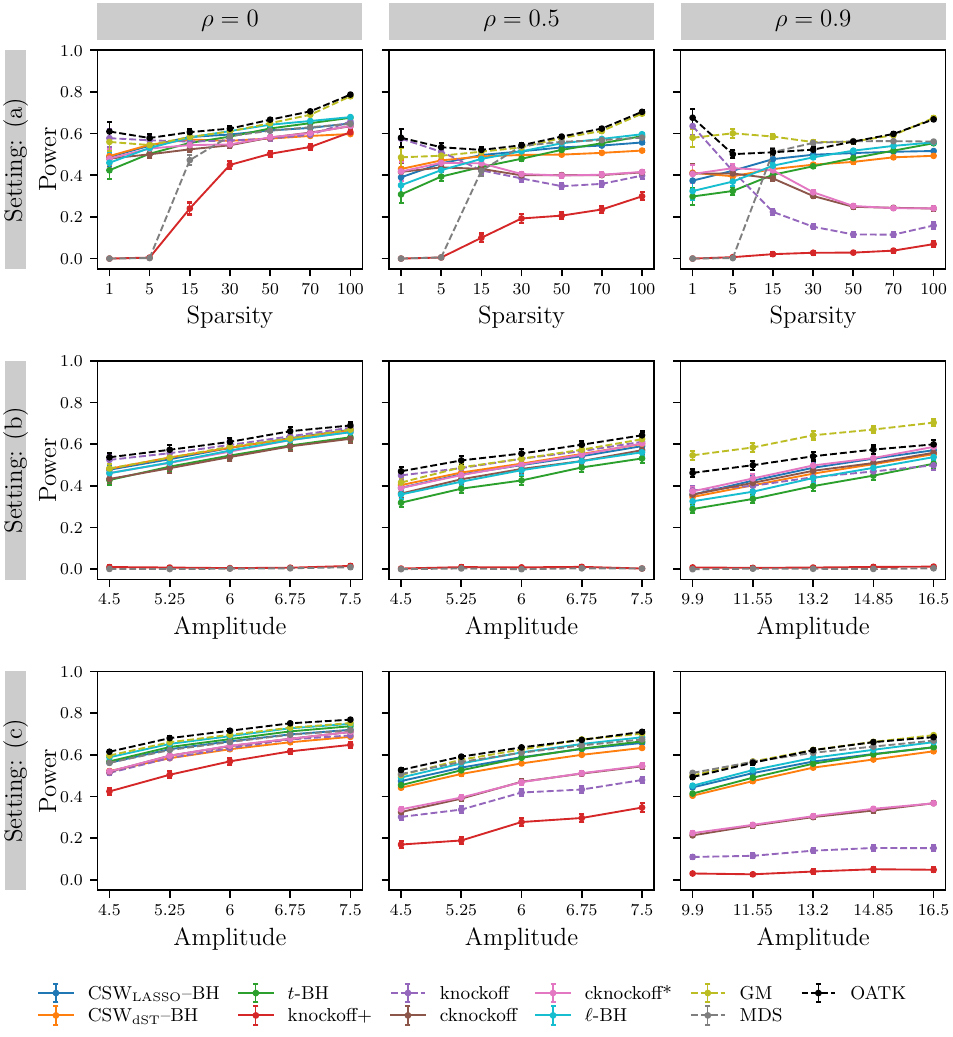}
    \caption{The empirical power of various FDR controlling procedures under continuous signal generation, $n/d = 3$, $d = 100$ and simulation regimes (a), (b) and (c). The error bars represent plus or minus two standard errors.}
    \label{fig:appendix_1_normal_power_fdr}
\end{figure}

\begin{figure}[h]
    \centering
    \includegraphics{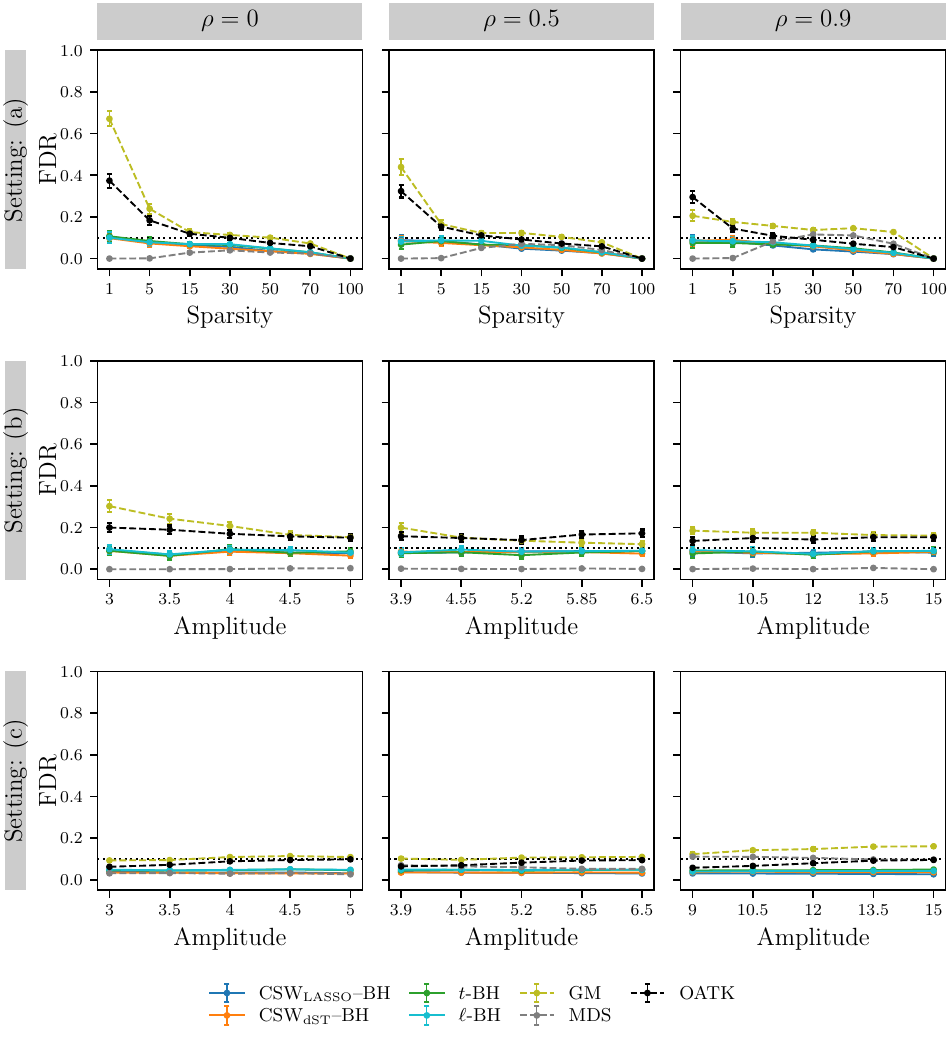}
    \caption{The empirical FDR of various FDR controlling procedures under discrete signal generation, $n/d = 1.5$, $d = 100$ and simulation regimes (a), (b) and (c). The error bars represent plus or minus two standard errors.}
    \label{fig:appendix_0_discrete_error_fdr}
\end{figure}

\begin{figure}[h]
    \centering
    \includegraphics{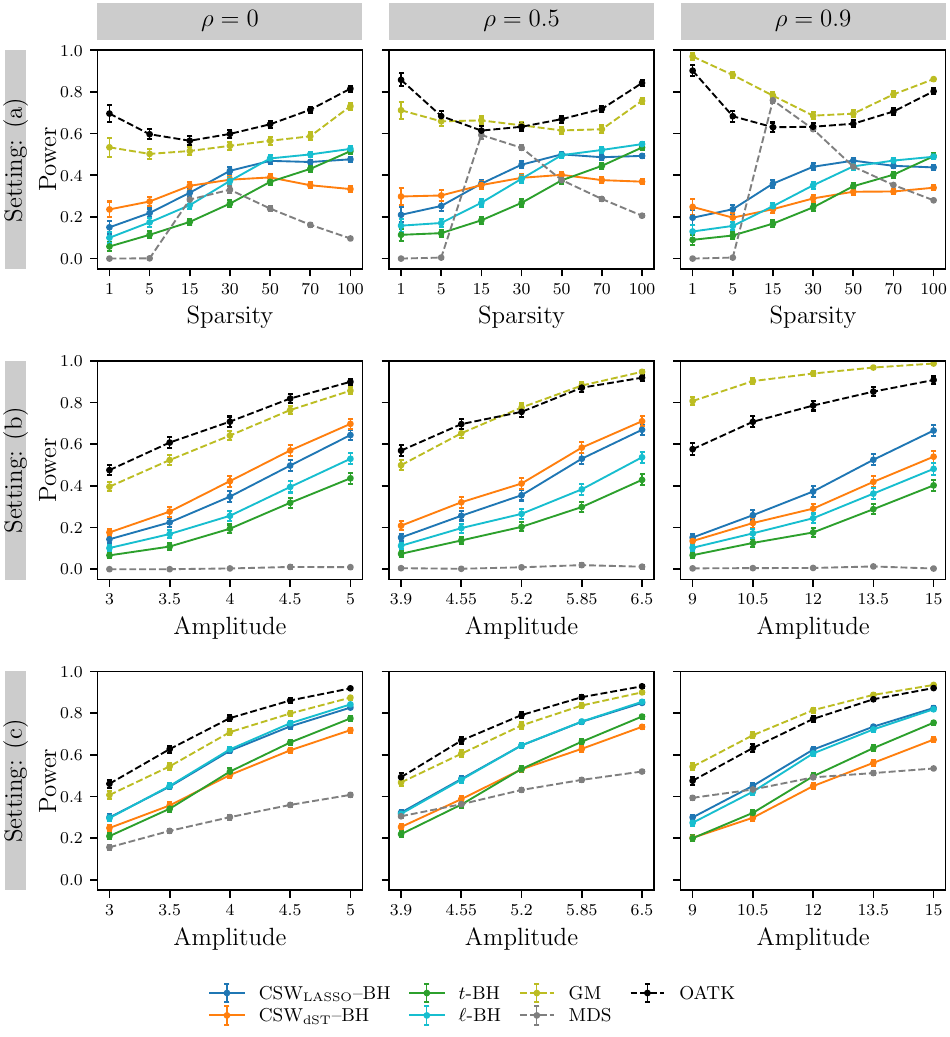}
    \caption{The empirical power of various FDR controlling procedures under discrete signal generation, $n/d = 1.5$, $d = 100$ and simulation regimes (a), (b) and (c). The error bars represent plus or minus two standard errors.}
    \label{fig:appendix_0_discrete_power_fdr}
\end{figure}

\begin{figure}[h]
    \centering
    \includegraphics{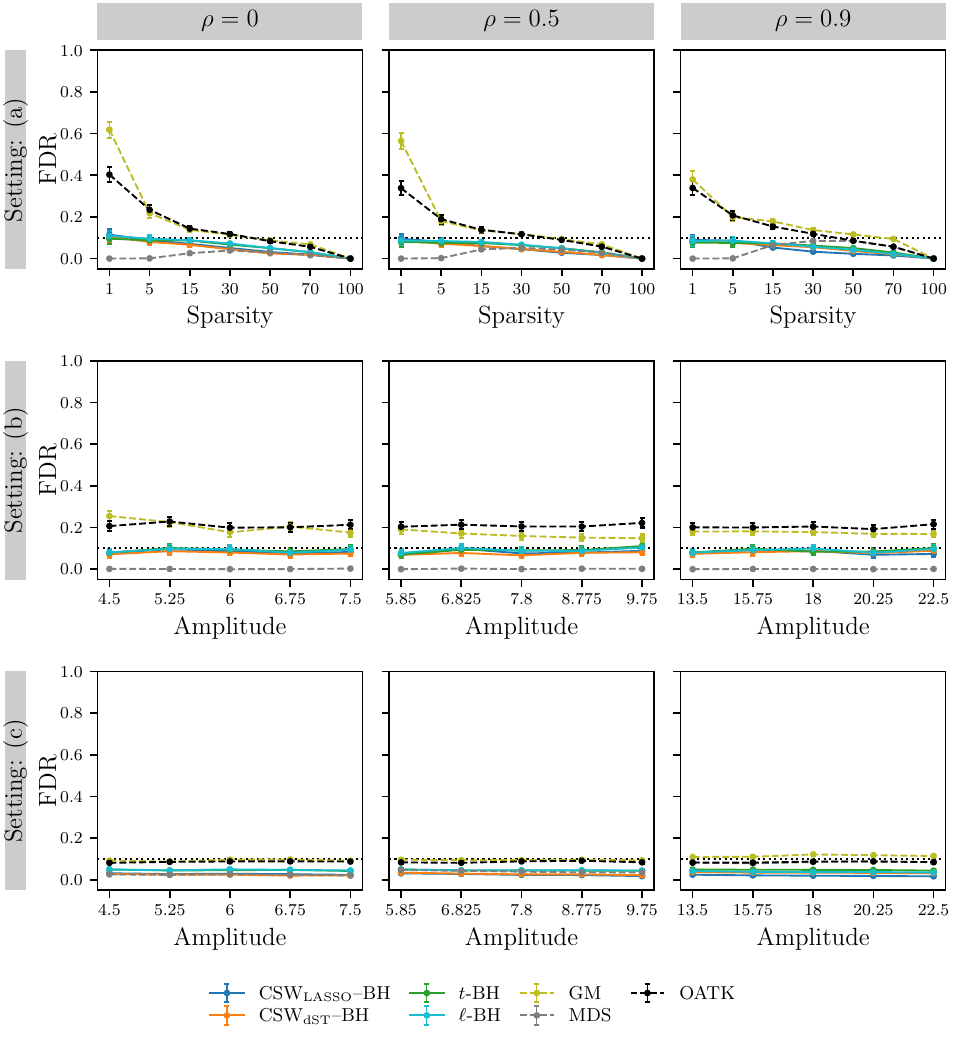}
    \caption{The empirical FDR of various FDR controlling procedures under continuous signal generation, $n/d = 1.5$, $d = 100$ and simulation regimes (a), (b) and (c). The error bars represent plus or minus two standard errors.}
    \label{fig:appendix_0_normal_error_fdr}
\end{figure}

\begin{figure}[h]
    \centering
    \includegraphics{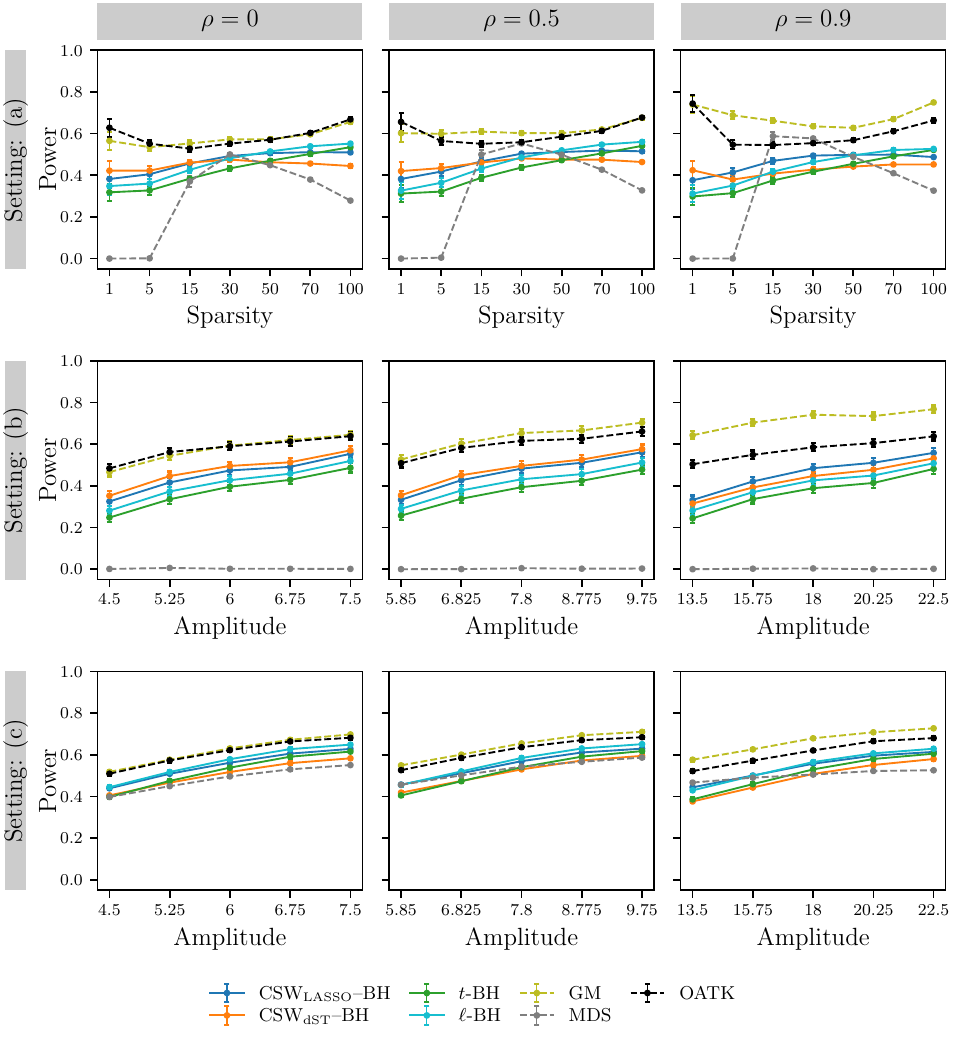}
    \caption{The empirical power of various FDR controlling procedures under continuous signal generation, $n/d = 1.5$, $d = 100$ and simulation regimes (a), (b) and (c). The error bars represent plus or minus two standard errors.}
    \label{fig:appendix_0_normal_power_fdr}
\end{figure}

\begin{figure}[h]
    \centering
    \includegraphics{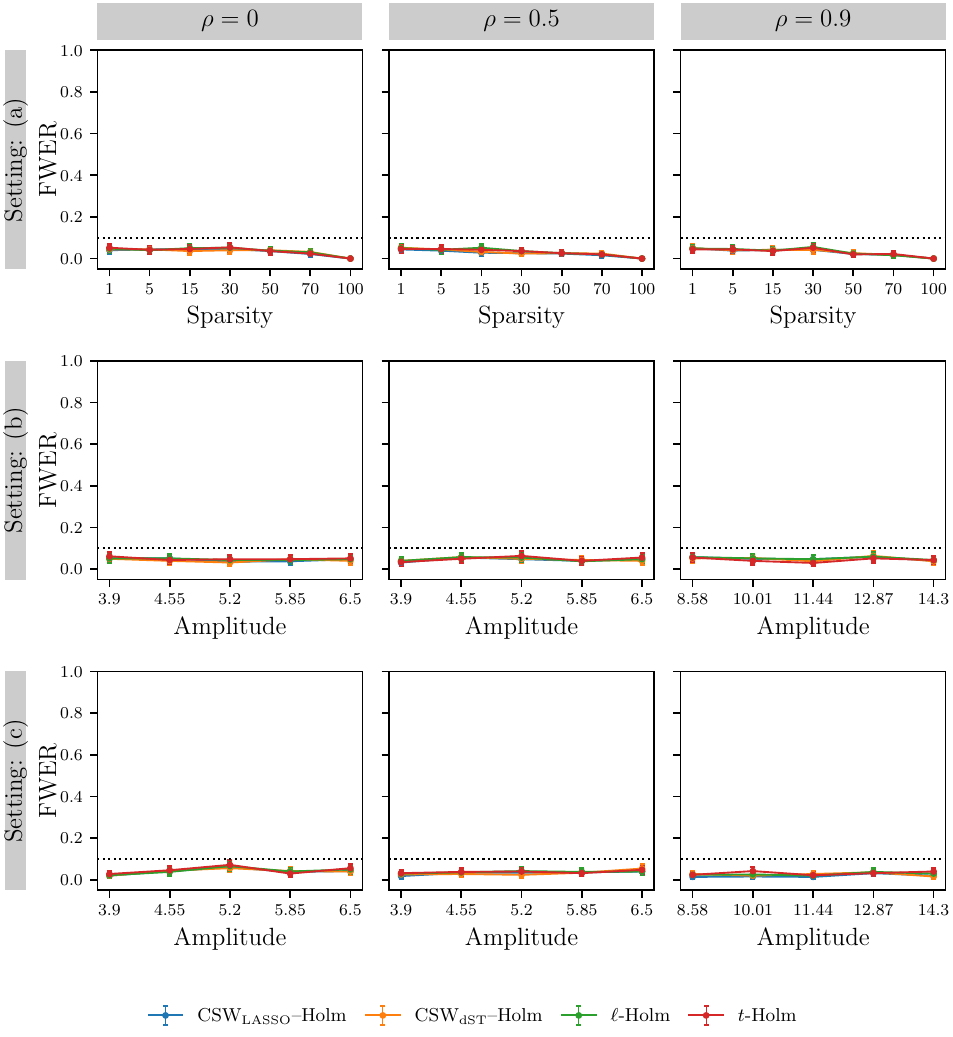}
    \caption{The empirical FWER of various FWER controlling procedures under discrete signal generation, $n/d = 3$, $d = 100$ and simulation regimes (a), (b) and (c). The error bars represent plus or minus two standard errors.}
    \label{fig:appendix_1_discrete_error_fwer}
\end{figure}

\begin{figure}[h]
    \centering
    \includegraphics{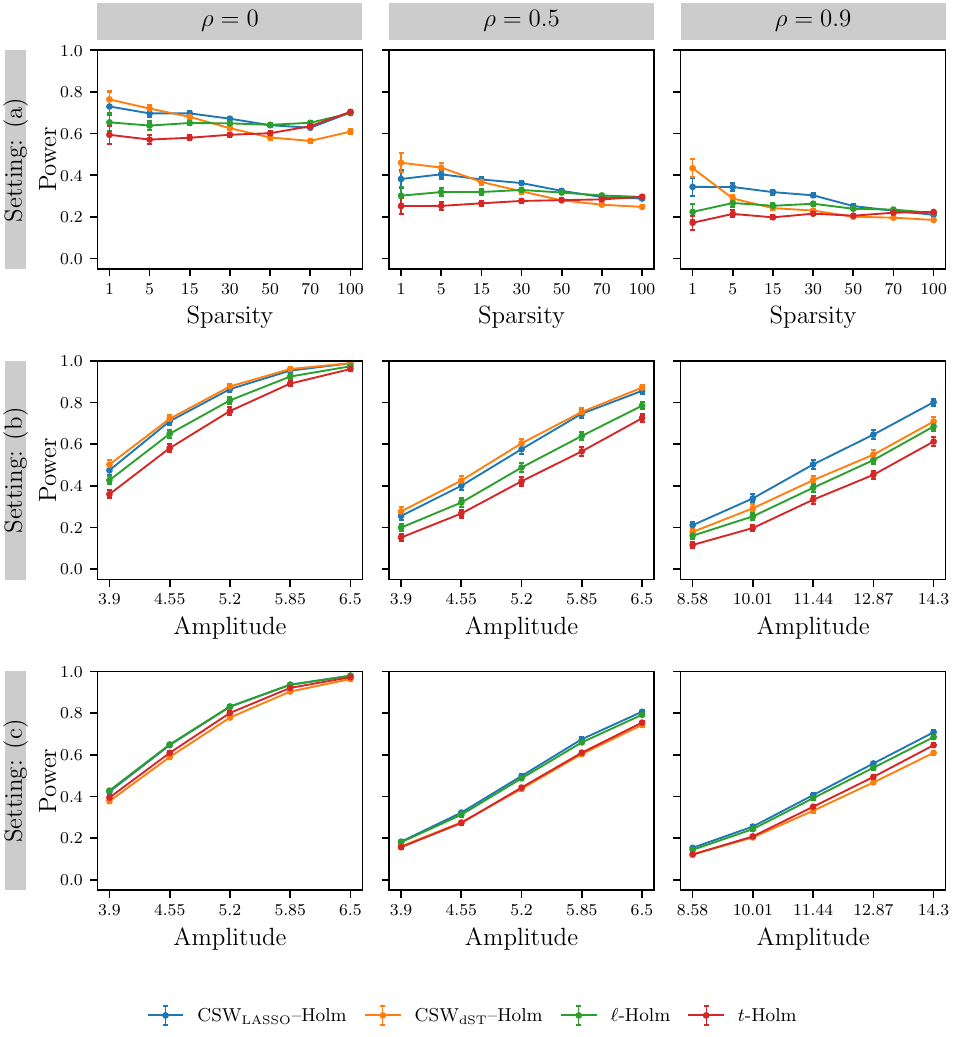}
    \caption{The empirical power of various FWER controlling procedures under discrete signal generation, $n/d = 3$, $d = 100$ and simulation regimes (a), (b) and (c). The error bars represent plus or minus two standard errors.}
    \label{fig:appendix_1_discrete_power_fwer}
\end{figure}

\begin{figure}[h]
    \centering
    \includegraphics{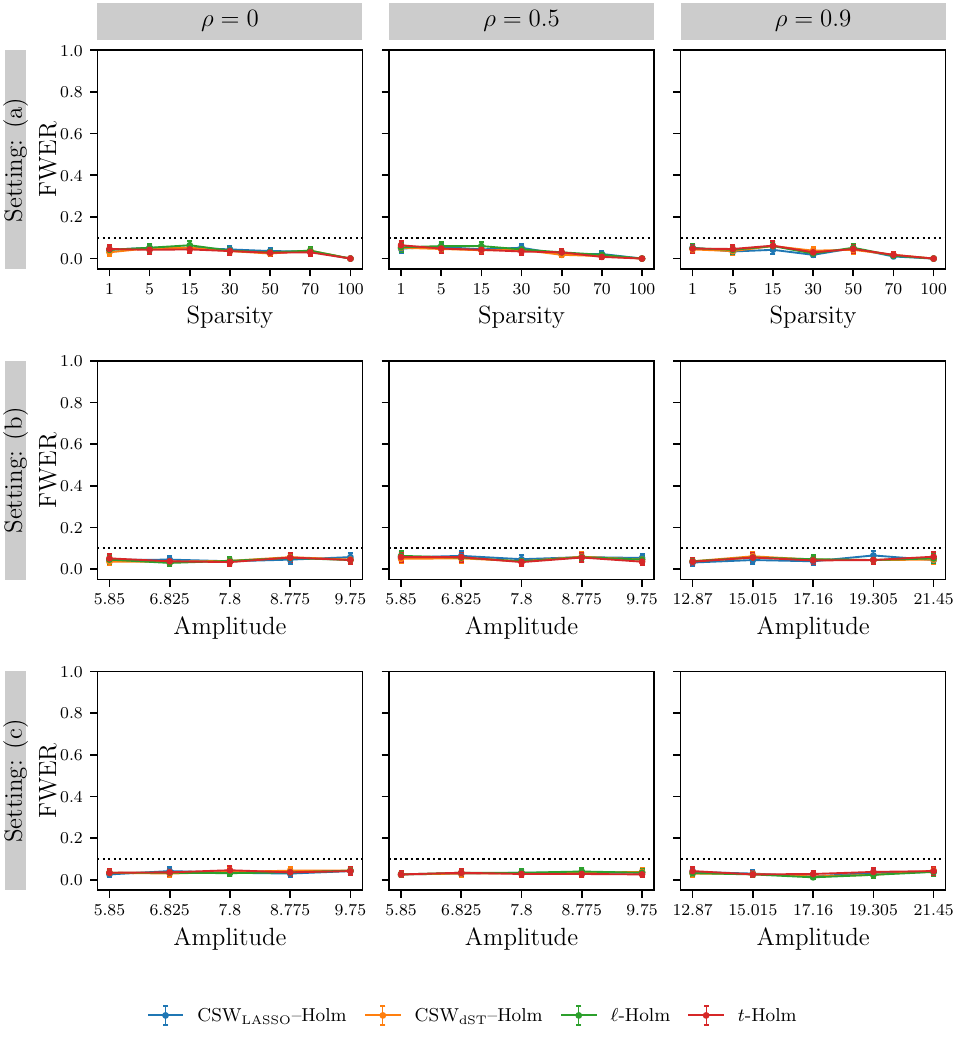}
    \caption{The empirical FWER of various FWER controlling procedures under continuous signal generation, $n/d = 3$, $d = 100$ and simulation regimes (a), (b) and (c). The error bars represent plus or minus two standard errors.}
    \label{fig:appendix_1_normal_error_fwer}
\end{figure}

\begin{figure}[h]
    \centering
    \includegraphics{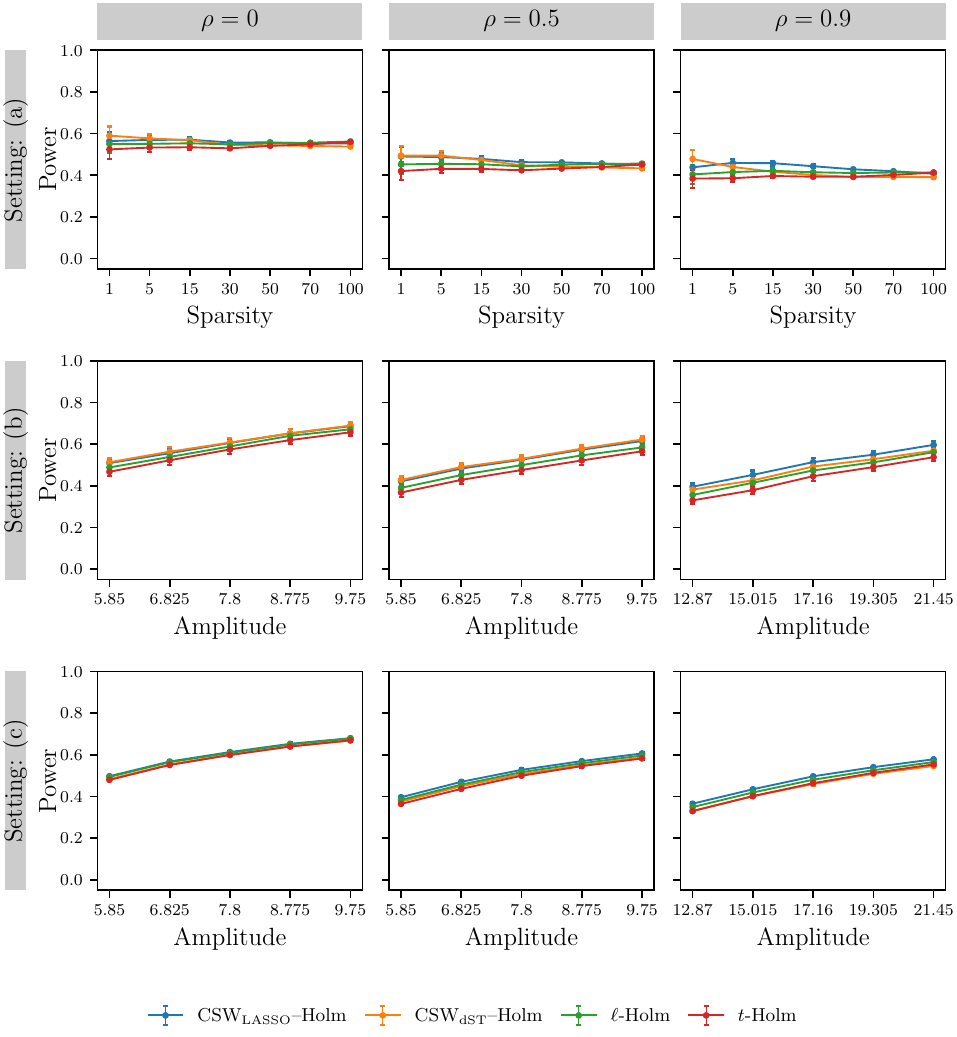}
    \caption{The empirical power of various FWER controlling procedures under continuous signal generation, $n/d = 3$, $d = 100$ and simulation regimes (a), (b) and (c). The error bars represent plus or minus two standard errors.}
    \label{fig:appendix_1_normal_power_fwer}
\end{figure}

\begin{figure}[h]
    \centering
    \includegraphics{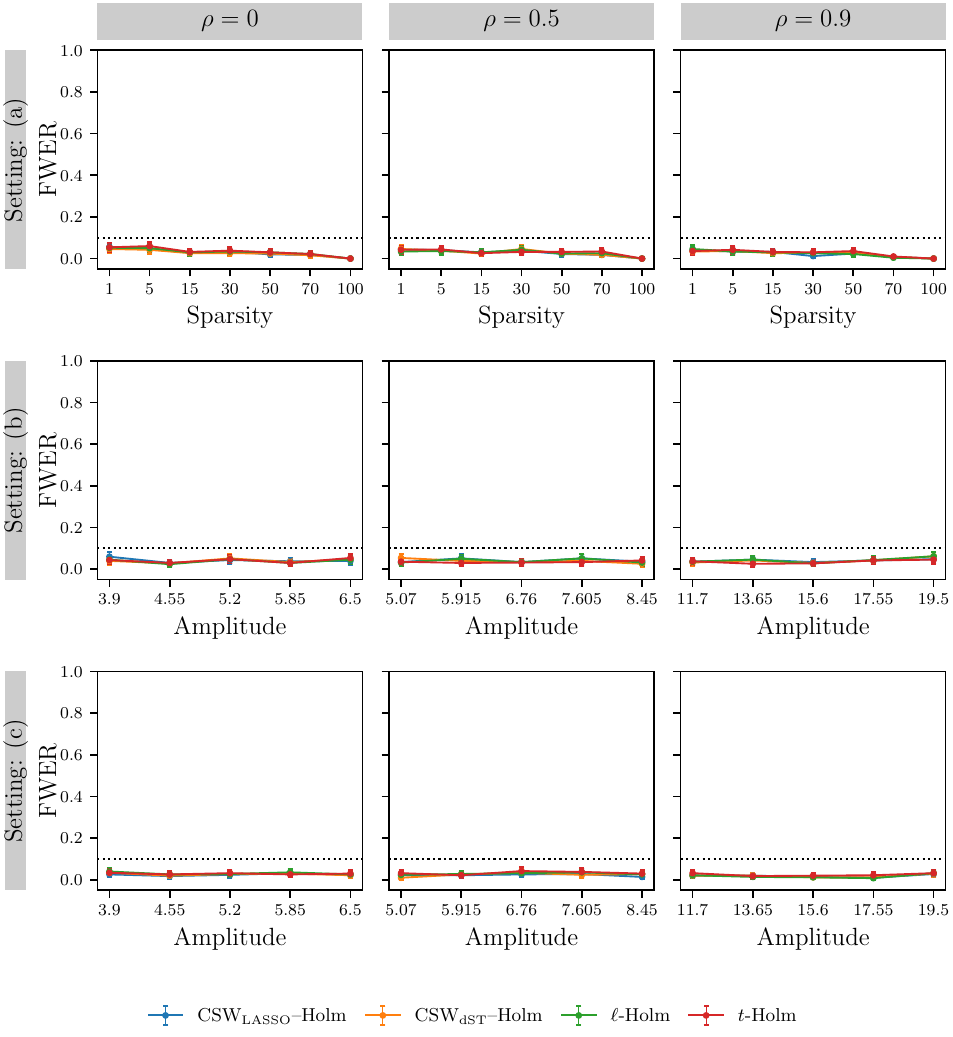}
    \caption{The empirical FWER of various FWER controlling procedures under discrete signal generation, $n/d = 1.5$, $d = 100$ and simulation regimes (a), (b) and (c). The error bars represent plus or minus two standard errors.}
    \label{fig:appendix_0_discrete_error_fwer}
\end{figure}

\begin{figure}[h]
    \centering
    \includegraphics{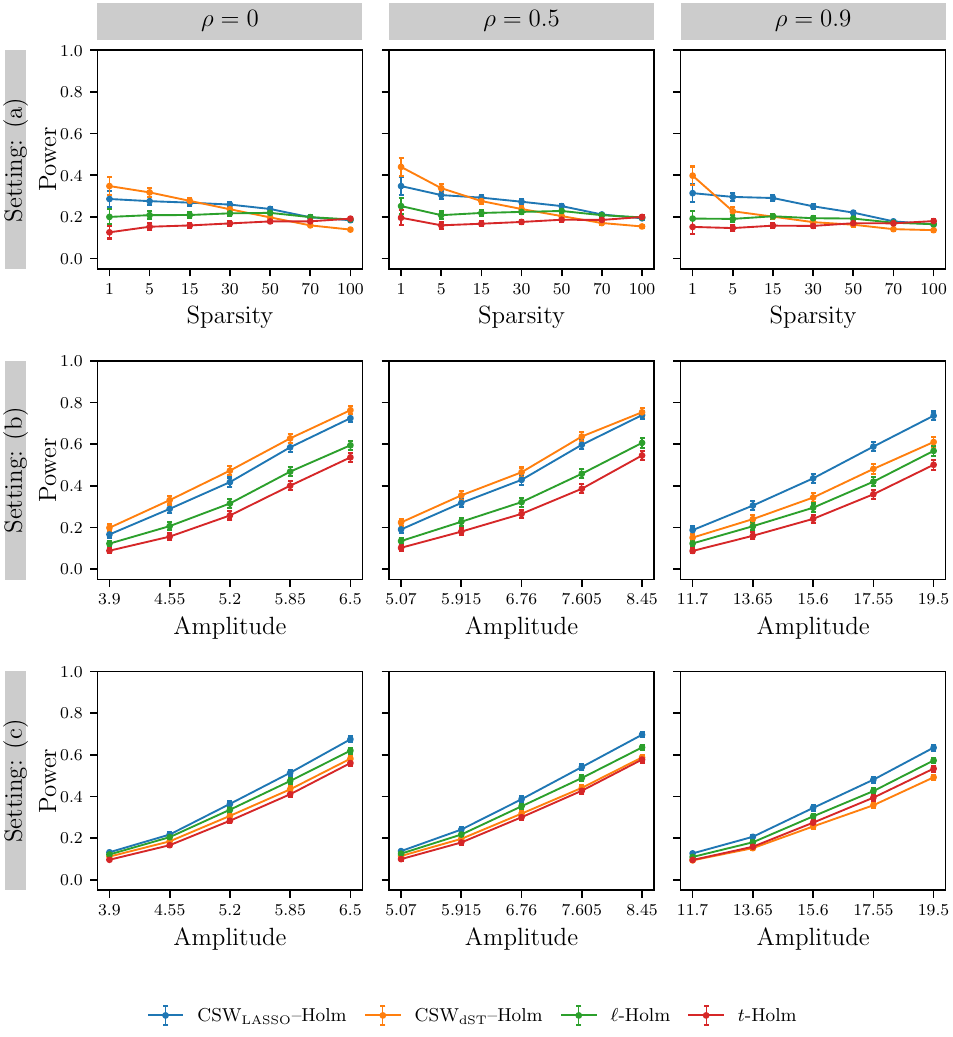}
    \caption{The empirical power of various FWER controlling procedures under discrete signal generation, $n/d = 1.5$, $d = 100$ and simulation regimes (a), (b) and (c). The error bars represent plus or minus two standard errors.}
    \label{fig:appendix_0_discrete_power_fwer}
\end{figure}

\begin{figure}[h]
    \centering
    \includegraphics{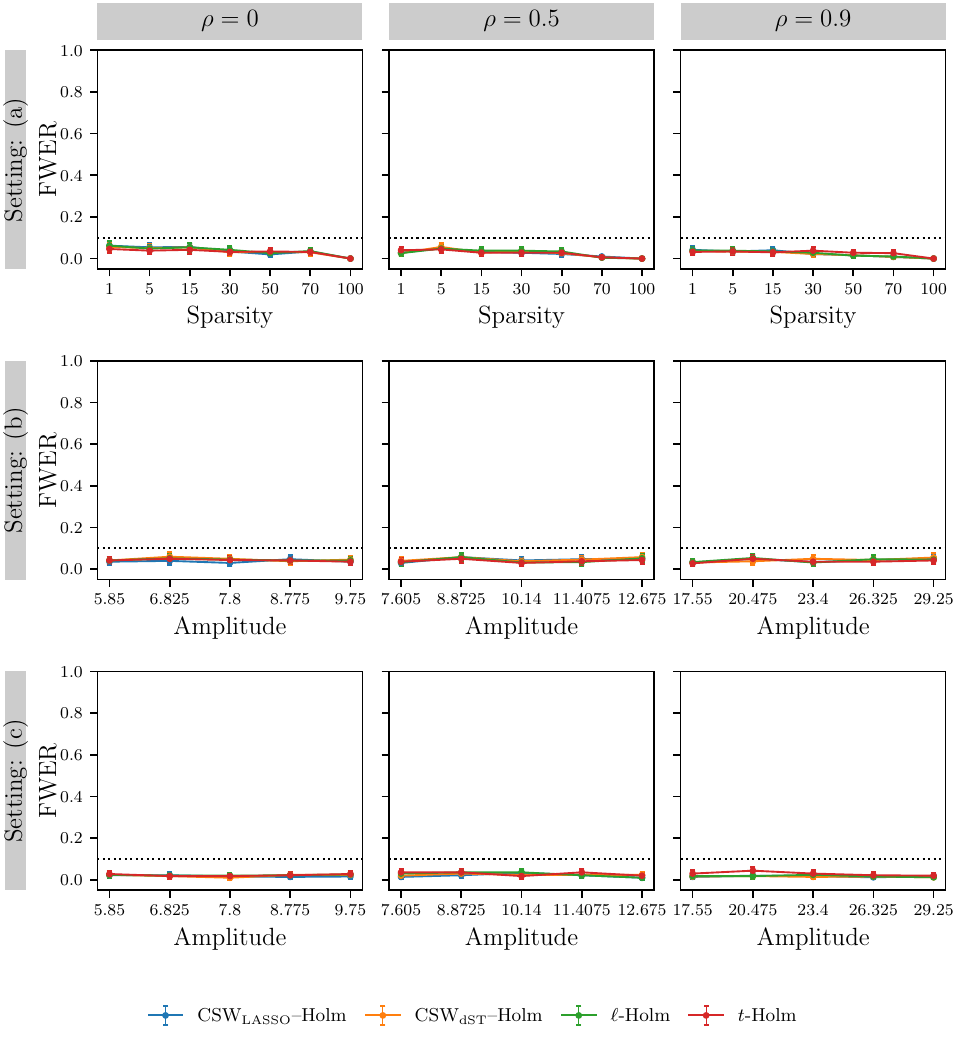}
    \caption{The empirical FWER of various FWER controlling procedures under continuous signal generation, $n/d = 1.5$, $d = 100$ and simulation regimes (a), (b) and (c). The error bars represent plus or minus two standard errors.}
    \label{fig:appendix_0_normal_error_fwer}
\end{figure}

\begin{figure}[h]
    \centering
    \includegraphics{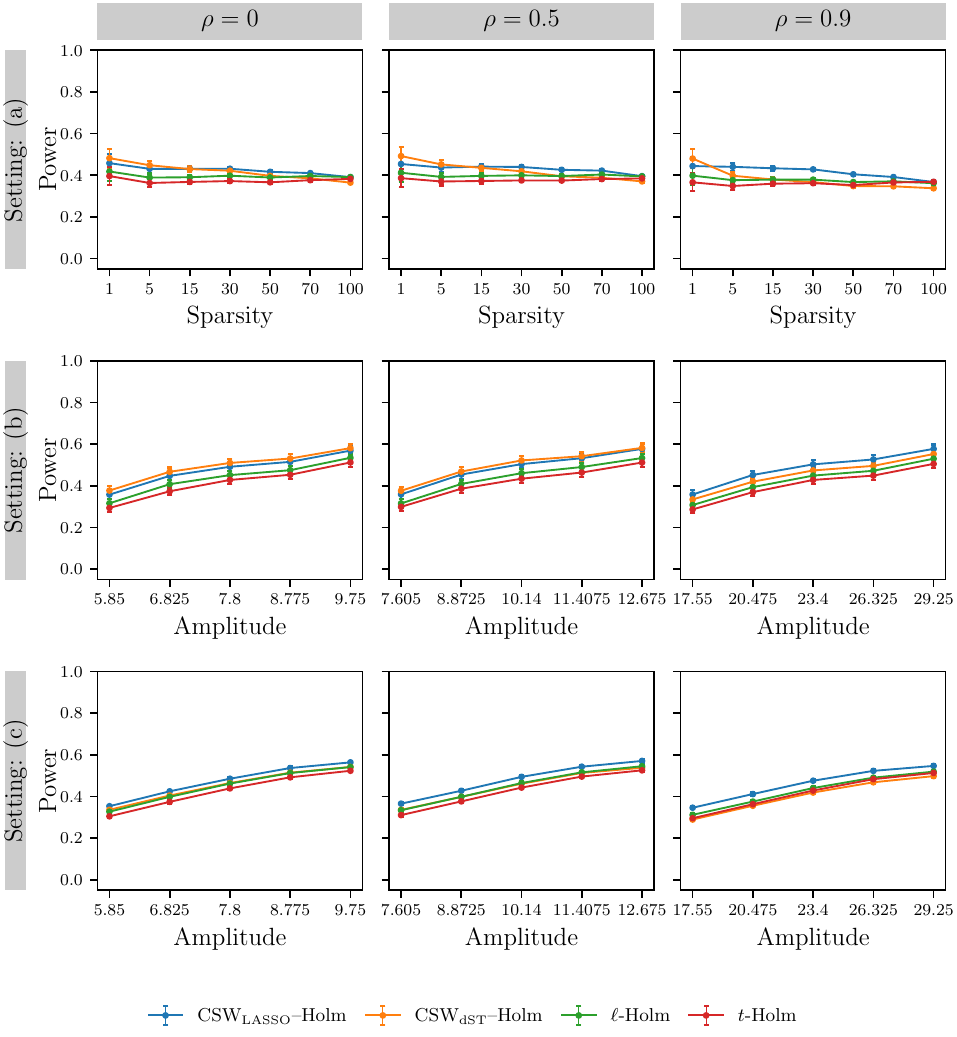}
    \caption{The empirical power of various FWER controlling procedures under continuous signal generation, $n/d = 1.5$, $d = 100$ and simulation regimes (a), (b) and (c). The error bars represent plus or minus two standard errors.}
    \label{fig:appendix_0_normal_power_fwer}
\end{figure}

\begin{figure}[h]
    \centering
    \includegraphics{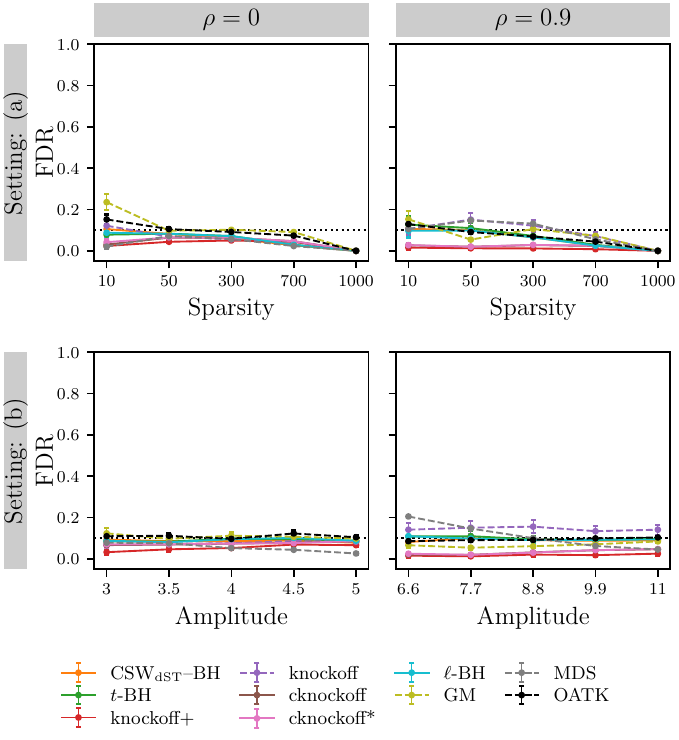}
    \caption{The empirical FDR of various FDR controlling procedures under discrete signal generation, $n/d = 3$, $d = 1000$, inter-variable correlations $\rho \in \{0, 0.9\}$ and simulation regimes (a) and (b). The error bars represent plus or minus two standard errors.}
    \label{fig:appendix_large_error_fdr}
\end{figure}

\begin{figure}[h]
    \centering
    \includegraphics{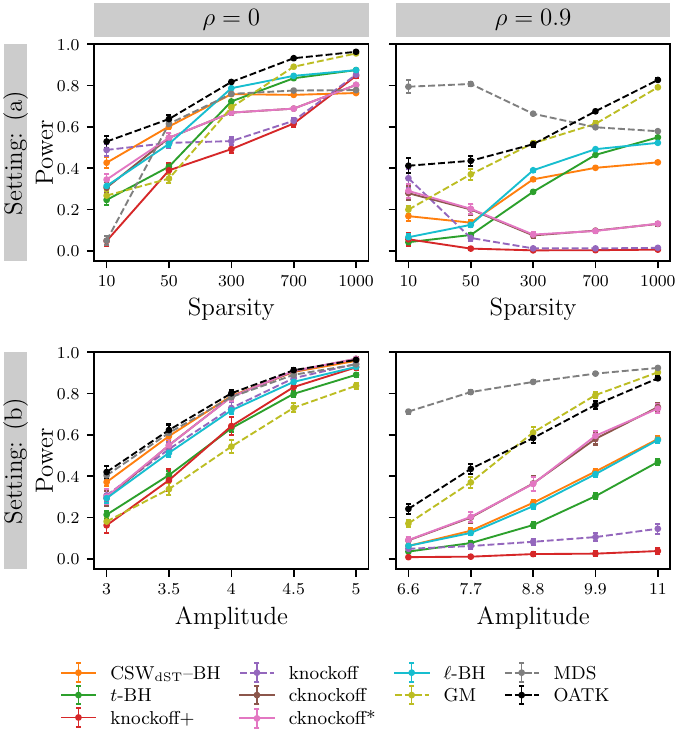}
    \caption{The empirical power of various FDR controlling procedures under discrete signal generation, $n/d = 3$, $d = 1000$, inter-variable correlations $\rho \in \{0, 0.9\}$ and simulation regimes (a) and (b). The error bars represent plus or minus two standard errors.}
    \label{fig:appendix_large_power_fdr}
\end{figure}

\begin{figure}[h]
    \centering
    \includegraphics{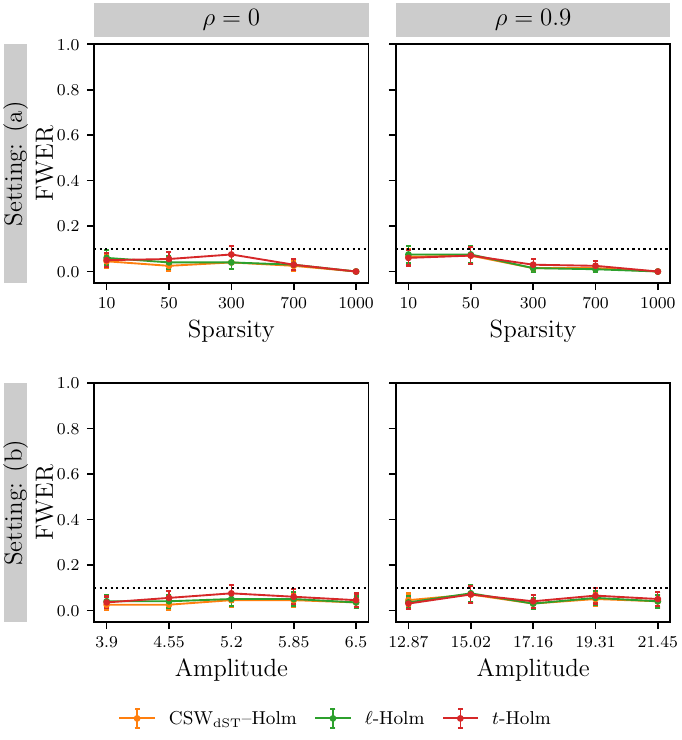}
    \caption{The empirical FWER of various FWER controlling procedures under discrete signal generation, $n/d = 3$, $d = 1000$, inter-variable correlations $\rho \in \{0, 0.9\}$ and simulation regimes (a) and (b). The error bars represent plus or minus two standard errors.}
    \label{fig:appendix_large_error_fwer}
\end{figure}

\begin{figure}[h]
    \centering
    \includegraphics{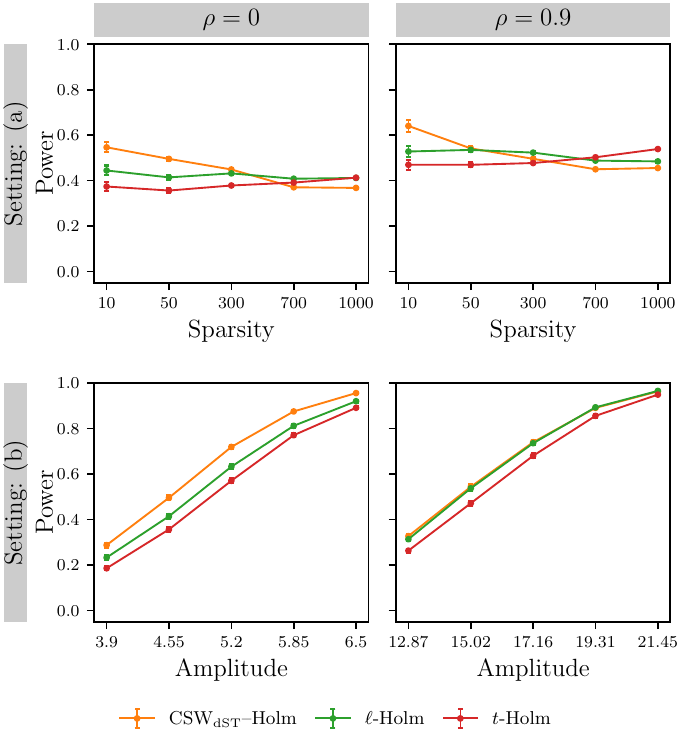}
    \caption{The empirical power of various FWER controlling procedures under discrete signal generation, $n/d = 3$, $d = 1000$, inter-variable correlations $\rho \in \{0, 0.9\}$ and simulation regimes (a) and (b). The error bars represent plus or minus two standard errors.}
    \label{fig:appendix_large_power_fwer}
\end{figure}

\begin{figure}[h]
    \centering
    \includegraphics{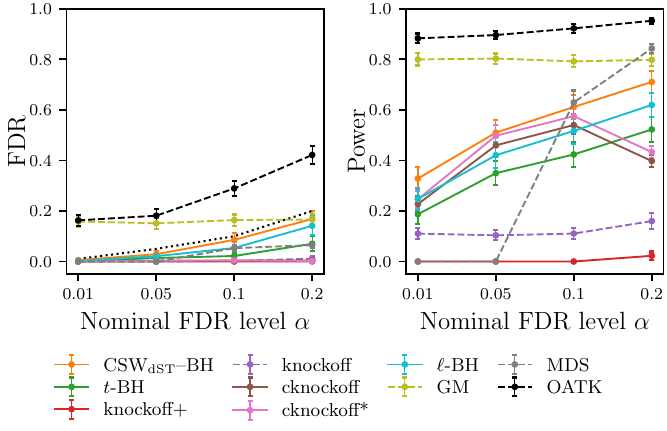}
    \caption{The empirical FDR (left) and power (right) of various FDR controlling procedures under the MCC design. The dotted line in the left panel marks FDR = nominal level. The error bars represent plus or minus two standard errors.}
    \label{fig:appendix_mcc_fdr}
\end{figure}

The full set of simulations are summarized in Figures~\ref{fig:appendix_1_discrete_error_fdr} to \ref{fig:appendix_mcc_fdr}. Under discrete signal generation, the full set of results supports the conclusions of Section~\ref{sec:expt}, which focused on that case. Under continuous signal generation, similar trends hold, but the power curves lie much closer together. Because the signal strengths now vary, every method readily discovers the strongest signals and struggles with the weakest, so the methods differ mainly on the subset of the signals with intermediate strength. 

%We argued that a similar situation occurred in the HIV data analysis presented in Sections~\ref{sec:real_data} and \ref{sec:weighting_hiv}

\subsection{Simulation results for exponentiated methods}
\label{sec:exponentiated_simulations}
\begin{figure}[h]
    \centering
    \includegraphics{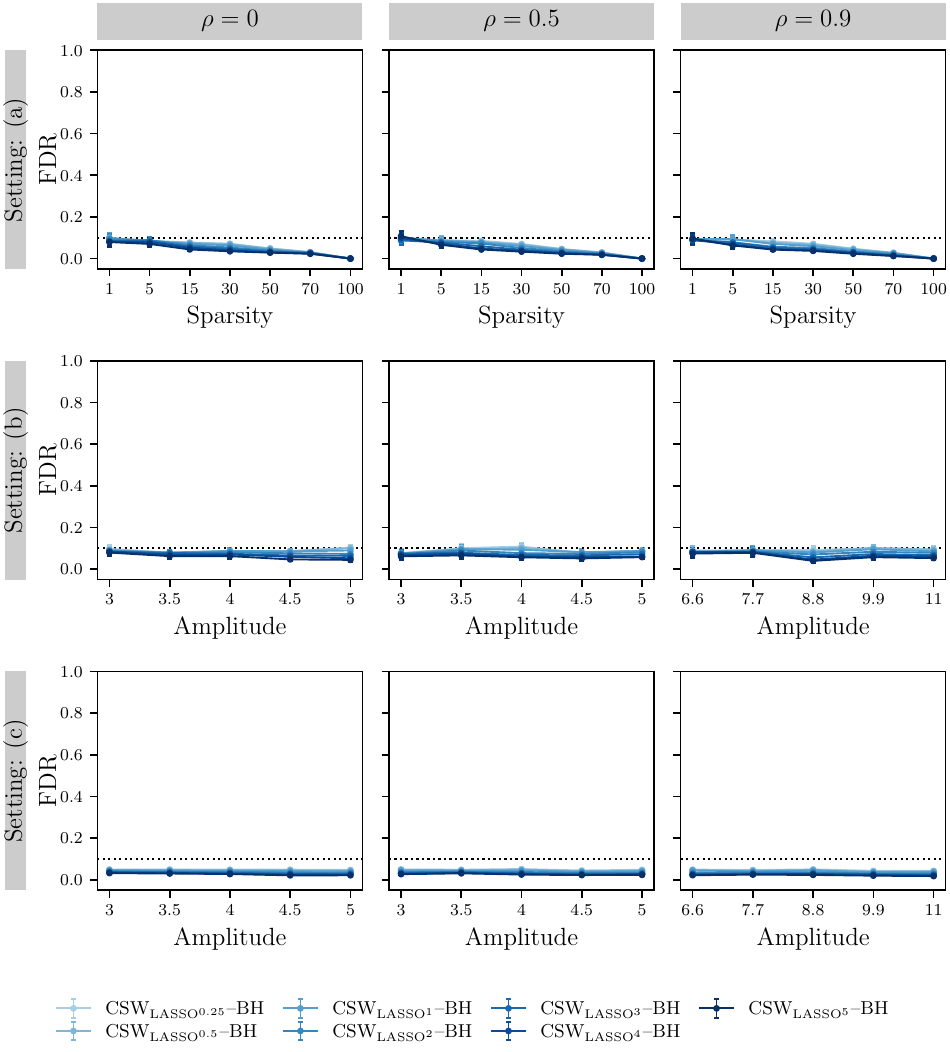}
    \caption{The empirical FDR of exponentiated versions of $\CSWLASSOBH$ under discrete signal generation, $n/d = 3$, $d = 100$ and simulation regimes (a), (b) and (c). The error bars represent plus or minus two standard errors.}
    \label{fig:appendix_1_discrete_error_fdr_powered}
\end{figure}

\begin{figure}[h]
    \centering
    \includegraphics{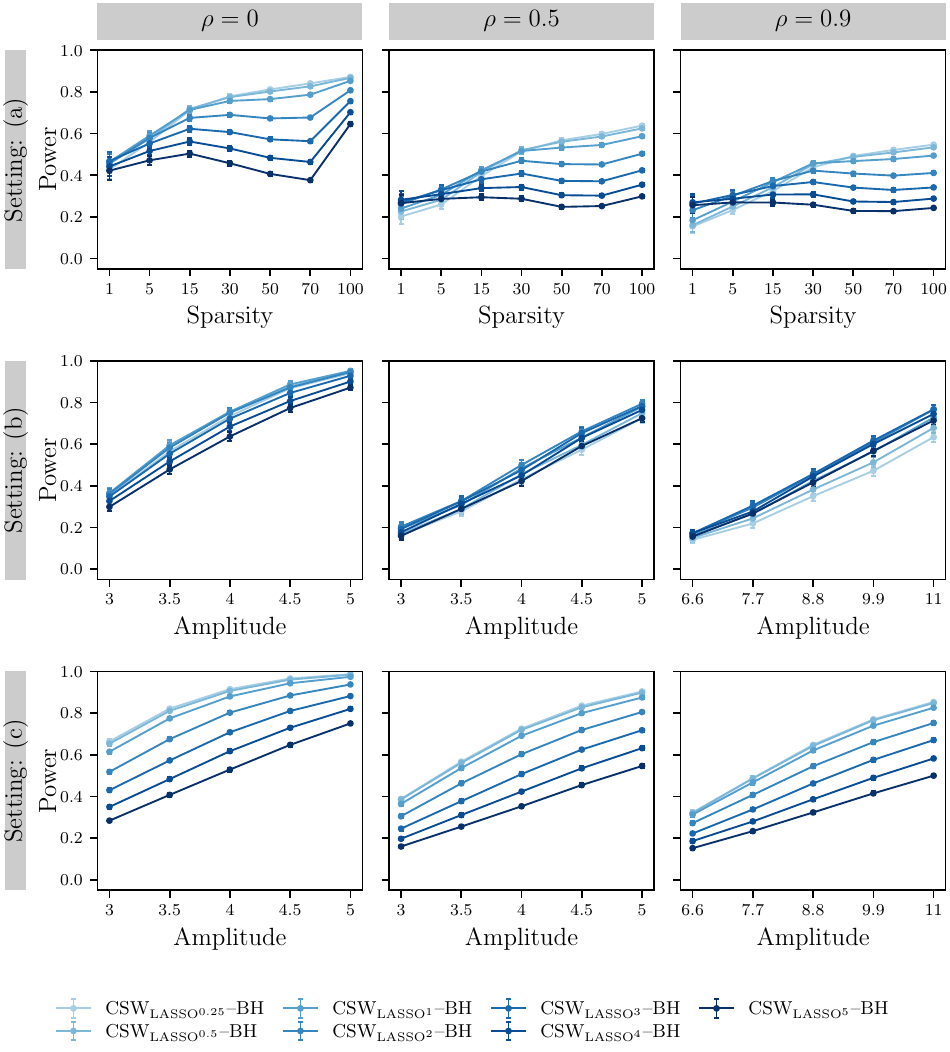}
    \caption{The empirical power of exponentiated versions of $\CSWLASSOBH$ under discrete signal generation, $n/d = 3$, $d = 100$ and simulation regimes (a), (b) and (c). The error bars represent plus or minus two standard errors.}
    \label{fig:appendix_1_discrete_power_fdr_powered}
\end{figure}

\begin{figure}[h]
    \centering
    \includegraphics{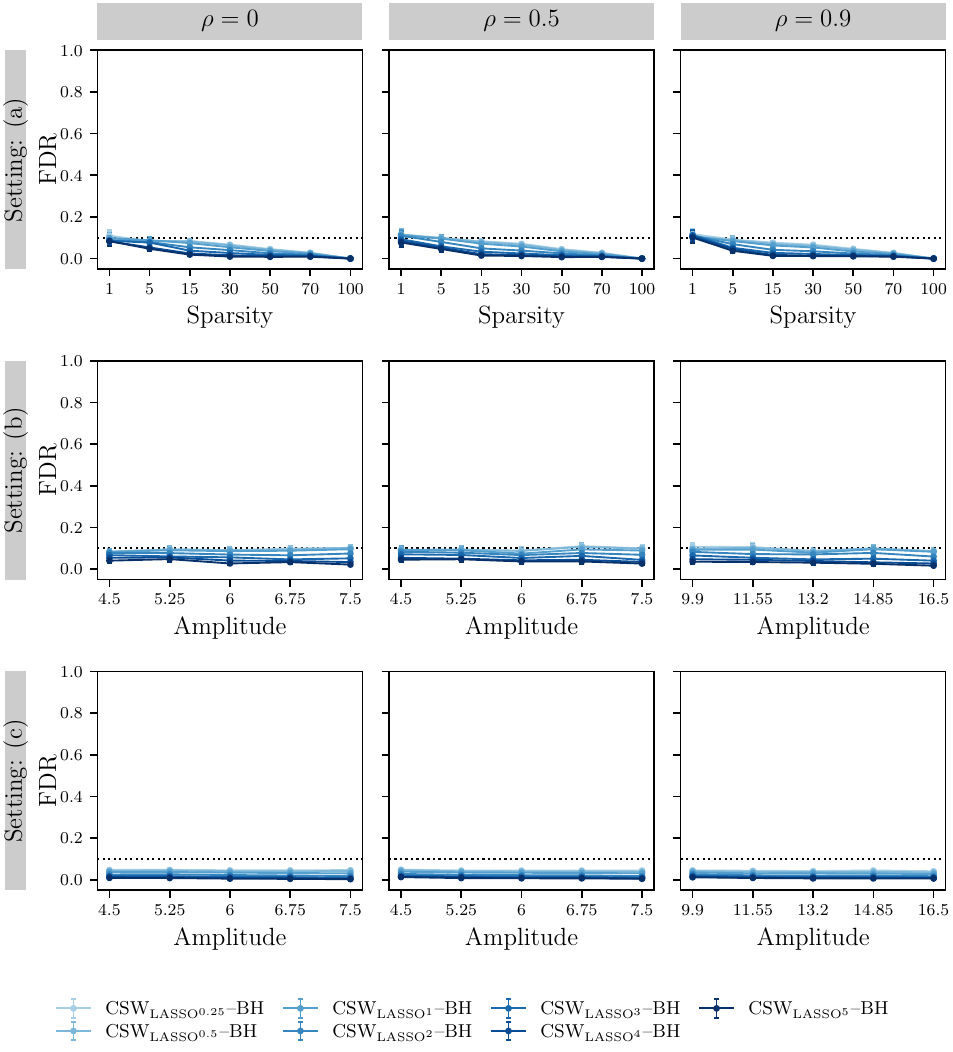}
    \caption{The empirical FDR of exponentiated versions of $\CSWLASSOBH$ under continuous signal generation, $n/d = 3$, $d = 100$ and simulation regimes (a), (b) and (c). The error bars represent plus or minus two standard errors.}
    \label{fig:appendix_1_normal_error_fdr_powered}
\end{figure}

\begin{figure}[h]
    \centering
    \includegraphics{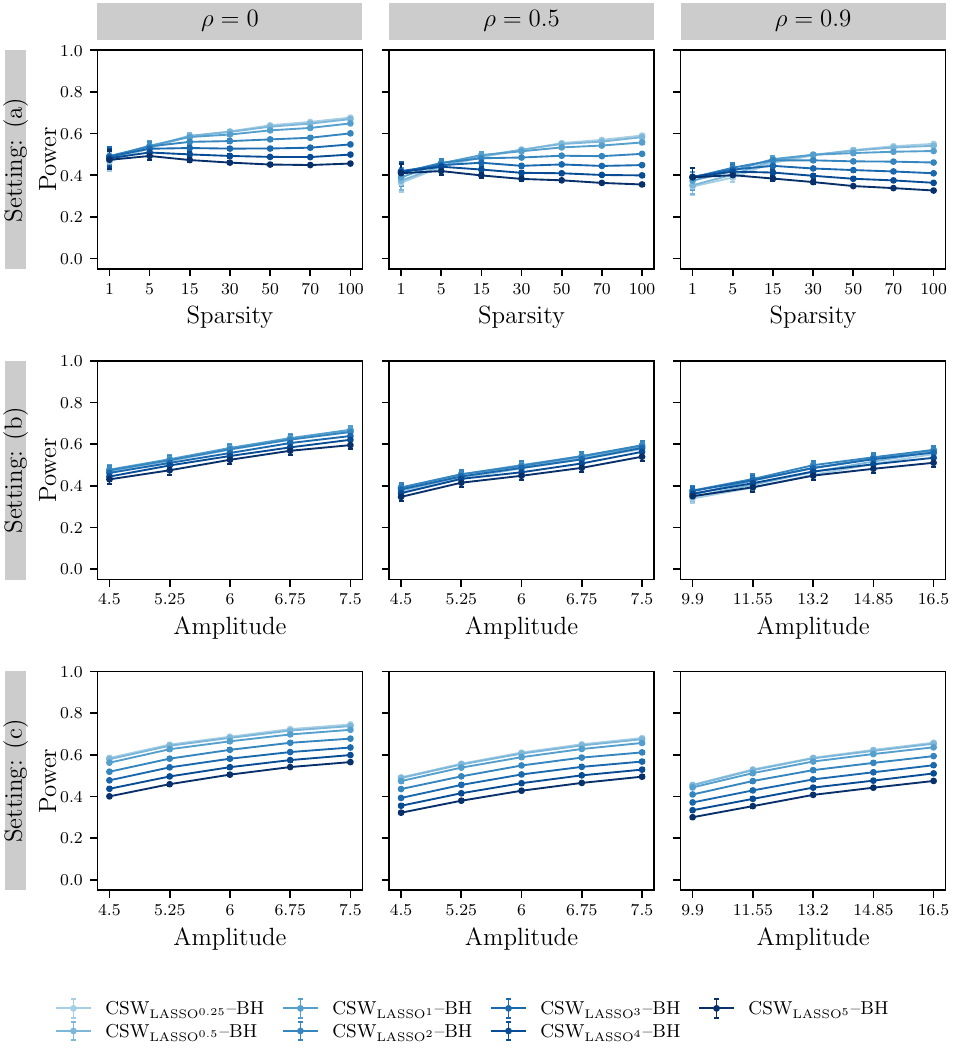}
    \caption{The empirical power of exponentiated versions of $\CSWLASSOBH$ under continuous signal generation, $n/d = 3$, $d = 100$ and simulation regimes (a), (b) and (c). The error bars represent plus or minus two standard errors.}
    \label{fig:appendix_1_normal_power_fdr_powered}
\end{figure}

\begin{figure}[h]
    \centering
    \includegraphics{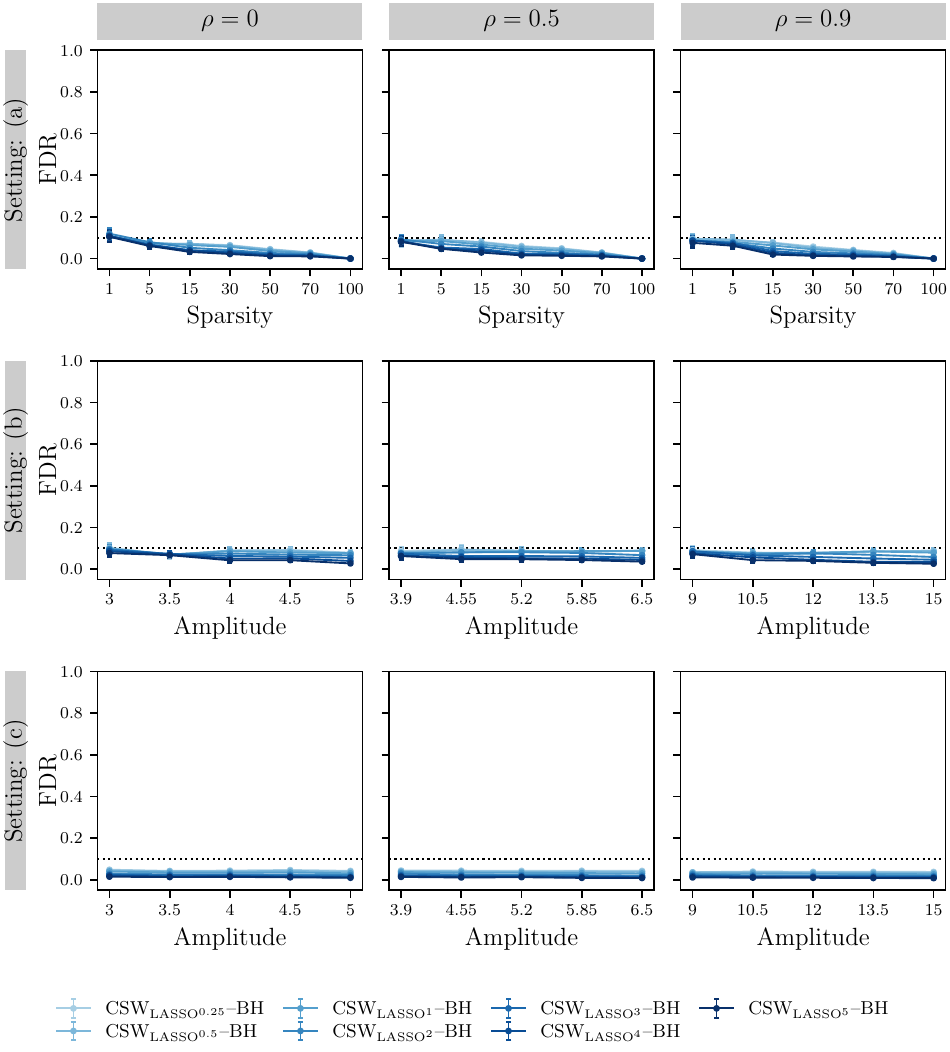}
    \caption{The empirical FDR of exponentiated versions of $\CSWLASSOBH$ under discrete signal generation, $n/d = 1.5$, $d = 100$ and simulation regimes (a), (b) and (c). The error bars represent plus or minus two standard errors.}
    \label{fig:appendix_0_discrete_error_fdr_powered}
\end{figure}

\begin{figure}[h]
    \centering
    \includegraphics{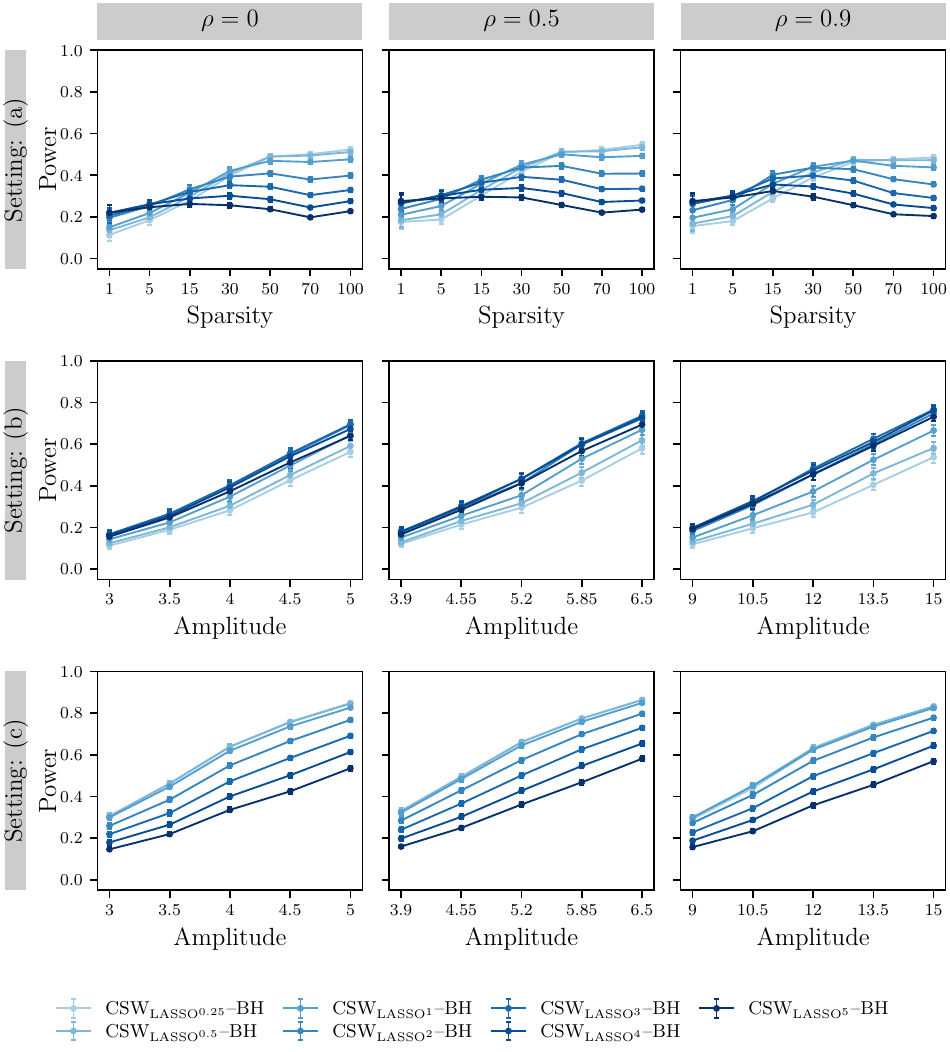}
    \caption{The empirical power of exponentiated versions of $\CSWLASSOBH$ under discrete signal generation, $n/d = 1.5$, $d = 100$ and simulation regimes (a), (b) and (c). The error bars represent plus or minus two standard errors.}
    \label{fig:appendix_0_discrete_power_fdr_powered}
\end{figure}

\begin{figure}[h]
    \centering
    \includegraphics{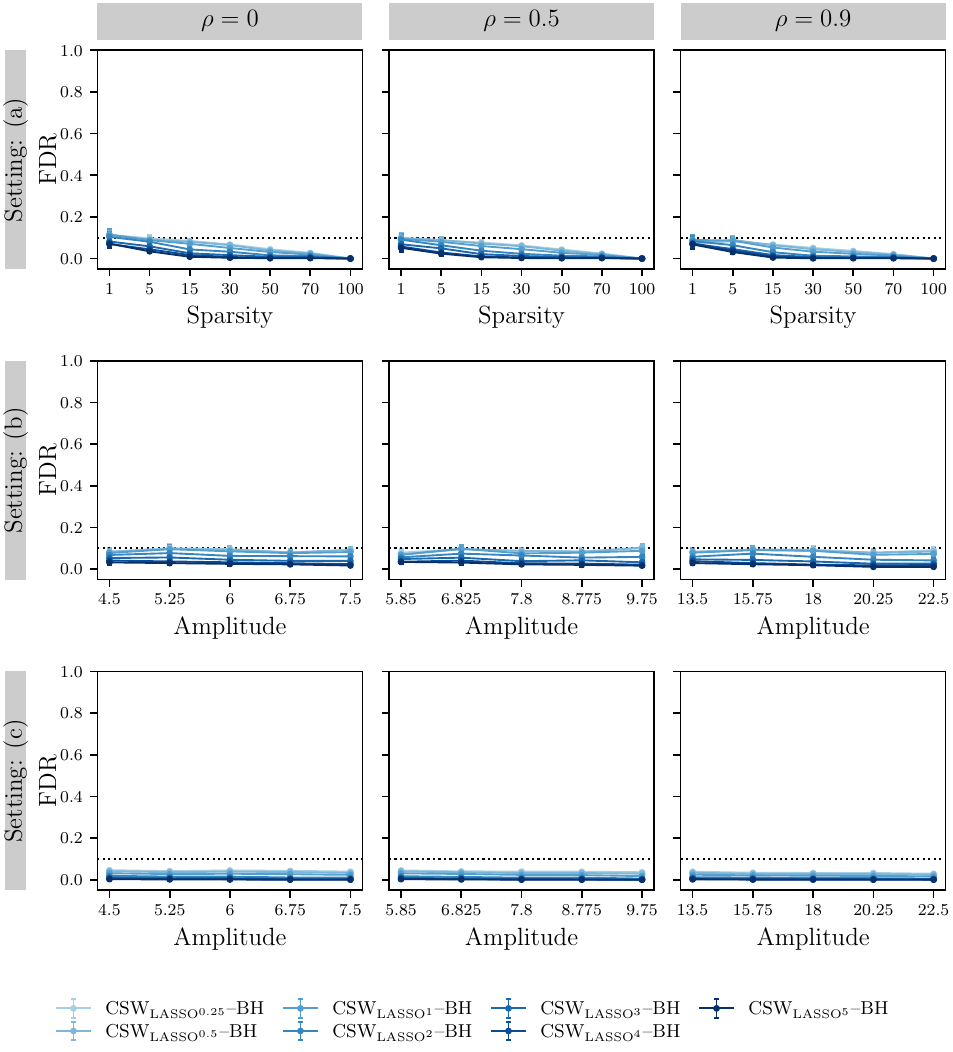}
    \caption{The empirical FDR of exponentiated versions of $\CSWLASSOBH$ under continuous signal generation, $n/d = 1.5$, $d = 100$ and simulation regimes (a), (b) and (c). The error bars represent plus or minus two standard errors.}
    \label{fig:appendix_0_normal_error_fdr_powered}
\end{figure}

\begin{figure}[h]
    \centering
    \includegraphics{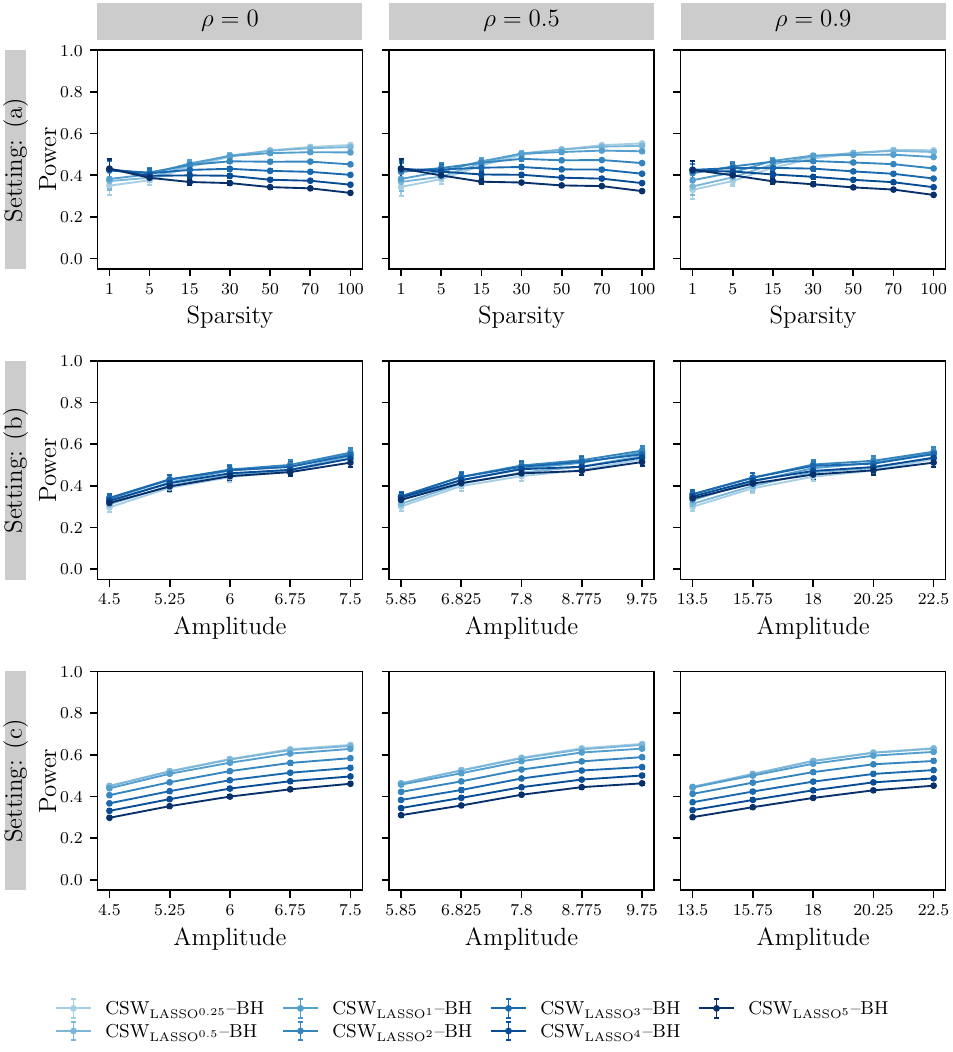}
    \caption{The empirical power of exponentiated versions of $\CSWLASSOBH$ under continuous signal generation, $n/d = 1.5$, $d = 100$ and simulation regimes (a), (b) and (c). The error bars represent plus or minus two standard errors.}
    \label{fig:appendix_0_normal_power_fdr_powered}
\end{figure}

\begin{figure}[h]
    \centering
    \includegraphics{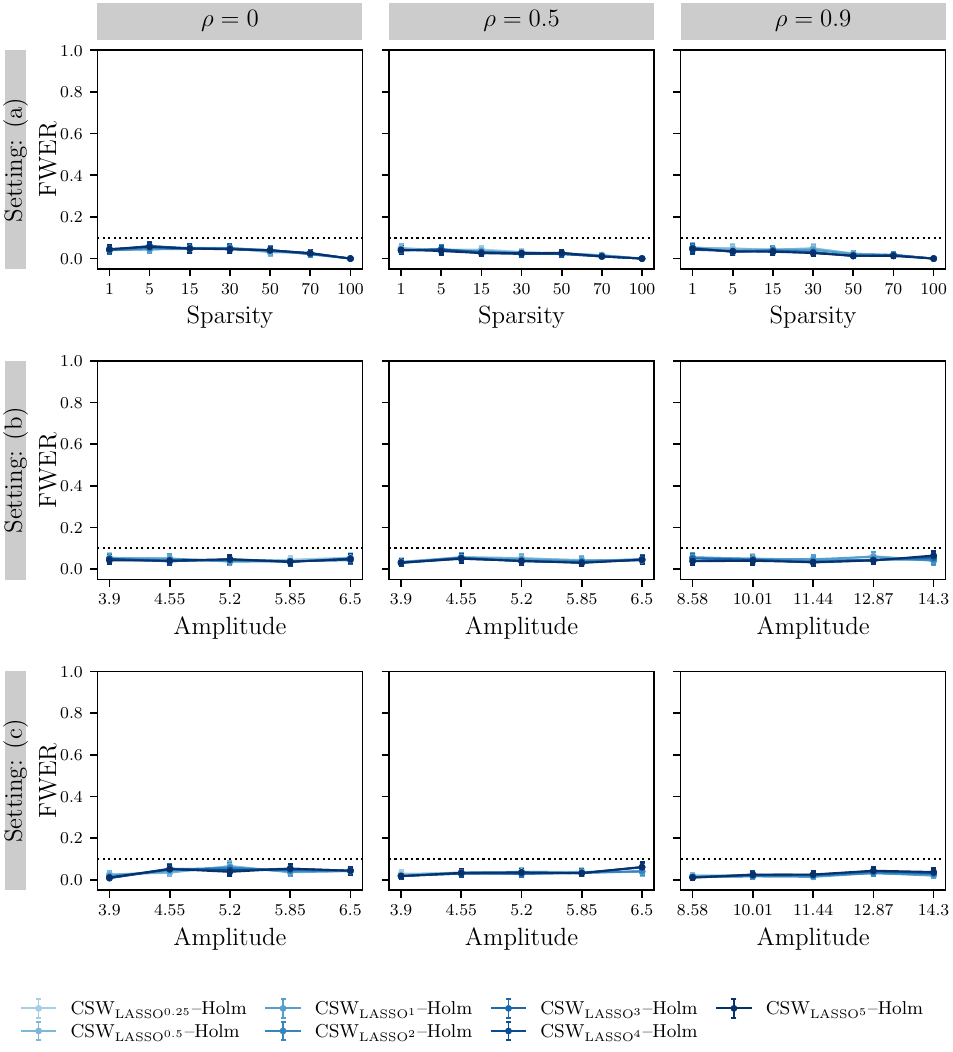}
    \caption{The empirical FWER of exponentiated versions of $\CSWLASSOHolm$ under discrete signal generation, $n/d = 3$, $d = 100$ and simulation regimes (a), (b) and (c). The error bars represent plus or minus two standard errors.}
    \label{fig:appendix_1_discrete_error_fwer_powered}
\end{figure}

\begin{figure}[h]
    \centering
    \includegraphics{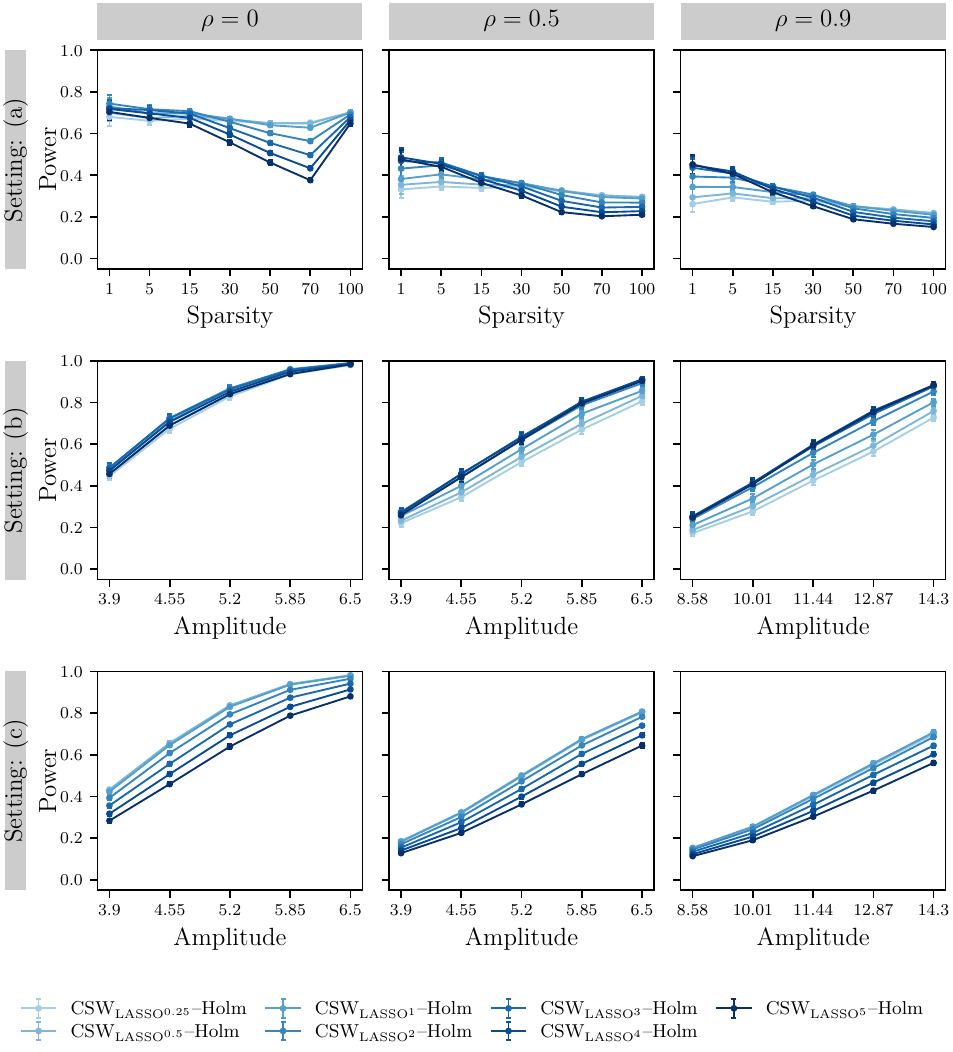}
    \caption{The empirical power of exponentiated versions of $\CSWLASSOHolm$ under discrete signal generation, $n/d = 3$, $d = 100$ and simulation regimes (a), (b) and (c). The error bars represent plus or minus two standard errors.}
    \label{fig:appendix_1_discrete_power_fwer_powered}
\end{figure}

\begin{figure}[h]
    \centering
    \includegraphics{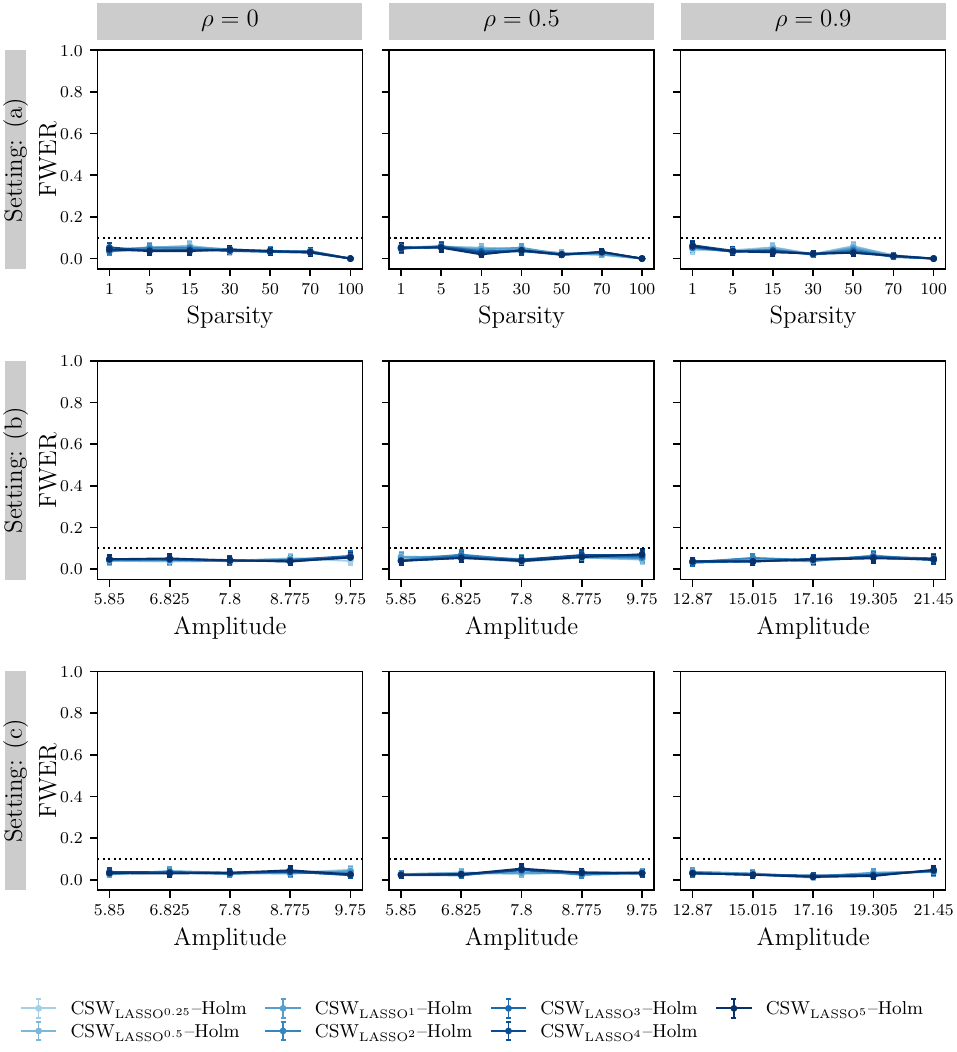}
    \caption{The empirical FWER of exponentiated versions of $\CSWLASSOHolm$ under continuous signal generation, $n/d = 3$, $d = 100$ and simulation regimes (a), (b) and (c). The error bars represent plus or minus two standard errors.}
    \label{fig:appendix_1_normal_error_fwer_powered}
\end{figure}

\begin{figure}[h]
    \centering
    \includegraphics{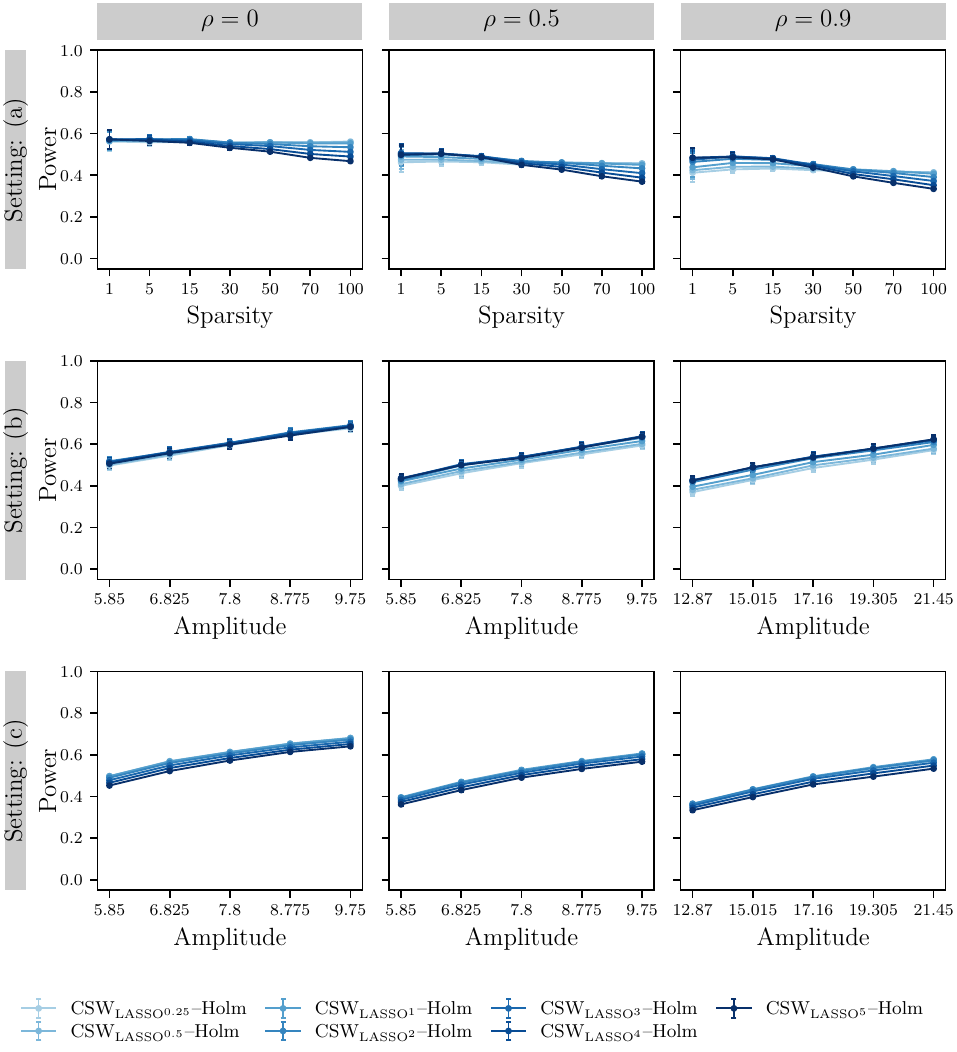}
    \caption{The empirical power of exponentiated versions of $\CSWLASSOHolm$ under continuous signal generation, $n/d = 3$, $d = 100$ and simulation regimes (a), (b) and (c). The error bars represent plus or minus two standard errors.}
    \label{fig:appendix_1_normal_power_fwer_powered}
\end{figure}

\begin{figure}[h]
    \centering
    \includegraphics{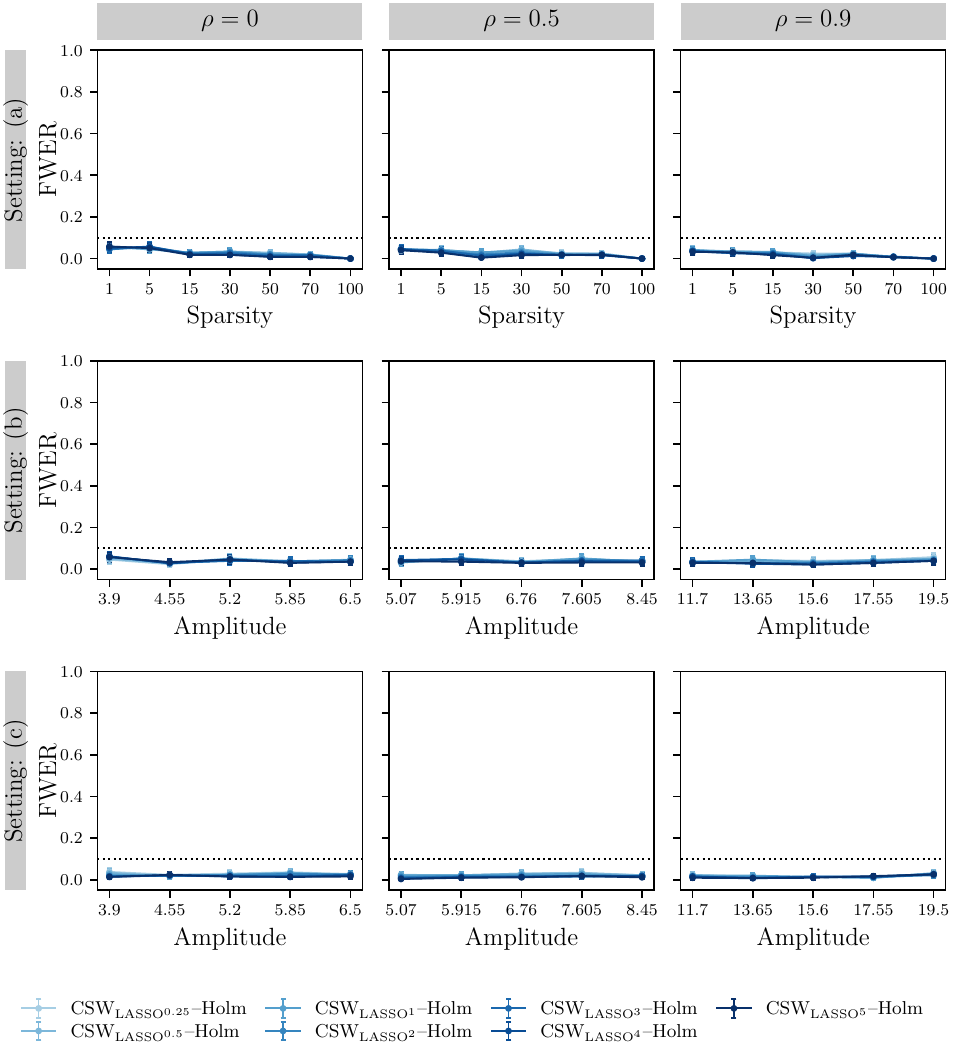}
    \caption{The empirical FWER of exponentiated versions of $\CSWLASSOHolm$ under discrete signal generation, $n/d = 1.5$, $d = 100$ and simulation regimes (a), (b) and (c). The error bars represent plus or minus two standard errors.}
    \label{fig:appendix_0_discrete_error_fwer_powered}
\end{figure}

\begin{figure}[h]
    \centering
    \includegraphics{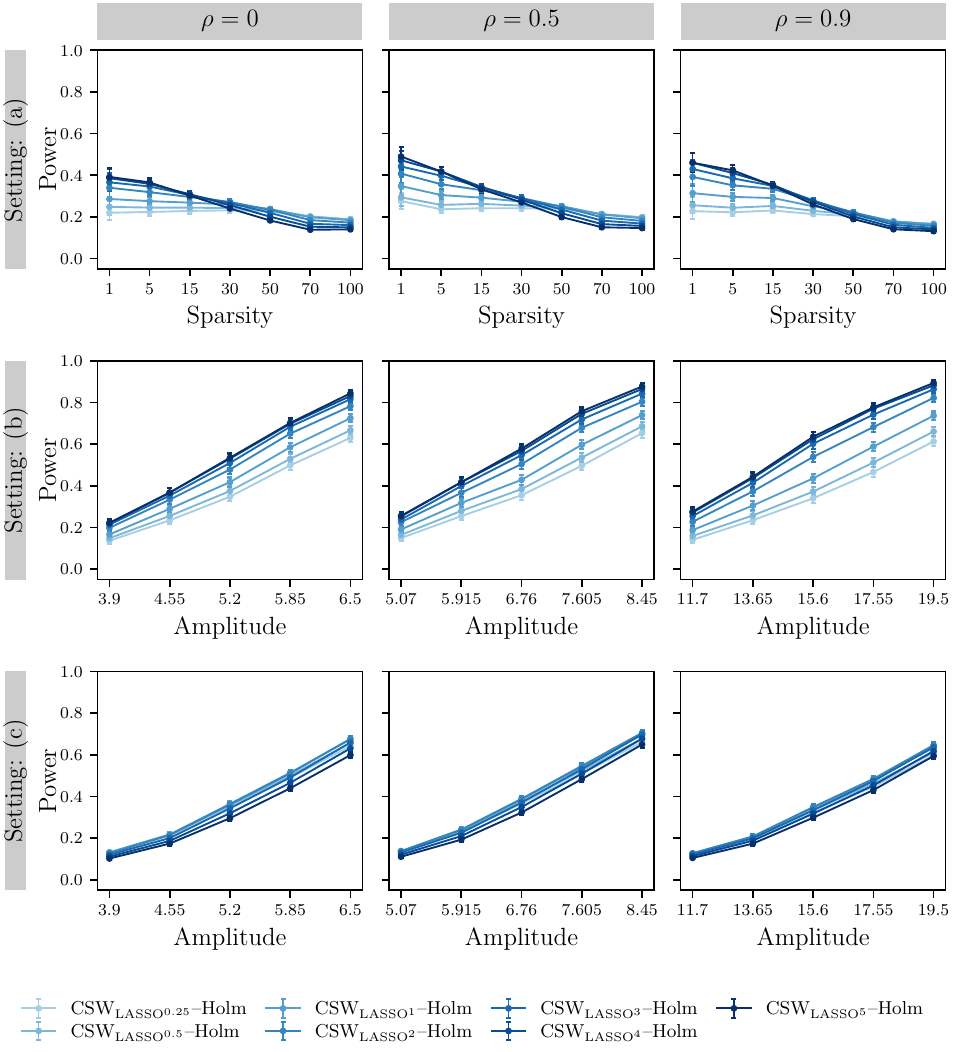}
    \caption{The empirical power of exponentiated versions of $\CSWLASSOHolm$ under discrete signal generation, $n/d = 1.5$, $d = 100$ and simulation regimes (a), (b) and (c). The error bars represent plus or minus two standard errors.}
    \label{fig:appendix_0_discrete_power_fwer_powered}
\end{figure}

\begin{figure}[h]
    \centering
    \includegraphics{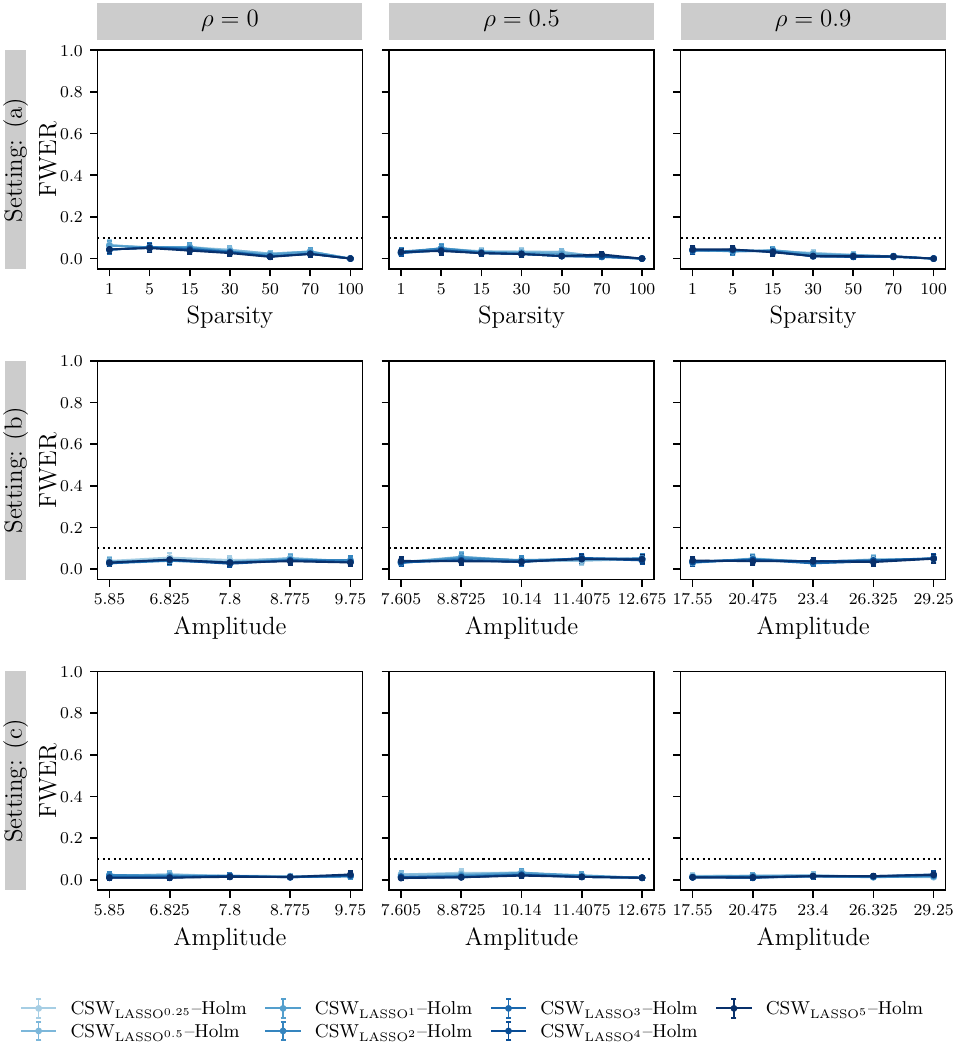}
    \caption{The empirical FWER of exponentiated versions of $\CSWLASSOHolm$ under continuous signal generation, $n/d = 1.5$, $d = 100$ and simulation regimes (a), (b) and (c). The error bars represent plus or minus two standard errors.}
    \label{fig:appendix_0_normal_error_fwer_powered}
\end{figure}

\begin{figure}[h]
    \centering
    \includegraphics{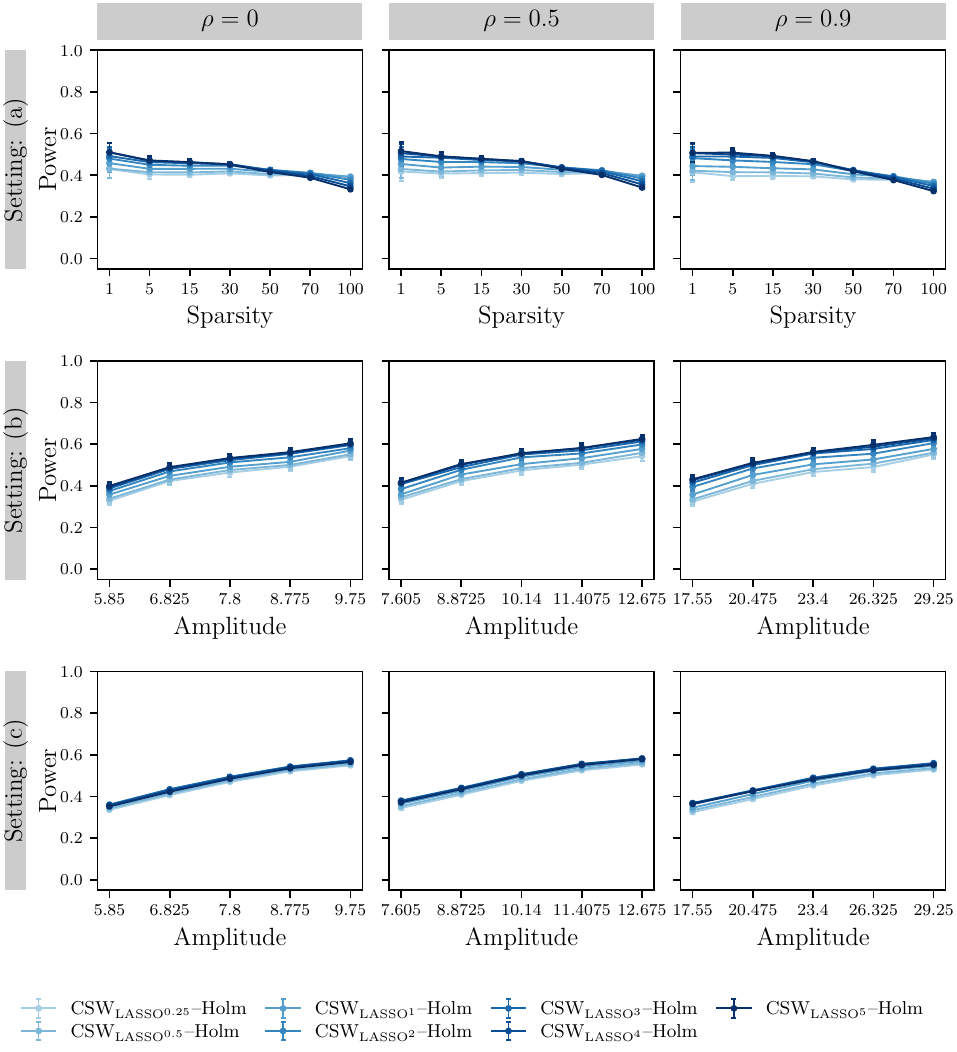}
    \caption{The empirical power of exponentiated versions of $\CSWLASSOHolm$ under continuous signal generation, $n/d = 1.5$, $d = 100$ and simulation regimes (a), (b) and (c). The error bars represent plus or minus two standard errors.}
    \label{fig:appendix_0_normal_power_fwer_powered}
\end{figure}

Analogous to Figures~\ref{fig:appendix_1_discrete_error_fdr} to \ref{fig:appendix_0_normal_power_fwer}, Figures~\ref{fig:appendix_1_discrete_error_fdr_powered} to \ref{fig:appendix_0_normal_power_fwer_powered} present the performance of $\CSWLASSOBH[m]$ and $\CSWLASSOHolm[m]$ for various values of $m$ in the $d=100$ simulations. As discussed in Section~\ref{sec:powered_methods}, these procedures use $\left(\mathcal A^{\rm LASSO}\right)^m$ as the variable importance measure.

First, all exponentiated versions continue to exhibit empirical FDR control across every setting considered, while their FWER control is retained theoretically. Next, these figures reveal a general trend. In sparse settings, larger values of $m$ tend to perform better, whereas in denser settings the trend reverses and smaller values of $m$ take the lead. Moreover, in the sparser regimes, the gap between procedures with larger and smaller $m$ appears to widen as the inter-variable correlation increases. Finally, as with the simulations in Section~\ref{sec:detailed_simulation_results}, the gaps between the different curves narrow under continuous signal generation.

% Consistent with our discussion in Section~\ref{sec:powered_methods}, these simulations demonstrate that alternative choices of variable importance measures, such as the exponentiated measures considered here, can improve the performance of the base CSW procedure in different settings.

% \subsection{High-dimensional simulations}
% \label{sec:detailed_simulation_high}
\end{document}